\pdfoutput=1
\documentclass{ws-ijmpd}
\usepackage[super,compress]{cite}

\usepackage{mathptmx}       % selects Times Roman as basic font
\usepackage{helvet}         % selects Helvetica as sans-serif font
\usepackage{courier}        % selects Courier as typewriter font
\usepackage{type1cm}        % activate if the above 3 fonts are
\usepackage{makeidx}         % allows index generation
\usepackage{graphicx}        % standard
\usepackage{multicol}
\usepackage[bottom]{footmisc}

\begin{document}

\newcommand{\pasj}{PASJ}
\newcommand{\aap}{A\&A}
\newcommand{\mnras}{MNRAS}
\newcommand{\apj}{ApJ}
\newcommand{\apjl}{ApJ}
\newcommand{\app}{APh}
\newcommand{\memsai}{MmSAIt}
\newcommand{\ssr}{Space Science Reviews}
\newcommand{\nat}{Nature}
\newcommand{\aapr}{Astronomy \& Astrophysics Rev.}
\newcommand{\araa}{Annual Rev. of Astronomy \& Astrophysics}
\newcommand{\apss}{Astrophysics and Space Science}
\newcommand{\jcap}{Journal of Cosmology and Astroparticle Physics}
\newcommand{\prd}{Physical Review D}
\newcommand{\physrep}{Physics Reports}
\newcommand{\na}{New Astronomy}
\newcommand{\aj}{Astronomical Journal}

\def\gtsim{\raise 2pt \hbox{$>$} \kern-1.1em \lower 4pt \hbox {$\sim$}}

\markboth{G. Brunetti \& T. W. Jones}{Cosmic rays \& nonthermal
emission in galaxy clusters}

\catchline{}{}{}{}{}
%
%%%%%%%%%%%%%%%%%%%%%%%%%%%%%%%%%%%%%%%%%%%%%%%%%%%%%%%%%%%%%%%%%%%%

\title{COSMIC RAYS IN GALAXY CLUSTERS AND THEIR NON-THERMAL EMISSION}

%\titlerunning{Cosmic rays in clusters}
\author{GIANFRANCO BRUNETTI}

\address{IRA-INAF, via P. Gobetti 101, 40129 Bologna, Italy \\
brunetti@ira.inaf.it}

\author{THOMAS W. JONES}

\address{School of Physics and Astronomy, University of Minnesota,
116 Church St SE, Minneapolis, MN 55455, USA \\
twj@msi.umn.edu}

\maketitle

\begin{history}
\received{Day Month Year}
\revised{Day Month Year}
\end{history}

\begin{abstract}
Radio observations prove the existence of relativistic particles and
magnetic field associated with the
intra-cluster-medium (ICM) through the presence of extended synchrotron emission in the
form of radio halos and peripheral relics.
This observational evidence has fundamental implications on the
physics of the ICM. Non-thermal
components in galaxy clusters are indeed unique probes of very
energetic processes operating within
clusters that drain gravitational and electromagnetic energy into
cosmic rays and magnetic fields. These
components strongly affect the (micro-)physical properties of the ICM,
including viscosity and electrical
conductivities, and have also potential consequences on the evolution
of clusters themselves.
The nature and properties of cosmic rays in galaxy clusters, including
the origin of the observed radio
emission on cluster-scales, have triggered an active theoretical
debate in the last decade. Only recently
we can start addressing some of the most important questions in this
field, thanks to recent observational
advances, both in the radio and at high energies. The properties of
cosmic rays and of cluster
non-thermal emissions depend on the dynamical state of the ICM,
the efficiency of particle
acceleration mechanisms in the ICM and on
the dynamics of these cosmic rays. In this review we discuss in some detail the
acceleration and transport of cosmic
rays in galaxy clusters and the most relevant observational milestones
that have provided important
steps on our understanding of this physics. Finally, looking forward to the 
possibilities from new generations
of observational tools, we focus on what appear to be
the most important prospects for the near future from radio and
high-energy observations.
\end{abstract}


\ccode{PACS numbers: 95.30.Cq; 95.30.Gv; 95.30.Qd; 98.65.Cw; 98.65.Fz;
98.65.Hb}

\section{Introduction}
\label{sec:1}

Clusters of galaxies and the filaments that connect them are the largest 
structures in the present universe in which the gravitational force 
due to the matter overdensity overcomes the expansion of the
universe. 
Massive clusters have typical total masses of the order of 
$10^{15} M_{\odot}$, mostly in the form of dark matter ($\sim 70-80\%$ of
the total mass), while baryonic matter is in the form of galaxies
($\sim {\rm few} \%$) and especially in the form of a hot ($T \sim 10^8 K$)
and tenuous ($n_{gas} \sim 10^{-1}-10^{-4} cm^{-3}$) gas ($\sim 15-20\%$),
the intra-cluster-medium (ICM)\cite{sarazinbook, borgani} (Figure 1). 
That ICM  emits thermal X-rays,
mostly via bremsstrahlung radiation, and also Compton-scatters the
photons of the cosmic microwave background, leaving an  
imprint in the mm-wavelengths band that provides information
complementary to the X-rays\cite{carlstrom02} (Figure 1).
In the current paradigm of structure formation, clusters are thought
to form via a hierarchical
sequence of mergers and accretion of smaller systems driven by 
dark matter that dominates the gravitational field.
Mergers, the most energetic phenomena since the Big Bang, 
dissipate up to $10^{63}-10^{64}$ergs during one cluster crossing time ($\sim$ Gyr).
This energy is dissipated primarily at shocks into heating of
the gas to high temperature, but also through large-scale ICM
motions\cite{borgani, normanbryan99, ryuetal08}.
Galaxy clusters are therefore veritable crossroads of cosmology and
astrophysics; on one hand they probe the physics that governs the 
dynamics of the large-scale structure in the Universe, while on the other 
hand they are laboratories to study the processes of dissipation 
of the gravitational energy at smaller scales.
In particular, a fraction of the energy that is dissipated during the hierarchical
sequence of matter accretion can be channeled into non-thermal plasma
components, i.e., relativistic particles (cosmic rays, or ``CRs'') and magnetic fields in the ICM.
Relativistic particles in the ICM are the main subject of our review.

The evidence for non-thermal particles in the ICM is 
routinely obtained from a variety of radio observations 
that detect diffuse synchrotron radiation from the ICM (Figure 1).
They also open fundamental
questions of their origins as well as their impact on both the physics of the ICM and
the evolution of galaxy clusters more broadly\cite{kaastrabook}.
Cosmological shock waves and turbulence
driven in the ICM during the process of
hierarchical cluster formation 
are obvious potential accelerators of cosmic ray
electrons (CRe) and protons (hadrons or CRp) 
\cite{norman95,sarazin99, loebwaxman00,ryu03, cassanobrunetti05,
brunetti08}.
In addition, clusters host other accelerators of CRs, ranging
from ordinary galaxies (especially as a byproduct of star formation) 
to active galaxies (AGN) \cite{voelk96, berezinsky97, ensslin97, blasicolafrancesco99}
and, potentially, regions of magnetic reconnection
\cite{lazarianbrunetti11}.
The long lifetimes of CRp (and/or nuclei) against energy losses
in the ICM and  their
likely slow diffusive propagation through the disordered ICM magnetic field, 
together with the large size of galaxy clusters, make clusters  
efficient storehouses for the hadronic component of CRs produced within 
their volume or within the individual subunits that later merged to make
each cluster \cite{voelk96, berezinsky97, ensslin97}. 
The consequent accumulation of CRs inside clusters occurs 
over cosmological times, with the potential implication
that a non-negligible amount of the ICM energy could be
in the form of relativistic, non-thermal particles.
An important result of trapped CRp above a few hundred MeV kinetic energy is that 
they will necessarily produce secondary
pions (and their decay products, including e$^{\pm}$ and $\gamma$-rays) through inelastic
collisions with thermal target-protons. Consequently, they 
can be traced and/or constrained by secondary-particle-generated radio 
and $\gamma$-ray
emission\cite{voelk96, berezinsky97, 
colafrancescoblasi98, blasicolafrancesco99, voelkatoyan99, miniati01, 
pfrommerensslin04, blasi07}.

\begin{figure}[ht!]
\centering
\includegraphics[width=130mm]{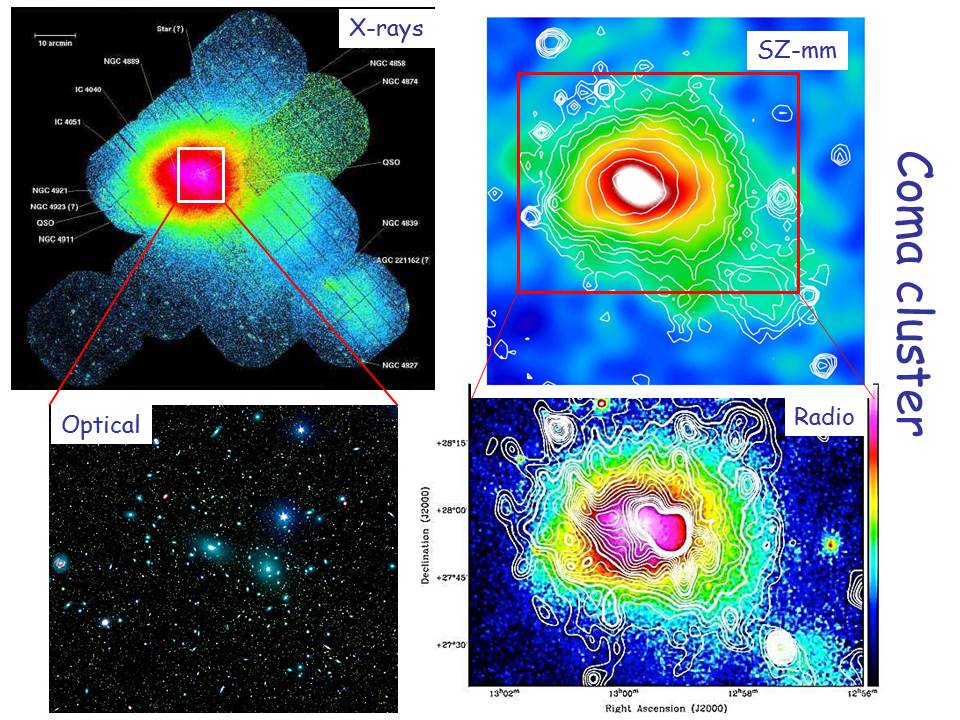}
\caption{
Multi-frequency view of the Coma cluster: the thermal ICM emitting in the
X-rays (top left, adapted from \cite{briel01}), the overlay
between thermal SZ signal (colors) and X-rays (contours) (top right, 
adapted from \cite{planck11}), the optical emission from
the galaxies in the central region of the
cluster (bottom left, from Sloan Digital Sky Survey, credits 
NASA/JPL-Caltech/ L.Jenkins (GSFC)), and the synchrotron (radio halo)
radio emission (contours) overlaied on the thermal X-rays (colors) 
(bottom right, adapted from \cite{brownrudnick11}).
}%\texttt{sidecapion} 
\label{fig:1}       % Give a unique label
\end{figure}

\noindent
The most direct way to pin-point CRp in galaxy clusters is through the detection of
$\gamma$-ray emission generated by the decay of secondary $\pi^0$ particles.
However to date, despite the advent of the orbiting Fermi-LAT and deep 
observations from
ground-based Cherenkov arrays, no ICM has been 
firmly detected in 
the $\gamma$-rays\cite{reimer03, perkinsetal06, kiuchietal09,
aharonian09a, ackermann10, aleksic12, Fermilat13, huberetal13}.
Cluster $\gamma$-ray upper limits, together with several constraints from complementary 
approaches based on radio observations \cite{reimer04, brunetti07, brown11}
suggest that the energy in the form of CRp is 
less than roughly a percent  
of the thermal energy of the ICM, at least if we consider
the central Mpc--size region.
This result contradicts several optimistic expectations derived in the
last decades, based mostly on estimates for CRp production in 
structure formation shocks,
and poses important constraints on the efficiency of CRp 
acceleration and transport in galaxy clusters.

\noindent
On the other hand, 
the existence of CRe and magnetic fields in the ICM of many clusters
is in fact demonstrated by radio observations.
CRe are indeed
very well traced to the ICM of clusters through their radio 
emission. Cluster-scale ($\sim$ Mpc-scale) diffuse synchrotron 
emission is frequently
found in merging galaxy clusters.
It appears in the form of so-called {\it giant radio halos},
apparently unpolarized synchrotron emission associated with 
the cluster X-ray emitting regions, and 
{\it giant radio relics}, elongated and often
highly polarized synchrotron sources typically seen
in the clusters' peripheral regions  with linear extents sometimes 
exceeding $\sim$ Mpc\cite{ferrari08,
feretti12}.
The locations, polarisation and morphological properties of radio relics suggest a
connection with large scale shocks that cross the ICM during
mergers and that may accelerate locally injected electrons or 
re-accelerate pre-existing energetic
electrons to energies where they emit observable synchrotron
emission\cite{bruggen12, brunetti11}.
Electrons responsible for giant radio halo emissions, on the other hand, 
require virtually cluster-wide generation, as we outline below.
There are good reasons to believe that
radio halos trace gigantic turbulent regions in the ICM, where
relativistic electrons can be re-accelerated through scattering
with MHD turbulence and/or injected by way of inelastic collisions
between trapped CRp and thermal protons \cite{sarazin04, blasi07,
brunetti11}.
As we will discuss in this review, one 
of the most interesting consequences of the present theoretical
scenario for CRe production in clusters is that cluster-scale radio
emission should be more common than presently seen, especially
at lower radio frequencies. Specifically, we may expect to find many more
clusters through radio emission in the frequency range that will
be explored in the next few years
by the new generation of  low frequency radio telescopes, such as
LOFAR\footnote{http://www.lofar.org/}, 
MWA\footnote{http://www.mwatelescope.org/} and 
LWA\footnote{http://www.phys.unm.edu/~lwa/}.
Another important consequence of current ICM CR models is that, 
despite a current dearth of detections, clusters should be
sources of high energy photons at a level that could be 
detectable by the next generation of X-ray 
and $\gamma$-ray telescopes. 
Successful, firm detection of galaxy clusters in the hard X-rays and in
$\gamma$-rays
would lead to a fundamental leap forward in our understanding, as it
will provide a unique way to measure the
energy
content of magnetic fields,  
as well as CRe and CRp in cosmic large scale structure.

\noindent
Since it seems likely that
radio halos and relics are signposts of the dissipation of
gravitational energy into non-thermal components and emission,
they can also be used as valuable (indirect)
probes of the merging rate of clusters 
at different cosmic epochs \cite{cassanobrunetti05, nuzaetal12}. 
In this respect the upcoming
radio surveys both at lower and higher radio frequencies, with 
LOFAR, MWA, and ASKAP\footnote{http://www.atnf.csiro.au/projects/askap/},
and on longer time-scales with the SKA\footnote{www.skatelescope.org}, have the
potential to provide important complementary data for cosmological studies.

\noindent
Diffuse synchrotron
radio emission on smaller scales, $\sim 100-300$ kpc,
known as {\it radio mini-halos}, is also found at the centers of
relatively
relaxed clusters with cool cores \cite{ferrari08, feretti12}.
The existence of mini halos indicates that mechanisms other than
major cluster-cluster mergers can power non-thermal emission 
in the ICM. As we will discuss in this review,
also in this case, gravity is likely to provide
the ultimate energy reservoir to power the non-thermal emission, 
potentially extracted, for example, from the sloshing of the gas
in response to motions 
of dark matter cores in and near the cluster.
However, AGNs that are 
usually found at the center of these sources and also 
frequently distributed over larger cluster volumes may be players. In addition, at
the present time any relationships 
between mini and giant halos are still poorly defined.

Starting from this background, the goals of this review are to discuss 
the most relevant aspects of the
origin and physics of CRs and non-thermal emission
in galaxy clusters and to elaborate on the 
present theoretical framework. We will place emphasis on
the most important current observational constraints, along with the 
observational prospects for the near future.
In Sect.2 we will discuss the physics of CRs acceleration by different
sources/mechanisms, whereas in Sect.3 we will discuss the relevant
energy losses and the dynamics of CRs in the ICM.
In Sect.4 we will discuss the most important observational properties
of diffuse radio sources in galaxy clusters, their origin and the main 
prospects for the near future. Specifically giant radio halos, mini
halos and relics are discussed in Sects. 4.2, 4.3, and 4.4, respectively,
whereas in Sect. 4.1 we will briefly discuss current observational 
constraints on the magnetic field in the ICM.
In Sect. 5 we will discuss current observational constraints on 
the high-energy 
emission from galaxy clusters, the most relevant
theoretical aspects and prospects for the near future.
Sect. 6 provides our Summary.

\section{Cosmic ray sources and acceleration}
\label{sec:2}

Consensus has been reached in the past decade that
shocks produced during the hierarchical formation of the large 
scale structure in the universe are likely sources of CRs
in galaxy clusters\cite{norman95, kang96, ryu03}, 
thus implying a direct connection between the 
generation of CRs and the formation and evolution of the hosting
clusters. Similarly, there is consensus 
on the fact that turbulence can be induced
in the ICM as a result of the same processes of clusters formation
and that such turbulence
affects the propagation of CRs, while also providing a potentially important
mechanism for re-acceleration of CRp and CRe \cite{fujita03,
cassanobrunetti05, brunettilazarian07, ensslin11}.

\noindent
Several additional sources can supply (inject) relativistic
particle populations (electrons, hadrons or both) into the ICM.
For instance, particles can be accelerated in 
ordinary galaxies as an outcome of supernovae (SN)
and then expelled into the ICM with a CRp luminosity as high as 
$\sim 3\times 10^{42} \rm erg~s^{-1}$ \cite{voelk96}.
Alternatively, high velocity outflows from AGNs may plausibly contribute up to
$\sim 10^{45}\rm erg~s^{-1}$ in CRs over periods 
of $\sim 10^8$ years \cite{ensslin97}. 

\subsection{Galaxies, Starbursts and Active Nuclei}

Individual normal galaxies are certainly sources of CRs 
as a consequent of current and past star formation. 
Massive clusters of galaxies contain more than a hundred galaxies where
SN and pulsars accelerate CRs.
The efficiency of CR acceleration at these sites is constrained from
complementary observations of Galactic sources. However, 
the amount of CR energy available to the ICM depends also on the way
these CRs are transported from their galactic sources into the ICM. 

\noindent
Voelk et al. 1996 \cite{voelk96}
pioneered the studies of the role of SN explosions, including starbursts,
in cluster galaxies.
The number of SNe experienced by a typical cluster since its
formation epoch, $N_{SN}$, can be estimated from the metal enrichment 
of the ICM,
assuming those metals are released by SNe. This gives 
a total energy budget in the form of CRp :

\begin{equation}
E_{CR}^{SN} = N_{SN} \eta_{CR}^{SN} E_{SN}
\leq
{{ [Fe]_{\odot} X_{cl} M_{cl,gas}}\over
{\delta M_{Fe}}} E_{SN} \eta_{CR}^{SN}
\label{voelk}
\end{equation}

\noindent
where $[Fe]_{\odot} X_{cl} M_{cl,gas}$ is the mass of iron in the ICM
($[Fe]_{\odot} \sim 4/10^5$ is the iron abundance, $X_{cl} \sim 0.35$ 
the typical metallicity measured in galaxy clusters, and 
$M_{cl,gas}$ the baryon mass of the cluster), $\delta M_{Fe}$ is the
iron mass available to the ICM from a single SN explosion, 
$E_{SN} \sim 10^{51}$erg is the SN kinetic energy and 
$\eta_{CR}^{SN}$ is the fraction of SN kinetic energy in the form
of CRp; $\eta_{CR}^{SN}\sim 0.2-0.3$ is constrained from observations
of SN in the Galaxy. Note that the efficiency for
acceleration of CRe in SNRs is apparently several orders of
magnitude smaller than for CRp \cite{jones11, morlinocaprioli12}.
Eq.\ref{voelk} implies a ratio between the CRp and thermal energy
budget in galaxy clusters $E_{CR}^{SN}/E_{gas} \sim 10^{-3}$, assuming
$\delta M_{Fe} \sim 0.1 M_{\odot}$ (appropriate for type II SN) and
$T_{gas} \sim 10^8$K.
This is an optimistic estimate of the expected energy content of
CRp in galaxy clusters from this source, because it does not account for
adiabatic losses in the likely event that CRp from SN are transported
into the ICM by SN-driven galactic winds. 

On the other hand, clusters of galaxies contain AGNs, which, by way of
their synchrotron-emitting jets and radio lobes, are known 
to carry CRe \cite{miley80}. The majority of cool-core clusters contain
central, dominant galaxies that are radio loud\cite{burns90, bestetal07,
mittaletal09}.
The radio lobes of these AGNs  are seen frequently to coincide with
X-ray dark volumes (``cavities'') that have turned out to be
the best calorimeters of the total energy deposited by AGN outflows.
The cavities, being filled with relativistic and
some amount of very hot thermal plasma at substantially lower
density than their surroundings, are poor thermal X-ray emitters.
Such cavities have been seen in something like $1/4$ of the
clusters observed by Chandra \cite{macnamara07}, 
despite the fact that they often exhibit low contrast. 
From an assumption of pressure balance between the cavity and 
the surrounding ICM the cavity energy contents
have been estimated generally in the range $\sim 10^{55} - 10^{61}$erg
\cite{macnamara07, gitti12}.
These approach $\sim 1$\%$E_{gas}$ for an entire ICM in some cases. Dynamical
estimates of cavity lifetimes are typically $\sim 10^{7}-10^{8}$yr, roughly
representing buoyancy timescales. These lead
to AGN power deposition estimates within an order of magnitude of the
X-ray cooling rate of the host cluster \cite{rafferty06}, at least 
while the AGN jets are active.  
The bubble forming duty cycles, estimated from the fraction of
cool-core clusters that harbor clear bubbles, range as high
as 70\% \cite{dunnfabian06}.
Simulations suggest that roughly $1/2$ of
the power of the AGN outflow is
immediately deposited irreversibly as ICM heat through
shocks and entrainment\cite{oneill10}. These
may also drive ICM turbulence that would contribute to CR acceleration
and to the dynamics of CRs (Sects. 2.2.2, 3.2).
Given their large energy inputs, AGN outflows are widely invoked to
account for heating needed to limit the effects of strong radiative 
cooling in cluster cores, e.g.\cite{macnamara07, gitti11, gitti12}.  
So, the total energy deposition into the ICM by AGNs is likely to 
be substantial.

\noindent
However, the energy in CRe and CRp is harder to establish in radio
lobes of cluster radio galaxies.  In a
few cases the absence of observed inverse Compton X-rays has been used to
establish that most of the energy filling the radio lobes must be in some form
other than radiating electrons \cite{hardcastlecroston10}, although CRe energy fractions
as high as 10\% are not ruled out. 
Meaningful CRp energy content estimates in the lobes do
not exist at present, although recent detailed comparisons of
internal-lobe and external pressure suggest that models in which
CRp transported by the jet dominate lobe energetics are
unlikely\cite{crostonhardcastle13}.
Remarkably, various theoretical arguments have been made 
suggesting that much of the direct
energy flux in AGN jets is carried by cold, non-radiating particles or
electromagnetic fields \cite{deyoung06}.
Even if much of the energy filling the cavities is carried by CRs,
it is not yet clear how efficiently those CRs can be dispersed
through diffusion and convective/turbulent mixing over the full
cluster volume, and how much of the energy would remain in CRs, after
accounting for adiabatic and other energy losses. Large scale magnetic
fields, for instance, can help confine lobe contents
\cite{dursipfrommer08, dongstone09, oneilletal09}.
A connected problem that will be discussed in Sect.4.3 
is the possible role of relativistic outflows from the central AGNs 
in the origin of radio mini halos in cool core clusters.

While most discussions of AGN energy deposition in clusters have
focused on central, dominant galaxies, there are other populations 
of AGNs in clusters that could contribute to the CR population, either 
directly or indirectly. Low luminosity
AGNs are quite commonly distributed throughout clusters. 
Stocke et al.\cite{stockeetal09} have argued, in fact,
that virtually all bright red sequence galaxies in rich
clusters are likely to be low-luminosity blazars with 
relativistic jets that could collectively dominate AGN energy inputs
to the ICM. In that case their CR outputs would
be more easily distributed across the cluster by way of ICM turbulence
and large scale ``weather'' (e.g., sloshing). 
In addition, tailed radio galaxies, quite common and widely
distributed in both relaxed and merging clusters, 
show clear evidence of strong interactions with the ICM that includes
entrainment and the generation of turbulence\cite{burns98,
hardcastlesakelliou04}.

\begin{figure}[ht!]
\centering
\includegraphics[width=130mm]{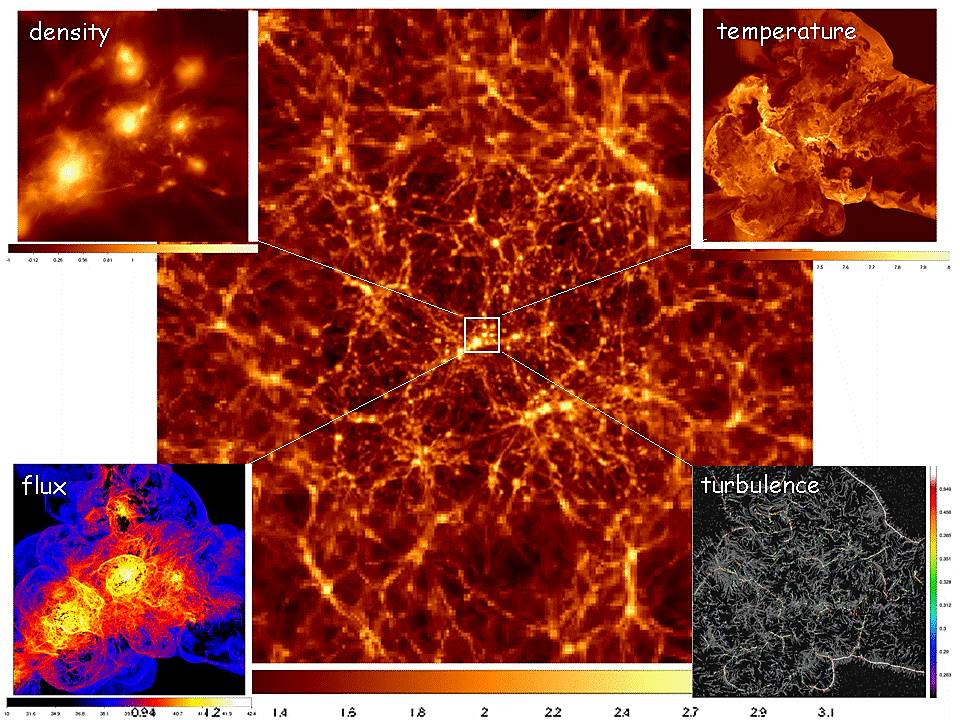}
\caption{
Projection of matter density for a volume 
of size 187 Mpc/h simulated with ENZO AMR (velocity and
density refinement technique) with peak spatial resolution 25 kpc/h.
The two panels on the left are 8x8x2 Mpc/h zooms of the central region
showing the projected density (top-left) and the kinetic energy-flux
(bottom-left). The two panels on the right are 8x8x0.025 Mpc/h zooms
of the same region showing the temperature distribution (top-right)
and the overlay of shock map and 
turbulent velocity-vectors (obtained with a filtering of laminar
motions on 300 kpc scale).
Images are obtained at $z=0.6$ from simulations 
presented in \cite{vazza10}.}
%\texttt{sidecapion}
\label{fig:1}       % Give a unique label % Also need a proper reference
\end{figure}

\subsection{Particle acceleration in the ICM}

The current prevailing view is that the 
process of structure formation may contribute directly 
and indirectly most of the
energetics of non-thermal components (CRs, magnetic fields and turbulence) 
in galaxy clusters 
\cite{sarazin99, roettiger99, 
takizawa00, miniati01, dolag02, ryu03, gabiciblasi03, fujita03, 
cassanobrunetti05, pfrommer06, blasi07, brunettilazarian07, 
ryuetal08, skillman08, vazza09b, cassano10a, bruggen12, vazza12CR}.
Mergers between two or more clusters are observed to heat
clusters through shocks 
\cite{markevitch01, markevitch10}.
Although more difficult to constrain observationally, 
the additional process of
semi-continuous accretion of material onto clusters, 
especially from colder filaments,
is expected to drive quasi-stationary, 
strong shocks and turbulent flows
at Mpc distances from cluster centers that should impact 
on the ICM physics and acceleration of CRs over wide volumes.

Particle acceleration during mergers should occur at shock
waves that are driven to cross the ICM. Particle acceleration 
is also expected to result from several mechanisms 
that may operate within turbulent regions also driven in the ICM during 
these mergers (e.g., turbulent acceleration and magnetic reconnection, 
etc)\cite{lazarianbrunetti11}.
The intricate pattern of 
shocks and large-scale turbulent motions and their interplay 
is still difficult to establish
observationally, but
can be traced in some detail by cosmological 
simulations of galaxy cluster formation.
Fig.2 provides a view of the complex dynamics of the ICM as seen
in simulations.  In particular, a complex pattern of strong and weak shocks is
naturally driven in the ICM, 
largely by gravity variations reflecting dark matter dynamics (Sect. 2.2.1).
This shock distribution, where most of the kinetic energy flux
is dissipated within clusters, is morphologically correlated
to some extent with the distribution of the turbulent motions
in the ICM, that are, indeed, partly driven by those shocks (Sect. 2.2.2).
In the following we will focus on the physics of shocks and particle
acceleration at shocks (Sect. 2.2.1), and on the physics of
turbulence and turbulent
acceleration in galaxy clusters (Sect. 2.2.2).

\begin{figure}[ht!]
\centering
\includegraphics[width=130mm]{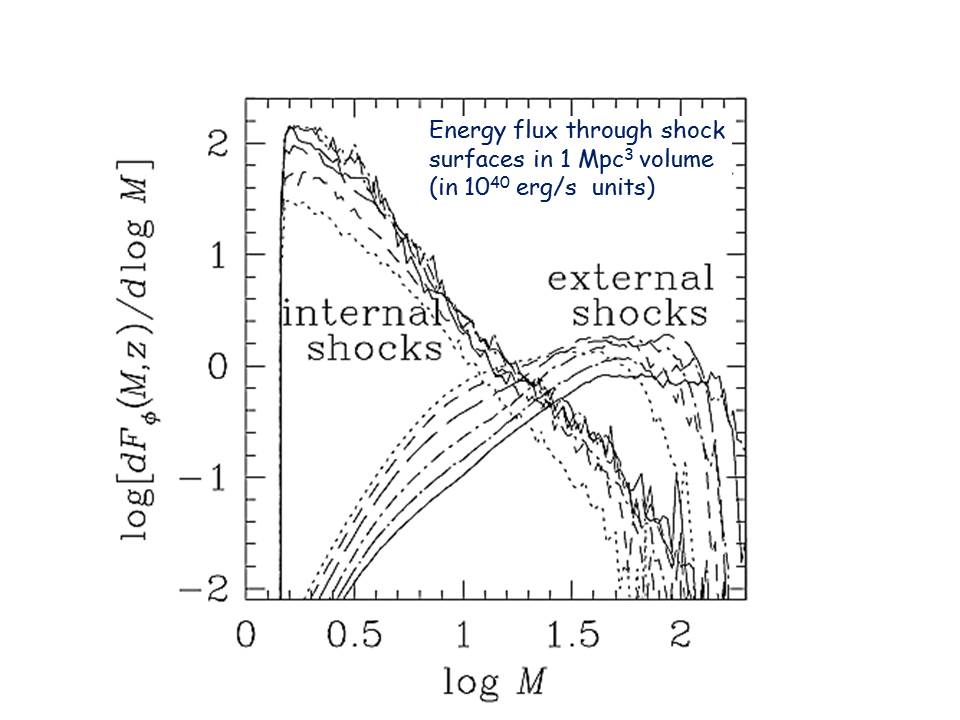}
\caption{
Distribution of the energy flux at shocks surfaces as a function
of the shock-Mach number from numerical (cosmological) simulations
(adapted from \cite{ryu03}).
Units are in $10^{40}$ erg$/$s $(1+z)^{3} h^3$Mpc$^{-3}$.
Shocks are divided into internal 
and external categories. External shocks are defined as shocks forming
when never-shocked, low-density, gas accreted onto nonlinear
structures, such as filaments etc.
Internal shocks form within the regions bounded by external shocks.
}
%\texttt{sidecapion}
\label{fig:1}       % Give a unique label
\end{figure}

\subsubsection{Shocks in galaxy clusters as CRs accelerators}

The total gravitational energy dissipated by baryonic matter in a merger 
of two clusters with roughly equal mass,
$M=10^{15}M_\odot$, is $E\approx 10^{64}$ erg. 
With the assumption that
the gaseous components of the initial clusters are at the associated virial
temperatures, it is easy to show that the merging components
approach each other at
slightly 
supersonic relative speed, therefore implying the formation of
weak, $M \sim 2$, shock waves \cite{sarazin99,gabiciblasi03,berrington03}. 
Those shocks will typically strengthen moderately as they emerge
into lower density and low temperature regions outside the cluster cores.
Additional, remote, accretion shocks that result from the
continuous accretion of matter at several Mpc-distances from
clusters center (also called ``external shocks'' when they
form due to the accretion of never-shocked gas), 
are typically much stronger; that is, 
they have higher Mach numbers,
since they develop in cold, un-virialised
external cluster regions. On the other hand, since gas densities
are also quite low in those environments, the energy available for
dissipation through such shocks
is relatively smaller than through lower Mach number shocks
that dissipate energy in higher density regions closer to cluster 
centers during mergers. 
Simple (analytical or semi-analytical) but accurate
estimates of the amount of kinetic energy associated with
accretion/external shocks are very challenging.
A leap forward in understanding in this area, however, has been 
achieved in the last decade 
through extensive cosmological simulations that allow one to study the 
formation of shocks in clusters, from their outskirts to more internal
regions with increasing detail\cite{ryu03, pfrommer06, skillman08,vazza09b, 
vazza10, kangetal07, vazzacomparison11}.

\noindent
Figure 3 illustrates these points. It shows the kinetic  
energy flux through shock surfaces, 
that is, $1/2 \rho V_{sh}^3 S$, measured in clusters formed 
during cosmological simulations. Here, $\rho$ is the upstream gas density,
while $V_{sh}$ and $S$ are shock velocity and surface area, respectively.
The distribution of energy fluxes shows that most of the 
gravitational energy is dissipated at relatively 
weak, ``internal'' (merger) shocks, with
Mach number $M \sim 2-3$. A modest fraction 
of the kinetic energy-flux passes through related stronger internal-shocks 
developed during merger activity as they propagate outwards 
after merging cores have their closest approach.
Figure 3 shows that only a few \% of the energy flux is 
dissipated at strong, ``external'' shocks.

\noindent
If even a small percentage of the merger and accretion
shock-dissipated energy can be converted
into non-thermal particles through a first order Fermi
process, then the ICM could be populated with an energetically significant
population of non-thermal, CR particles \cite{miniati01gamma, ryu03,
blasi07, kangetal07, pfrommer07, vazza12CR}. 
This prospect
has drawn considerable focus to potential consequences of shock generated
CRs and the refinement of early estimates of CR production in ICMs.

\

\noindent
{\it 2.2.1.1. \, Shock acceleration of CRs}
\vskip 2mm

The acceleration of CRs at shocks is customarily described 
according to the diffusive shock acceleration (DSA) theory 
\cite{bell78, drury83, blandford87, jones91}.
In effect diffusing particles are temporarily trapped 
in a converging flow across
the shock if their scattering lengths across 
the shock are finite but much greater
than the shock thickness.  Particles escape eventually
by convection downstream. Until they do, they gain energy each time they are
reflected upstream across the shock,
with a rate determined by the velocity change they encounter across the 
shock discontinuity and a competition between convection and 
diffusion on both sides of the shock.
The hardness (flatness) of the resulting spectrum reflects the
balance between energy gain and escape rates. In other words, it depends on  
the energy gain in each shock crossing combined with the
probability that particles remain trapped long enough to reach
high energies.
Mathematically this balance can be conveniently described through the
diffusion-convection equation for a pitch angle averaged CRs distribution
function $f(p,t)$ in a compressible 
flow\footnote{$f$ is the number of CRs per unit
phase-space volume, $d^3p dV$} that is \cite{blandford87}:

\begin{equation}
{{\partial f}\over{\partial t}} + 
( {\bf V} \cdot \nabla ) f 
-
\nabla \cdot \big \{ {\bf n} \, D \, ( {\bf n}\cdot \nabla ) f \big \}
=
{ 1 \over 3} (\nabla \cdot {\bf V} ) p {{\partial f}\over{\partial p}}
\label{eq:equationshock}
\end{equation}

\noindent
where $p$ is the modulus of the particle's momentum,
${\bf V}$ is the velocity of the background medium
(assuming $c >> V >> V_A$, with $V_A$ the Alfv\'en velocity), while
${\bf n}$ is the unit vector parallel to the local magnetic field,
and $D$ is the particle spatial diffusion 
coefficient  (see Sect. 2.2.2). 
The 2nd and 3rd terms 
account for convection and diffusion, respectively, while the right hand
side takes account of the adiabatic energy gains (losses) suffered by 
particles in a converging (expanding) flow. 
As written, Eq.\ref{eq:equationshock} omits 
non adiabatic losses, such as from radiation,
that can be important especially for CRe (CRs energy losses are
discussed in Sect. 3.1), 
momentum diffusion and effects such as CR
energy transfer to wave amplification/dissipation (turbulent-CRs
coupling is discussed in Sect. 2.2.2).

\begin{figure}[ht!]
\centering
\includegraphics[width=130mm]{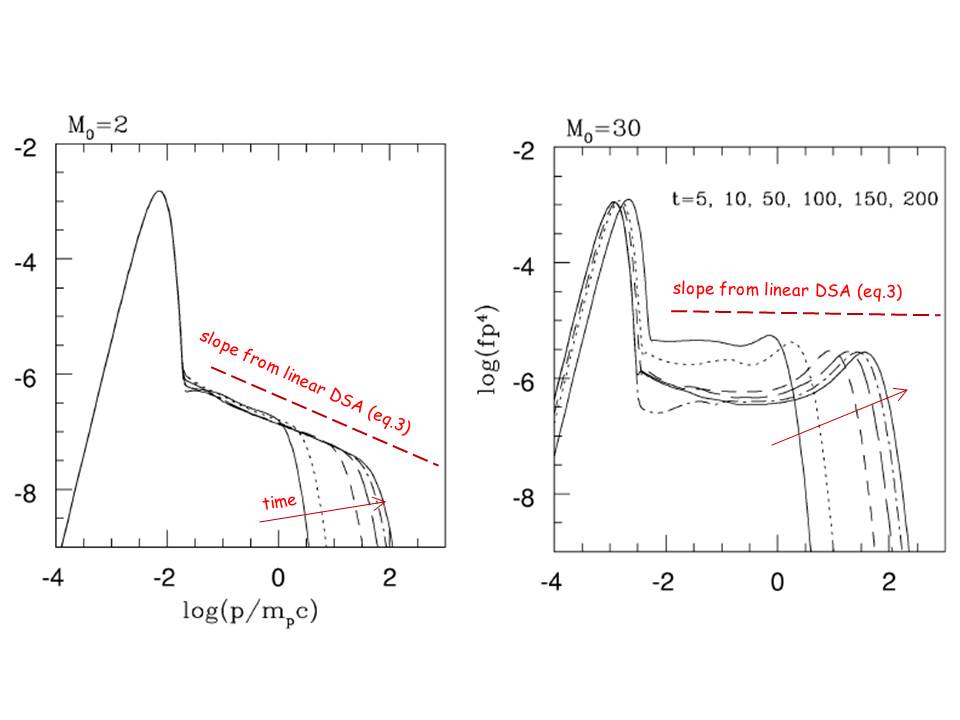}
\caption{Time evolution of the spectrum of protons
accelerated from a Maxwellian upstream distribution at shocks with Mach
number 2 (left) and 30 (right) (adapted from \cite{kangjones02}).
The spectral slope predicted by (test-particle) DSA is
represented as a red-dashed line.}
%\texttt{sidecapion}
\label{fig:1}       % Give a unique label
\end{figure}

\noindent
Under these conditions, if all particles are injected at low energies and 
``see'' the same velocity change across the shock, the steady state 
spectrum of test-particle 
CRs at a
plane shock is a power law
in momentum, $f(p)= K p^{-(\delta_{inj} +2)}$, where the slope is

\begin{equation}
\delta_{inj} =
2 {{M^2 +1}\over{M^2 -1}}.
\label{test_slope}
\end{equation}

\noindent
$M = V_{sh} / c_s$ is the Mach number of the shock.
For strong shocks, $M \rightarrow \infty$, this slope tends to 
$\delta_{inj} \rightarrow 2$.
Thus, in the strong shock limit, the energy and pressure in the resulting
CRs are broadly distributed towards the highest energies that are achieved.
On the other hand, for weak shocks, 
$M^2 \approx 1 + \epsilon$ with $\epsilon \ll 1$,
this tends to $\delta_{inj} \approx 2 + 4/\epsilon \gg 2$.
In this case the fractional velocity jump across the shock
is small, so the energy in
CRs accelerated from suprathermal values is concentrated in the lowest energy
CRs. That is, the CRs gain relatively little energy before they 
escape downstream. Consequently, for the same number of CRs and 
the same kinetic energy
flux through the shock, $\sim \rho V_{sh}^3$, 
the energy input to locally injected CRs through DSA
is much greater in strong shocks than in weak shocks.

As a consequence of this theory, 
it is apparent that even a modest injection of particles
at a strong shock can lead to a substantial
fraction of the kinetic
energy flux into strong, initially purely hydrodynamical shocks going
into CRs. Those, in turn backreact on and modify the structure
of the shocks themselves.
Under these conditions the process of particle acceleration
is described using nonlinear theory
\cite{drury83,jones91,malkov01,blasi02,
kang02, kang05}.
The main outcome of that development is the
formation of a compressive
precursor to the shock, leading to an increase in the total shock
compression, upstream turbulence and magnetic field amplification,
followed by
an actual weakening of the fluid shock transition
(the so-called `sub-shock'').
In a highly CR-modified shock a large part of the DSA process at
high CR energies actually
takes place in the precursor when the spatial diffusion coefficient, 
$D$ is an increasing function of particles momentum. Then, the subshock 
is responsible mostly for
the acceleration process at low energies \cite{kang09} and injection 
of seed DSA particles.

\noindent
The importance and detailed outcomes 
of nonlinear evolution in {\it strong} shocks depend on the
size of the CR population at the shock, the hardness of the CR
spectrum being accelerated at the shock, the efficiency and distribution 
of turbulent magnetic field amplification
upstream of the shock and the geometry of the shock \cite{kang13}.
These physical details are important, since they
regulate how much energy is extracted from the flow into the shock
and, accordingly how much pressure will develop from these CRs and 
amplified magnetic field within the shock transition.
On the other hand, unless they include much larger total CR populations or interact with
a pre-existing CR population with a hard spectrum,
{\it weak shocks} are minimally affected by nonlinear effects,
because of the steeper CR spectra generated in these shocks.
Fig.4 shows the time evolution of CRp spectra accelerated at simulated weak and
stronger shocks. In the case of weak shocks the spectrum agrees
with the prediction of test particle DSA theory, while the spectrum becomes
concave and 
flatter than test particle DSA for stronger shocks, due to the non-linear
back-reaction of CRp. 
The quality of comparisons between real and theoretical strong, 
DSA-modified shocks
is still an open question (see the discussion at the end of this section).

\noindent
The acceleration time-scale at the shock (i.e. the time necessary
for CRs of energy $E$ to double that energy) depends on the 
time interval between shock crossings for the CR, $\sim 4D/(V_p V_{sh})$,  $V_p$ is the particle
velocity, and on the ratio of
the CR velocity to the fluid velocity
change across the shock, so $\sim V_p/\Delta V_{sh}$. 
Thus, it primarily depends on
the spatial
diffusion coefficient of particles, and inversely on the shock 
velocity, $V_{sh}$, and the compression
through the shock.
In the simplified case that the spatial diffusion coefficient
does not change across the shock,
the mean acceleration time 
to a given momentum in an unmodified shock can be written as :

\begin{equation}
\tau_{acc}(p) \simeq {{4 D(p)}\over{(c_s M)^2}} 
{{M^2 (5M^2 +3)}\over{(M^2+3)(M^2-1)}}
\label{tauaccshock}
\end{equation}
that approaches $\tau_{acc} \approx 20 D/(c_s M)^2$ for strong shocks.

\begin{figure}[ht!]
\centering
\includegraphics[width=130mm]{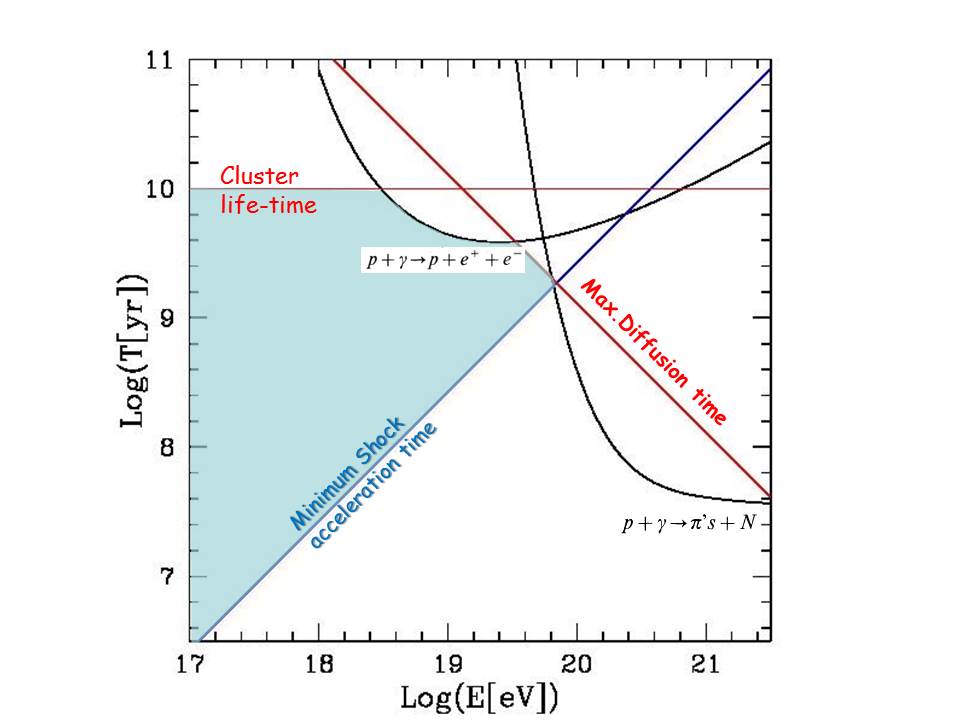}
\caption{The life-time of CRp in the ICM as a function of energy due 
to photo-pion and photo-pair production (solid curves) are
compared to a reference
life-time of clusters $\sim 10$ Gyrs, 
to the maximum diffusion time-scale of CRp (assuming
Bohm-diffusion on 3 Mpc scales and using an optimistic value
$B= 5 \mu$G) and to the minimum acceleration time-scale of
CRp by shocks (assuming an optimistic configuration with
Bohm-diffusion with $B= 5 \mu$G, and a 
shock velocity $= 4000$ km/s).
The allowed region, marking the energies of CRp that can be
obtained via shock acceleration, is highlighted in blue.}
\label{figure}
\end{figure}

\noindent
In order to derive a maximum energy of the accelerated CRs in a given
time interval we
assume a Bohm diffusion coefficient, 
$D(p) = (1/3)r_L(p) c$, (see also Sect 2.2.2)
where, $r_L(p)$, is the particle Larmour radius.  This is optimistically
small, thus giving us an optimistic upper energy bound, 
since it assumes a mean free path equal to the CR gyroradius.  
The spatial diffusion 
coefficient for relativistic particles then becomes in practical terms,

\begin{equation}
D(p) \sim 3 \times 10^{22} {{ (c p /{\rm GeV}) }\over{(B / \mu G)}}
\,\,\,\,\, {\rm cm}^2 {\rm s}^{-1},
\label{eq:bohm}
\end{equation}

\noindent
resulting in an acceleration time scale to GeV energies of the order
of 1 yr, if we assume a typical shock velocity in galaxy clusters,
$V_{sh} = c_s M \sim {\rm few}~10^3$ km/s, and $B \sim 1 \mu$G.
Then, from eqs.4-5, $p_{max} \sim (\tau/yrs)$ GeV, implying for realistic 
available acceleration times ($\gg$ years)
that the power law distribution of the accelerated particles
should extend up to very high energies, where energy losses or diffusion
from the acceleration region quenches the acceleration
process (spatial diffusion in the ICM is discussed in Sect. 3.2). 

\noindent
The energy losses for CRe, due especially to synchrotron and
inverse Compton processes (see Sect. 3), are much more significant 
than those for CRp. For CRe the maximum energy accelerated
at shocks in galaxy clusters can be of the order of several tens of
TeV \cite{blasi01, kang12}.
Once these CRe are advected downstream of the shock, radiative cooling 
due to inverse Compton scattering (ICS) with the CMB
(and if the magnetic fields are strong enough, synchrotron, Sect. 3) 
will steadily reduce the maximum electron energy, so that 
it scales asymptotically as $\gamma_{e,max} \propto 1/x$, where
$x = V_d t$ is the propagation distance downstream from the shock reached over
a time, $t$. This
causes the ``volume integrated'' electron spectrum to steepen by one in
the power law index, $\delta = \delta_{inj} + 1$ above energies reflecting 
this loss over the lifetime of the accelerating shock \cite{ensslin98,
hoeft07}.
Consequently, the ``spatially unresolved'' 
synchrotron spectrum ``from the downstream
region'' will show a steepened spectral slope, 
$\alpha = (\delta - 1)/2= \alpha_{inj} + 1/2$.

\noindent
By contrast, CRp in these shocks are not subjected to significant energy 
losses until
they reach extremely high energies where they suffer inelastic collisions
with CMB photons. CRp with energies above a few hundred PeV will produce
$e^{\pm}$ when they collide with CMB photons, limiting their life-times
to a period below a few Gyr \cite{berezinskygrigoreva88}(Figure 5).  
Most important, at energies above about $5\times 10^{19}$eV, such collisions
produce pions, and each such collision extracts a significant fraction
of the CRp energy (this is the same physics that determines
the so-called
Greisen-Zatsepin-Kuzmin (GZK) cutoff in the ultra-high energy CR
spectrum). For example, for energies above $\sim 10^{20}$eV the 
CRp life-time drops rapidly below $\sim 10^8$ yr (Fig. 5).
In Figure 5 we compare 
these relevant time-scales with the shock-acceleration
time-scale (eqs. 4-5) and with the life-time of shocks and clusters
themselves. It follows that CRp acceleration at cluster shocks 
can reach at most maximum energies of a few $10^{19}$ eV\cite{kang96, blasi01}.
We note that this is an {\it optimistic} estimate, as we are
implicitly assuming that such high-energy 
CRp are still effectively confined in galaxy
clusters\footnote{e.g., confinement of CRp with energies $10^{19}-10^{20}$eV
for a Hubble time would require conditions comparable to 
Bohm diffusion at distances of several 100 kpc from the shock} 
(CRs diffusion/confinement is discussed in Sect. 3.2).

\

\noindent{\it 2.2.1.2. \, Shock acceleration in the ICM \& open questions}
\vskip 2mm

Particle acceleration efficiency at strong shocks is becoming
constrained
by studies of SN-driven shocks in our Galaxy \cite{jones11,
morlinocaprioli12}. Those
shocks
transfer $\sim 10$\% or more of the energy flux though them into CRp.
It is important to keep in mind that the shocks mostly responsible
for acceleration of observable Galactic CRs are very strong,
with Mach numbers upwards of $10^3$, and that they are found in 
low beta-plasma, $\beta_{pl}= P_{gas}/P_{B}$,
environments. By contrast, and as discussed
above, the ICM is a high-$\beta_{pl}$ environment and 
most of the kinetic energy
flux penetrating galaxy cluster shocks is associated with much weaker
shocks where, the
acceleration efficiency is probably much less, although still
poorly understood \cite{blasi04, kang05, vink14}.
In this respect galaxy clusters are special environments, as they
are unique laboratories for constraining the physics
of particle acceleration at (Mpc-scale) weak shocks.
It remains an open issue whether these weak shocks can accelerate
CRs in the ICM at meaningful levels. 

A critical, unresolved ingredient in shock acceleration theory is the minimum
momentum of the seed particles that can be accelerated by DSA; i.e.,
the minimum momentum that leads to diffusive particle
transport across the shock. This, along with the detailed processes that
control this minimum momentum are 
crucial in determining the efficiency with which thermal 
particles are injected into the population of CRe and CRp.
Particles
must have momenta at least several times the characteristic
postshock thermal ion momenta in order to be able to successfully
recross into the preshock space. Quasi-thermalized particles
are inherently less likely to recross in the upstream direction
in weak shocks than strong shocks, because of the weaker
dissipation in weak shocks; i.e., the ratio of the postshock thermal
speed to the postshock convection speed is relatively smaller. 
Injection is expected to depend
sensitively on the charge/mass ratio of the injected species, since
that determines the rigidity ($\propto p/q$) of particles at
a given energy. For this reason, nonrelativistic 
electrons appear to be very difficult to inject from the thermal
population (because $p = \sqrt{2 m E}$), so are likely to be far fewer 
than injected protons. Typically
some kind of upstream, pre-injection process, often involving
protons reflected by the shock that generate upstream waves
that can resonate with nonrelativistic electrons, is invoked to enable 
electron
injection at shocks \cite{amato06,burgess06,amano09,spitk11}.
The processes ``selecting'' the particles that can
recross define so-called {\it thermal-leakage injection}. It is important to
realize that they are poorly
understood, especially in the relatively weak shocks with large 
$\beta_{pl}$ expected in cluster media. 
Existing collisionless shock simulations, hybrid and PIC simulations, 
have focused on strong shocks or low beta-plasma
\cite{spitk11, gargate12, capriolispitkovsky13}. 
The orientation of the magnetic field
with respect to the shock normal is also important, since it strongly 
influences the physics of the shock structure \cite{spitk11, parketal12}.

A connected, unresolved ingredient in nonlinear CR shock theory
is the level of amplification of the magnetic
field and its distribution  within the shock due to CR-driven instabilities. 
The evolution of the magnetic field
through the full shock structure is important, since the magnetic
field self--regulates the diffusion process of supra--thermal
particles on both sides of the shock and also affects the 
injection process \cite{bell78b, kang02}.
There are several proposed models to amplify
magnetic fields  significantly within 
the CR-induced shock precursor \cite{bell00, bell04, amato09}; none 
of them applies until some degree of shock modification already takes 
place. 
This means they only apply in strong shocks, so probably not in cluster merger
shocks. Some magnetic field generation 
and/or amplification downstream of curved or intersecting shocks may 
result when the electron density and pressure gradients are 
not parallel (the so-called Biermann Battery effect \cite{biermann50}), 
due to the Weibel filamentation instability \cite{weibel59,
gedalinetal10} or when the downstream 
total plasma and pressure gradients are not parallel (so-called baroclinic 
effects that amplify vorticity \cite{davieswidrow00, ryuetal08}). 
Amplification of turbulence and magnetic fields only downstream, however, 
have minimal impact on DSA, which depends essentially on those properties 
on both sides of the shock.

As a final remark we mention that there is still discussion on the
spectrum of CRs resulting from shock acceleration. 
Although calculations in the last decade agreed on the conclusion that the 
spectra of CRs accelerated at strong shocks are concave (flat) as 
a result of the dynamical
back-reaction of CRs (Figure 4), current data for SNRs 
seem to favour steeper spectra and suggest a 
partial revision of the theory \cite{caprioli12}.  
Specifically, the absence of CR spectral
concavity in modified shocks requires a relative 
reduction in acceleration efficiency or
increased escape probability
for the highest energy CRs\cite{kang13}.

\subsubsection{Turbulence in the ICM and CR reacceleration}

Galaxy clusters contain many potential sources of turbulence.
These include cluster galaxy motions \cite{jaffe77, deiss96},
the interplay between ICM and the outflowing 
relativistic plasma in jets and lobes of AGNs \cite{heinz06,bruggen09},  
and buoyancy instabilities such as the magnetothermal
instability (MTI) in the cluster outskirts \cite{parrish07}.
However, the most important potential source of turbulent motions 
on large scales 
is the process that leads directly to the formation of galaxy 
clusters \cite{roettiger99, ricker01, cassanobrunetti05, subramanian06}.
Mergers between clusters deeply stir and rearrange the cluster
structure. In this case turbulence is expected from core sloshings,
shearing instabilities, 
and especially from the complex patterns of interacting shocks that form
during mergers and structure formation more generally \cite{dolag05, 
iapichinoniemeyer08, vazza09a, vazza10, keshet10, iapichino11, 
paul11, zuhone11, hallman11, vazzaetal11, vazza12, nagaietal13,
miniati13} (Figure 2).
Such a complex ensemble of mechanisms should drive in the
ICM both compressive and incompressive turbulence, as also supported 
by the analysis of numerical simulations \cite{beresnyaketal13, miniati13};
in Figure 6 we report a sketch of the turbulent properties of the ICM.

Large scale turbulent motions that are driven during
cluster-cluster mergers and dark matter sub-halo motions are expected on scales 
comparable
to cluster cores scales, $L_o \sim 100-400$ kpc, and might have typical 
velocities around $V_o \sim 300-700$ km/s, e.g. \cite{subramanian06}.
These motions are sub--sonic, typically with $M_s = V_o/c_s \approx
0.2-0.5$, 
but they are strongly super-Alfv\'enic, with $M_A=V_o/V_A \approx 5-10$,
e.g. \cite{brunettilazarian07}.
This implies a situation in which magnetic field lines in the ICM are 
continuously advected/stretched/tangled on scales larger than the Alfv\'en
scale; that is, the scale where the velocity of turbulent eddies
equals the Alfv\'en speed,
$l_A \sim L_0 (V_0/V_A)^{ {2\over{a-1}} }$ ($a$ is the slope of
the turbulent velocity power-spectrum, $W(k) \propto k^{-a}$).
Below this scale, turbulent, ``Reynolds stresses'' are insufficient to
bend field lines and turbulence becomes MHD (Fig. 6).
Under these conditions the effective particle mean-free-path 
in the ICM should be $l_{mfp} \sim l_A$ rather than the value
of the classical Coulomb ion-ion mean free path, 
$l_C \sim 10-100$ kpc\cite{lazarian06, brunettilazarian07} (Fig.
6)\footnote{the effective mfp should be the smaller of
the two scales, however under typical ICM conditions 
$l_A < l_C$, see\cite{brunettilazarian07}}.

\noindent
However the ICM is a ``weakly collisional'' plasma and can be very 
different from collisional counterparts, because it is subject to various
plasma instabilities, e.g. \cite{schekochihin05, schekochihin10}.
In many cases, as a result of plasma instabilities
the (relatively weak) magnetic field is perturbed on very small scales. 
That provides the potential to strongly reduce the effective
thermal particle (and CRs) mean free path
\cite{levinson92,pistinner96,brunettilazarian11a,yanlazarian11} 
and also the effective viscosity of the fluid, 
below the classical Braginskii viscosity
determined by thermal ion-ion Coulomb collisions.

\noindent
All these considerations about the velocities of large-scale motions and 
the effective particle mean-free-path in the ICM 
allow us to conclude that the effective Reynolds
number in the inner ICM is ${\cal R}e >> 10^3$; that is,
much larger than it would be if it were determined by
the classical ion--ion mean free path (${\cal R}e \sim 100$).
Theoretically this suggests that a 
cascade of turbulence and a turbulent inertial range
could be established from large to smaller scales (Fig. 6).
In addition plasma/kinetic instabilities in the ICM generate waves at small,
resonant, scales (Fig. 6). Among the many types of waves that can be excited
in the ICM we mention the slab/Alfv\'en 
modes that may be excited for example 
via streaming instability\cite{wentzel74} and gyro-kinetic 
instability\cite{yanlazarian11}, and the whistlers that may be excited 
for example via heat-flux driven instabilities\cite{pistinner96,
pistinner97}.
Also coherent wave phenomena in MHD turbulence can 
generate non-linear electrostatic waves; for example
lower hybrid electrostatic waves in the ICM might be excited through
the non-linear modulation of density in large amplitude Alfv\'en
wave-packets\cite{watheralleilek99}.
Both large-scale motions (and their cascading at smaller scales) and the
component of self-excited turbulent waves at small scales have a strong
role in governing the micro-physics of the ICM through the scattering
of particles and the perturbation of the magnetic field.

\noindent
Current X-ray observations do not allow one to derive stringent
constraints on the turbulent motions in
dynamically active (i.e. merging or non cool core)
clusters\cite{churazov12, sandersscience13} (see however the pioneering
attempt by \cite{shuecker04}) (constraints on cool core clusters are
discussed in Sect. 4.3).
This will hopefully change in the next years thanks to the 
ASTRO-H satellite\footnote{http://astro-h.isas.jaxa.jp/en/} that
will allow measurement of ICM turbulence %(or put stringent constraints) 
through the Doppler broadening and shifting of metal lines induced by 
turbulent motions and the effect of turbulence on resonant lines 
properties\cite{sunyaev03, dolag05, vazzaghellerbrunetti10,
takahashi12, zhuravleva12, zhuravleva13, nagaietal13}.

\begin{figure}[ht!]
\centering
\includegraphics[width=130mm]{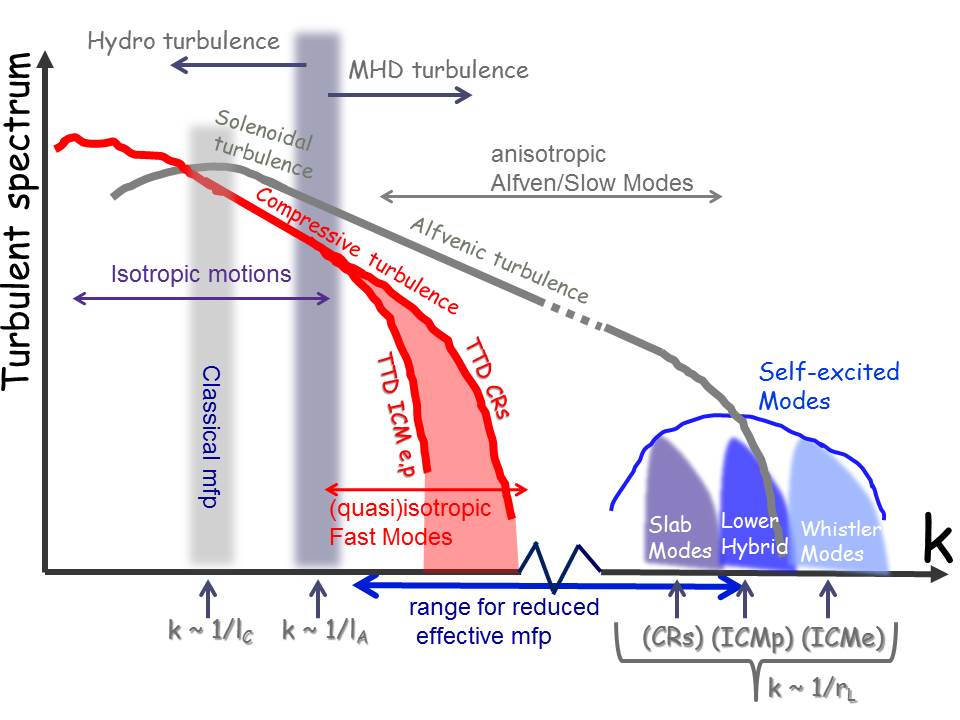}
\caption{
A schematic view of turbulence in the ICM.
The transition from hydro- to MHD turbulence is marked (see text).
The expected spectral features of both solenoidal and compressive
turbulence generated at large scales are illustrated :  
solenoidal turbulence develops an Alfv\'enic cascade at
small (micro-) scales whereas the compressible part (fast modes in the
MHD regime) is presumebly dissipated via TTD resonance with electrons
and protons in the ICM (or eventually via TTD resonance with CRs in case
of reduced effective mean free path, see text).
A schematic illustration of relevant examples of
``self-excited'' modes, excited via CRs- or turbulent-induced 
instabilities, is also shown together with the relevant scales:  
slab modes, 
lower hybrid electrostatic waves, and 
whistler waves (see text).
A schematic illustration of the scales of the Alfv\'en scale, $l_A$,
classical mean-free-path due to Coulomb ion-ion collisions,
$l_C$, and reduced particles mean-free-path is also given.
}
%\texttt{sidecapion} 
\label{fig:1}       % Give a unique label
\end{figure}

\

\noindent{\it 2.2.2.1 \, Turbulent Acceleration}
\vskip 2mm

Turbulence in the ICM can potentially trigger several mechanisms
of particle acceleration.
The non-linear interplay between particles and turbulent waves/modes
is a stochastic process that drains energy from plasma turbulence to
particles \cite{melrose80, schlickeiser02}.
In addition, reconnection of magnetic fields may be faster in turbulent
regions than in stationary media or laminar flows\cite{lazarian99}, 
potentially providing an additional source 
of particle acceleration in the ICM
\cite{lazarianbrunetti11}. In this Section we will focus 
on stochastic re-acceleration of CRs due to resonant
interaction with turbulence, in particular with low-frequency
waves, because this turbulent-acceleration
mechanism has the best developed theory and is the most commonly 
adopted in the current literature.
Acceleration of CRs directly from the thermal pool to relativistic energies
by MHD turbulence in the ICM is very inefficient and faces 
serious problems due to associated energy arguments
\cite{petrosianeast08, chernyshov12}.
Consequently, turbulent acceleration in the ICM is rather a matter
of re-acceleration of pre-existing (seed) CRs  rather than {\it ab initio} 
acceleration of CRs, e.g. \cite{brunetti01, petrosian01}

The (re)acceleration of CRs by turbulence is customarily described
according to the quasi-linear-theory (QLT), where 
the effect of linear waves on particles is studied by calculating
first-order corrections to the particle orbit in the uniform/background 
magnetic field ${\bf B_0}$, and then ensemble-averaging over the
statistical properties of the turbulent modes \cite{jokipii66,
schlickeiser93, miller95}.
In QLT one works in the coordinate system in which the space coordinates are measured
in the Lab system and the particle momentum coordinates are measured
in the rest frame of the background plasma that supports the turbulence
and in which turbulence is homogeneous. 
Then the gyrophase-averaged particle density distribution, 
$f(x,p,\mu,t)$, $\mu$ is the cosine of the particle pitch angle,
evolves in response to electromagnetic turbulence according 
to the Fokker-Planck equation \cite{schlickeiser02}:

\begin{eqnarray}
{{d f}\over{d t}} = {{\partial  }\over{\partial 
\mu }} \left[ \left( D_{\mu \mu} {{\partial
}\over{\partial \mu }} + D_{\mu p} {{\partial
}\over{\partial p}} \right) f(p,\mu , t) \right]
+ {1 \over{p^2}} {{\partial
}\over{\partial p}} \Big[ p^2 \times \nonumber\\ 
\Big(D_{\mu p} {{\partial }\over{\partial \mu }} 
+ D_{pp} {{\partial }\over{\partial p}} \Big) f(p,\mu , t) \Big]
\end{eqnarray}

\noindent
where $D_{pp}$, $D_{\mu \mu}$ and $D_{p \mu}$ 
are the fundamental transport coefficients describing 
the stochastic turbulence--particle 
interactions. These are determined by the electromagnetic
fluctuations in the turbulent field.

\noindent
Much attention has been devoted to the interaction with
the low-frequency Alfv\'en 
and magnetosonic MHD waves\footnote{Following \cite{footekulsrud79}
here we define low frequency waves as those having frequency $\omega
<< \Omega_i/\beta_{pl}$,  where $\Omega_i$ is the 
Larmor frequency of nonrelativistic ions}.
For these waves 
the relevant Fokker-Planck coefficients
are given by \cite{schlickeiser93, yanlazarian04}:

\begin{eqnarray}
\left(
\begin{array}{c}
D_{\mu\mu}\\
 \\
D_{pp}
\end{array}\right)
=
{\frac{\Omega^{2}(1-\mu^{2})}{2B_{0}^{2}}}
{\mathcal{R}}e\sum_{n=-\infty}^{n=\infty}\int_{\bf k_{min}}^{\bf
k_{max}}d^{3}k  
\left( \begin{array}{c}
\left(1-\frac{\mu \omega}{v k_{\parallel}} \right)^{2}\\
\\
\left( \frac{p \, c}{v} \right)^{2}
\end{array}
\right)
\int_{0}^{\infty}dt
e^{-i(k_{\parallel}v_{\parallel}-\omega+n\Omega)t} 
\nonumber\\
\Big\{
J_{n+1}^{2}(x) \left(
\begin{array}{c}
P_{{{RR}}}^{\mathbf{k}}\\
\\
R_{{{RR}}}^{\mathbf{k}}
\end{array}
\right) + J_{n-1}^{2}(x)
\left( \begin{array}{c}
P_{{{LL}}}^{\mathbf{k}}\\
\\
R_{{{LL}}}^{\mathbf{k}}
\end{array}
\right)+ J_{n+1}(x)J_{n-1}(x)
\Big[e^{2i\Psi}
\left( \begin{array}{c}
-P_{{{RL}}}^{\mathbf{k}}\\
\\
R_{{{RL}}}^{\mathbf{k}}
\end{array}
\right)
\nonumber\\
+ e^{-2 i\Psi}
\left(
\begin{array}{c}
-P_{{{LR}}}^{\mathbf{k}}\\
\\
R_{{{LR}}}^{\mathbf{k}}
\end{array}
\right)
\Big]
\Big\} 
\label{DppDmumu}
\end{eqnarray}

\noindent
where $\omega$ is the wave frequency, $k$ the wave-number, $k_{\perp}$ and
$k_{\parallel}$ the wave-number components perpendicular and parallel
to the background field, $\Psi=\arctan(k_x/k_y)$, $\Omega = (q/|q|)
\Omega_0/\gamma$ ($\Omega_0 = q B/(m \, c)$ is the 
non-relativistic gyrofrequency) and 
where we define $x= k_{\perp} v_{\perp}/\Omega$ as the argument 
for the Bessel functions, $J_n$.
The relevant electromagnetic fluctuations are :

\begin{equation}
<B_{\alpha}(\mathbf{k})B_{\beta}^{*}(\mathbf{k'})>=
\delta(\mathbf{k}-\mathbf{k'})P_{\alpha\beta}^{\mathbf{k}}
\label{bb}
\end{equation}

and

\begin{equation}
<E_{\alpha}(\mathbf{k})E_{\beta}^{*}(\mathbf{k'})>=
\delta(\mathbf{k}-\mathbf{k'})R_{\alpha\beta}^{\mathbf{k}}
\label{ee}
\end{equation}

\noindent
where $\alpha$ and $\beta = R,L$ indicate right-hand and
left-hand wave polarizations.

Under conditions of negligible damping $\omega = \omega_r + i \Gamma
\rightarrow \omega_r$ and the integral in eq.(7) is

\begin{equation}
\int d^{3}k
\int_{0}^{\infty}dt
e^{-i(k_{\parallel}v_{\parallel}-\omega+n\Omega)t}
\big( {\ldots}  \big) \rightarrow
\pi \int d^{3}k \delta (k_{\parallel}v_{\parallel}-\omega+n\Omega)
\big( {\ldots}  \big),
\end{equation}

\noindent
where $\delta (k_{\parallel}v_{\parallel}-\omega+n\Omega)$ selects
the resonant conditions between particles and waves; namely,
$n = \pm 1, ..$ (gyroresonance that is important
for Alfv\'en waves) and $n=0$
({\it Transit Time Damping, TTD},
or wave surfing that is the most important
for magnetosonic waves) \cite{melrose80, schlickeiser93}.
In the MHD approximation the polarisation and dispersion 
properties of the waves are relatively simple.
For Alfv\'en waves,
$P_{RR}^{\mathbf{k}}=P_{LL}^{\mathbf{k}}=-P_{RL}^{\mathbf{k}}
=-P_{LR}^{\mathbf{k}}$,
$R_{RR}^{\mathbf{k}}=R_{LL}^{\mathbf{k}}=R_{RL}^{\mathbf{k}}
=R_{LR}^{\mathbf{k}}$ and $\omega = v_A k_{\parallel}$,
while for (fast) magnetosonic waves, 
$P_{RR}^{\mathbf{k}}=P_{LL}^{\mathbf{k}}=P_{RL}^{\mathbf{k}}
=P_{LR}^{\mathbf{k}}$, $R_{RR}^{\mathbf{k}}=R_{LL}^{\mathbf{k}}=
-R_{RL}^{\mathbf{k}} =- R_{LR}^{\mathbf{k}}$ and $\omega = V_f k$
(see \cite{melrose80, schlickeiser02} for details); $V_f = c_s$ in
high beta-plasma such as the ICM.

In the case of low-frequency MHD waves with phase velocities, $V_{ph}$,
much less than the speed of light the magnetic-field component is much
larger than the electric-field component, $\delta B \sim c/V_{ph} \delta E$.
Then the particle distribution function, $f$, adjusts very rapidly to
quasi-equilibrium via pitch-angle scattering, approaching a quasi-isotropic
distribution.
In this case the Fokker-Planck equation (eq. 6) simplifies to a 
diffusion-convection equation \cite{dung94, kirk88}:

\begin{equation}
{{\partial f(p,t)}\over{\partial t}} =
{1 \over{p^2}} {{\partial  }\over{\partial p}} \left(
p^2 {\cal D}_{pp} {{\partial f }\over{\partial p}} 
- p^2 \big| {{d p}\over{dt}} \big|_{loss} f \right)
+
{{\partial  }\over{\partial z}} \left(
D {{\partial f }\over{\partial z}} \right) + Q(p,t)
\end{equation}

\noindent
where the momentum diffusion coefficient parallel to
the magnetic field, ${\cal D}_{pp}$, is:

\begin{equation}
{\cal D}_{pp} = {1\over 2} \int_{-1}^1 d\mu D_{pp}
\end{equation}

\noindent
and the spatial diffusion coefficient, $D$, is :

\begin{equation}
D = {{{\rm v}^2}\over 8} \int_{-1}^1 d\mu
{{(1-\mu^2)^2}\over{D_{\mu \mu}}}.
\end{equation}

\noindent
and where we added two terms in eq. (11), $f p^2 |dp/dt|$
and $Q$, that account for energy losses (see Sect. 3.1) and injection of CRs, 
respectively.
To avoid confusion, we mention that the basic physics behind the two
diffusion convection equations (2) and (11) is similar. The
principal differences are that eq. (11), which targets CR interactions
with local turbulence,
ignores large-scale spatial variations in the background velocity, $V$, while
eq.(2) ignores the momentum diffusion coefficient, $D_{pp}$, because
in strongly compressed flows at shocks it is sub-dominant (note that
eq.2 also omits energy losses and injection of CRs).

\noindent
From eqs. (7)-(9) and (12) and (13) it is clear that the momentum and spatial 
diffusion coefficients
depend on the electric field and magnetic field fluctuations, respectively.
Simple, approximate forms for these coefficients can be written
in some circumstances that are useful in several astrophysical
situations, including galaxy clusters. 

\noindent
For instance, if one assumes isotropic pitch angle scattering
by resonant (linearly polarized and undamped)
Alfv\'en waves with $k \sim r_L^{-1}$, the spatial diffusion coefficient
from eq. (13) can be written for relativistic CRs as \cite{skilling75}:

\begin{equation}
D \approx A c r_L \frac{B_0^2}{(\delta B)^2},
\end{equation}

\noindent
where $\delta B$ represents the net amplitude of (resonant) magnetic field
fluctuations defined in eq. (8) and where $A \sim 1$ (in Sect.3.2 we
will give a equivalent formula in terms of the Alfv\'en wave spectrum, 
eq. 25).
We note that for $\delta B \sim B_0$ the result is equivalent
to the classical Bohm diffusion formula, $D \sim (1/3) c r_L$, that
has been used in Sect. 2.2.1.

\noindent
Similarly, if we assume momentum diffusion due to 
Transit-Time-Damping, TTD, ($n = 0$) interactions
with isotropic magnetosonic waves, we can write approximately for relativistic
CRs \cite{brunettilazarian07}:

\begin{equation}
{\cal D}_{pp} \approx A_1 p^2 \frac{c_s^2}{c l}\frac{(\delta B)_f^2}{B_0^2},
\end{equation}

\noindent
where $A_1$ ($A_1 \sim 5$) depends on details of the turbulence, 
$l$ is the scale on which magnetosonic waves are dissipated, 
and now $(\delta B)_f$ represents magnetic field fluctuations associated 
with those waves through eq.(8) ($\delta B_f / B_0 \sim V_t / c_s$, e.g.
\cite{brunettilazarian07}, where $V_t$ is the velocity of large-scale
turbulent eddies).

\begin{figure}[ht!]
\centering
\includegraphics[width=130mm]{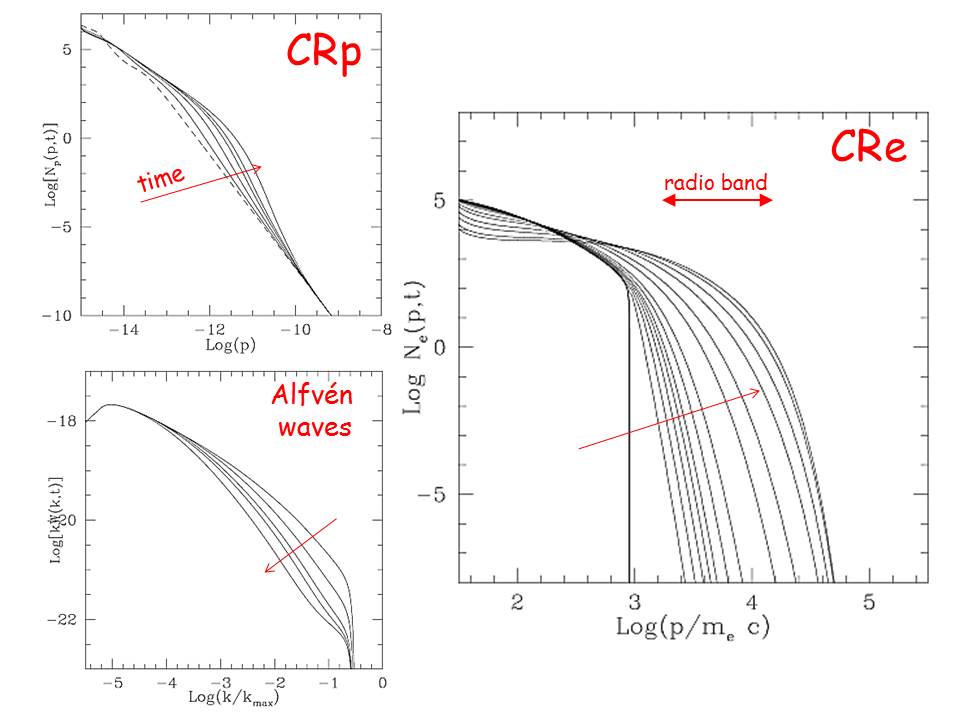}
\caption{
The coupled evolution with time of the spectra of CRp (upper-left; $p$
is in cgs units), CRe (right) and
Alfv\'en waves (bottom-left); Alfv\'en waves are continuously injected
assuing an external source (adapted from \cite{brunettietal04}).
Panels highlight the non-linear interplay between the
acceleration of CRs
and the evolution of waves 
that, indeed, are increasingly damped with time as they transfer an
increasing amount of  energy to CRs.
Saturation of CRe acceleration at later times is due to the combination
of radiative losses and the damping of the waves that limits
acceleration efficiency.}
\label{fig:1}       % Give a unique label
\end{figure}

\

\noindent{\it 2.2.2.2. \, Application to the ICM \& open questions}
\vskip 2mm

To account for the turbulence--particle interaction properly, one must know
both the scaling of turbulence down to resonant interaction lengths
(Eqs. \ref{bb}--\ref{ee}), the changes with time of the turbulence spectrum 
on resonance scales due to the most relevant damping processes,
and the interactions of turbulence with various waves produced by CRs. 
This is extremely challenging.  However, in the last decade several
modeling efforts attempted to study turbulent acceleration in astrophysical
environments, including galaxy clusters, 
by using physically motivated turbulent scalings and the relevant
collisionless damping of the turbulence.

\noindent
Several calculations suggest that there is room for turbulence
in galaxy clusters to play an important role in the
acceleration of CRs. This outcome depends on 
the fraction of the turbulent energy that goes into
(re)acceleration of CRs in the ICM.

\noindent
Presumably the many types of waves generated/excited in the
ICM, both at large and very small scales (Fig. 6), should jointly contribute to
the scattering process and (re)acceleration of CRs.
Much attention has been devoted to CR acceleration due to 
compressible (fast mode)
turbulence that is driven at large scales in the ICM from cluster
mergers and that cascades
to smaller scales. 
Under this hypothesis, and as expected for typical conditions in galaxy 
clusters (i.e., high temperature, high beta plasma and with likely forcing and 
dissipation scales), it is the compressible fast modes that are the
most important in the acceleration of CRs in ICMs. Most of the energy
of these modes is converted into the heating of the
thermal plasma via the TTD resonance. However current calculations
show that TTD can drain as much as a few
to $\sim$10 percent of the total turbulent--energy flux into the
CR component of the ICM plasma\cite{cassanobrunetti05, brunettilazarian07, 
brunettilazarian11b}.
In this case CRe in the ICM can be re-accelerated up
to energies of several GeV, provided that the energy budget of
fast modes on scales of tens of kpc 
is larger than about 3-5 percent of the local thermal energy 
budget\footnote{We 
note that if compressive turbulence
is generated at larger (e.g. a few 100 kpc) scales, this condition
in terms of energy budget, essentially implies that a large fraction 
of the energy of such turbulence is transported to smaller scales.
It implies, for example, that weak shocks generated by compressive
turbulence and viscosity
do not dissipate most of the turbulent energy budget into heat.}.
We expect
that this conclusion is important for understanding the origin 
of radio halos (Sect. 4).

\noindent
Moreover there are several circumstances under which the fraction of the
turbulent energy that is transferred into CRs may be much larger than
a few \%, thus making the acceleration process also more efficient than in
the previous case. For example, if we consider the scenario discussed
above, where compressive (fast mode) turbulence is
generated at large scales, a large fraction of the energy
of the fast modes can be converted into the acceleration of
CRs if the CR particle collision frequencies in the ICM are
much larger than those due to the classical process of
ion-ion Coulomb collisions\cite{brunettilazarian11a}.
Potentially this may occur if the 
interactions between particles are mediated
by magnetic perturbations generated by plasma instabilities.
Instabilities may be
driven by compressive turbulence and CRs in the ICM
\cite{brunettilazarian11a, yanlazarian11}.
Another case where
a large fraction of the turbulent energy is potentially dissipated
into the (re)acceleration of CRs in galaxy clusters is that of
incompressible turbulence, where
acceleration is driven by Alfv\'en modes via
gyro-resonance ($n = \pm 1$)\cite{ohno02,fujita03,brunettietal04}.
In this case the important caveat is that
Alfv\'en modes develop an anisotropic cascade toward smaller scales, 
below the Alfv\'en scale,
that quenches the efficiency of the gyro-resonance, $n = \pm 1$, 
scattering (acceleration)
process \cite{chandran00, yanlazarian02, yanlazarian04}. 
Consequently models of
Alfv\'enic acceleration assume that waves are generated in the ICM
at small (quasi-resonant) scales (Fig. 6), although it remains still rather
unclear whether such small-scale waves can be efficiently
generated in the ICM.

\noindent
In all these cases, where a fairly large fraction of
the turbulent energy is drained into CRs, the efficiency
of acceleration is essentially fixed by the damping of turbulence
by the CRs themselves\cite{brunettietal04, brunettiblasi05, brunettilazarian11a}.
Figure 7 shows an example of CRp and CRe acceleration under these
conditions.
As time proceeds during the
acceleration, the CRs gain energy and extract an increasing energy 
budget from the turbulent cascade. Consequent modifications are induced
in the spectrum of the turbulence, causing a decrement in the acceleration 
efficiency.
This differs from the cases where a smaller fraction of turbulent
energy is dissipated into CRs acceleration. In these cases indeed the
turbulent properties and the efficiency of turbulent acceleration
depend only weakly on the CRs properties and energy 
budget\cite{brunettilazarian07, brunettilazarian11b}.

\subsection{Generation of secondary particles}

If we assume that CRp remain in galaxy clusters for a time-period,
$\tau$, the grammage they encounter during their propagation is 
$X_g \sim n_{ICM} m_p c \tau \sim 1.6 \times {{n_{ICM}}\over{{10^{-3}}}}\times 
{{\tau}\over{{\rm Gyr}}}$g cm$^{-2}$.
On cosmic time scales 
that would often be comparable in the ICM with the nuclear grammage required
for an inelastic collision, $X_{nuc} \approx 50$ g
cm$^{-2}$.
This implies that 
the generation of secondary particles due to inelastic collisions
between CRp and thermal protons in the ICM is an important source of CRe.

\noindent
The decay chain for the injection
of secondary particles is \cite{blasicolafrancesco99}: 

$$p+p \to \pi^0 + \pi^+ + \pi^- + \rm{anything}$$
$$\pi^0 \to \gamma \gamma$$
$$\pi^\pm \to \mu^\pm + \nu_\mu(\bar{\nu}_\mu) \,\,\,\,\, , \,\,\,\,\,
\mu^\pm\to e^\pm + \bar{\nu}_\mu({\nu}_\mu) + \nu_e(\bar{\nu}_e).$$

\noindent
A threshold reaction requires CR protons with kinetic
energy just larger than $T_p \approx 300$ MeV to produce $\pi^0$.
The injection rate of pions is given more generally by \cite{moskalenko98,
blasicolafrancesco99}:

\begin{equation}
Q_{\pi}^{\pm,o}(E,t)= n^p_{th} c
\int_{p_{*}} dp N_{CRp}(p,t)  {{ F_{\pi}(E_{\pi},E_p)
\sigma^{\pm,o}(p)}\over
{\sqrt{1 + (m_pc/p)^2} }},
\label{q_pi}
\end{equation}

\noindent
where $N_{CRp}$ is the CRp spectrum,
$F_{\pi}$ is the spectrum of pions from the individual collisions of
CRp (of energy $E_p$) and thermal
protons\cite{stecker70, badhwar77, stephens81, dermer86a, dermer86b,
moskalenko98, brunettiblasi05, kamae05, kelner06} and
$\sigma^{\pm,o}(p)$ is the pp cross section for $\pi^o$, $\pi^+$ and
$\pi^-$\cite{dermer86a, dermer86b, kelner06}.

\noindent
Neutral pions decay into $\gamma$--rays with spectrum\cite{dermer86a,
dermer86b, blasicolafrancesco99, kelner06} :

\begin{equation}
Q_{\gamma}(E_{\gamma})
= 2 \int_{E_{min}}^{E_p^{max}}
{{ Q_{\pi^o}(E_{\pi^o}) }\over{
\sqrt{E_{\pi}^2 - m_{\pi}^2 c^4} }}
d E_{\pi} \, ,
\label{gamma}
\end{equation}

\noindent
where $E_{min} = E_{\gamma} + {1/4} m_{\pi}^2 c^4 / E_{\gamma}$.
Neutral pions produced near threshold will decay into a pair of 
$\gamma$s with average
energy $E_{\gamma} \simeq 67$ MeV. 
This provides a rough measure of the low energy end of the
expected $\gamma$-ray spectrum.
Charged pion decays produce muons, which then produce 
secondary electrons and positrons (as well as neutrinos) as they decay.
The injection rate of relativistic electrons/positrons then becomes:

\begin{equation}
Q_{e^{\pm}}(p,t)=
\int_{E_{\pi}}
Q_{\pi}(E_{\pi^{\pm}},t) dE_{\pi} \int dE_{\mu} 
F_{e^{\pm}}(E_{\pi},E_{\mu},E_e) F_{\mu}(E_{\mu},E_{\pi}),
\label{qepm1}
\end{equation}

\noindent
where $F_e^{\pm}(E_e,E_\mu,E_\pi)$ is the spectrum of
electrons and positrons from the
decay of a muon of energy $E_\mu$ produced in the decay of a pion with
energy $E_\pi$ \cite{blasicolafrancesco99, kelner06},
and $F_{\mu}(E_{\mu},E_{\pi})$ is the muon spectrum
generated by the decay of a pion of energy $E_{\pi}$
\cite{moskalenko98, blasicolafrancesco99, brunettiblasi05}.

Secondary electrons continuously generated in the ICM are subject to
energy losses (Sect. 3.1).
If these secondaries are not accelerated by other
mechanisms, their spectrum approaches a stationary distribution because of
the competition between injection and energy losses \cite{dolagensslin00}:

\begin{equation}
N_e^{\pm}(p)=
{1 \over
{\Big| \left[ {{dp}\over{dt}} \right]_{\rm L} \Big| }}
\int_{p}^{p_{\rm max}}
Q_e^{\pm}(p) dp \, ,
\label{sec_stat}
\end{equation}

\noindent
where ${{dp}\over{dt}}_L$ accounts for CRe energy losses (see Sect. 3.1). 
Assuming a power law distribution of CRp, $N_p(p) = K_p p^{-s}$,
the spectrum of secondary electrons at high energies,
$\gamma > 10^3$, is $N_e(p) \propto p^{-\delta}$, with $\delta=
s+1-\Delta$, where $\Delta \sim 0.05$ approximately accounts for
the log--scaling of the p-p cross--section at high
energies \cite{brunettiblasi05,kelner06,kamae06}.
The radio synchrotron emission from these electrons ($e^\pm$ actually) 
would have a spectral slope, $\alpha = (\delta - 1)/2$.

\section{The life-cycle of CRs in galaxy clusters}
\label{sec:3}

In this Section we discuss the energy-evolution and dynamics of
relativistic CRp and CRe that are released in the cluster volume
from the accelerators described in the previous Section.
In this Section 
we also discuss the constraints on CR acceleration efficiency
and propagation that come from the current $\gamma$-ray
and radio observations.

\subsection{Energy Losses}

Cosmic rays, especially electrons for energies above $\sim$ GeV,
are subject to energy losses that limit their
life-time in the ICM and the maximum energy at which they can
be accelerated by acceleration mechanisms.

\subsubsection{Electrons}

The energy losses of ultra-relativistic electrons in the ICM are
essentially
dominated by ionization and Coulomb losses at low energies\cite{sarazin99}

\begin{equation}
\left[ {{ d p }\over{d t}}\right]_{\rm i}
=- 3.3 \times 10^{-29} n_{\rm th}
\left[1+ {{ {\rm ln}(\gamma/{n_{\rm th}} ) }\over{
75 }} \right],
\label{ion}
\end{equation}

\noindent
where $n_{\rm th}$ is the number density of the thermal plasma protons, and 
and by synchrotron and inverse Compton losses at higher
energies\cite{sarazin99},

\begin{equation}
\left[ {{ d p }\over{d t}}\right]_{\rm rad}
=- 4.8 \times 10^{-4} p^2
\left[ \left( {{ B_{\mu G} }\over{
3.2}} \right)^2 
+ (1+z)^4 \right],
\label{syn+ic}
\end{equation}

\noindent
where $B_{\mu G}$ is the magnetic field strength in
units of $\mu G$, and we assumed isotropic magnetic fields and distributions
of CRe momenta.
The factor in the square brackets can alternatively be expressed 
as $B^2_{IC} + B^2$, where $B_{IC} = 3.2 (1+z)^2~\mu$G is the
equivalent magnetic field strength for energy losses due to ICS 
with CMB photons.

\noindent
The life-time of CRe, $\tau_{l} \sim p/(dp/dt)$, from
Eqs. \ref{ion}--\ref{syn+ic}, is :

\begin{eqnarray}
\tau_{\rm e}({\rm Gyr}) \sim
4 \times \Big\{
{1\over 3}
\Big( {{\gamma}\over{300}} \Big)
\left[ \left( {{ B_{\mu G} }\over{
3.2}} \right)^2 + (1+z)^4 \right]
\nonumber\\
+
\Big( {{ n_{\rm th} }\over{
10^{-3} }} \Big)
\Big({{ \gamma }\over{
300 }} \Big)^{-1}
\left[1.2 + {1\over{75}}
\ln \Big( {{\gamma/300}\over{
n_{\rm th}/10^{-3} }} \Big) \right]
\Big\}^{-1}.
\label{tau1/2}
\end{eqnarray}

\noindent
This depends on 
the number density of the thermal medium, which can be estimated from 
X-ray observations, on the IC-equivalent magnetic field (i.e.,
redshift of the cluster), and on
the magnetic field strength, which is important only in the
case $B^2 >> B_{IC}^2$ and eventually can be constrained from 
Faraday rotation measures (Sect. 4.1).

\subsubsection{Protons}

For relativistic CRp, the main channel of energy losses
in the ICM is provided by inelastic p-p collisions (Sect. 2.3).
This sets a CRp life--time 

\begin{equation}
\tau_{pp}(p)\simeq
{1\over{c \, n_{th} \sigma_{pp}}}
\label{tpp}
\end{equation}

\noindent
$\sigma_{pp}$ is the inclusive p--p cross-section \cite{dermer86a,
dermer86b}.

\noindent
For trans-relativistic and mildly relativistic CRp, energy
losses are dominated by ionization and Coulomb scattering.
CRp more energetic than the thermal electrons have\cite{schlickeiser02}

\begin{equation}
\Big( {{ d p }\over{dt}} \Big)_i \simeq
- 1.7 \times 10^{-29}
\left( {{n_{\rm th}}\over{10^{-3}}} \right)
{{
\beta_p }\over{
{{3}\over{4}} \sqrt{\pi} \beta_e^3
+ \beta_p^3 }} \,\,\,\,\,\,\,\,\, ({\rm cgs})
\label{coulomb_p}
\end{equation}

\noindent
where $\beta_e = v_e/c \simeq 43 \beta_p \simeq 0.18 (T/10^8 K)^{1/2}$ 
is the RMS velocity of the
thermal electrons, while $\beta_p$ is the corresponding thermal proton velocity.

\begin{figure}[ht!]
\centering
\includegraphics[width=130mm]{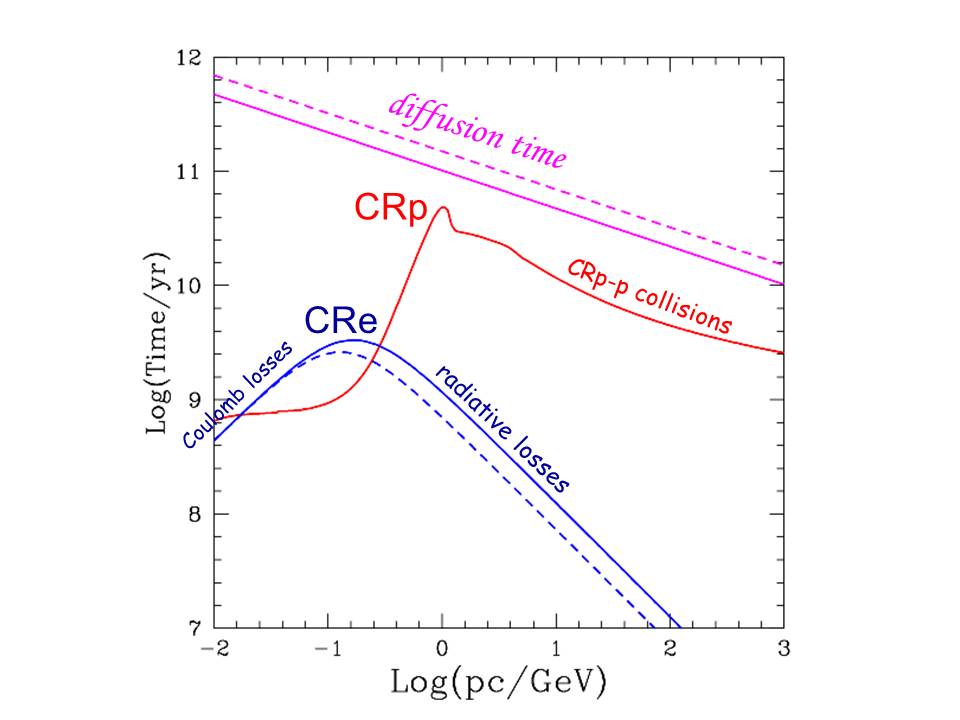}
\caption{
Life-time of CRp (red) and CRe (blue, lower curves) in the ICM at
redshift
$z=0$, compared with the CR diffusion time on Mpc scales (magenta,
upper curves) (adapted from \cite{blasi07}).
The most relevant channels of CR energy losses at different energies are
highlighted in the panel.
Adopted physical parameters are : $n_{th} = 10^{-3}$cm$^{-3}$,
$B= 1$ (solid) and $3 \mu$G (dashed). Diffusion is calculated
assuming a Kolmogorov spectrum of magnetic fluctuations with
$L_{max}=100$ kpc and $f=1$.}
\label{fig:1}       % Give a unique label
\end{figure}

\subsubsection{General energy loss considerations}

Figure 8 shows the (total) time scales for losses 
of CRe and CRp. 
CRp with energy 1 GeV -- 1 TeV are long-living particles with life-times 
in the cores of galaxy clusters $\sim$ several Gyrs.
At higher energy the CRp time-scale gradually drops below 1 Gyr, while
at very high energy, in the regime of ultra high energy CRp, 
the life-time is limited by inelastic $p-\gamma$ collisions with 
CMB photons, as discussed in Section 2.2.1 (Fig. 5).

On the other hand, CRe are short-lived particles at the
energies where they radiate observable emissions, due to the unavoidable
radiation energy-losses (mainly ICS and synchrotron).
The maximum life-time of CRe,  about 1 Gyr, is
at energies $\sim 100$ MeV, where radiative losses
are roughly equivalent to Coulomb losses. 
On the other hand, CRe with energy $\sim$several GeV that emit synchrotron 
radiation
in the radio band (GHz), have shorter life-times, $\sim$0.1 Gyrs.
The life-times of CRe at high energies do not vary much from cluster
cores to periphery, because for weak magnetic fields they are determined
by the unavoidable losses from ICS off CMB photons.
On the other hand, CRe ICS lifetimes 
will scale strongly and inversely with cluster 
redshift according to $(1+z)^{-4}$ (eqs. 21-22).

\subsection{Dynamics of CRs in the ICM}

The propagation of CRs injected 
in the ICM is mainly determined by diffusion and convection.

The time necessary for CRs to diffuse over distances $L$ is $\tau_{diff}
\sim (1/4) L^2/D$. This implies that the spatial diffusion 
coefficient necessary for
diffusion of CRs over Mpc-scales within a few Gyrs is extremely large, 
$D > 2 \times 10^{31}$cm$^2$s$^{-1}$. For CRs with GeV energy this 
is several orders 
of magnitude larger than that in our Galaxy.
This simple consideration suggests 
that galaxy clusters are efficient containers of CRs.

More specifically, according to QLT the 
diffusion coefficient for gyro-resonant scattering of particles with 
Alfv\'enic perturbations of the magnetic field is \cite{wentzel74,
bhattacharjee00} (see also Sect. 2.2.2):

\begin{equation}
D(p) = 
\frac{1}{3}r_L c \frac{B^2}{\int_{2\pi/r_L}^\infty dk P(k)}, 
\label{eq:d}
\end{equation}

\noindent
where $P(k)$ is the power spectrum 
of turbulent field-perturbations on a scale $k$ (that interacts
resonantly with particles with momentum $p\propto 1/k$), such that
$\int_{k_{min}}^\infty dk P(k) = f B^2$ and $f \leq 1$; $k_{min}$ is
the minimum wavenumber (maximum scale) of turbulence\footnote{We note 
that if turbulent field-perturbations are on scales $\leq r_L$,
i.e. $k_{min} \geq 2\pi/r_L$, and $f \approx 1$ eq. (25) is
the coefficient of parallel Bohm diffusion, $D \sim 1/3 r_L c$}.
In Figure 8 we show a comparison between the life-time of CRs and their
diffusion-time on Mpc scales assuming a Kolmogorov spectrum
of the Alfv\'enic fluctuations, $f B^2 \propto k^{-5/3}$,
with a maximum scale $L_{max} =
2 \pi k_{min}^{-1} = 100$ kpc and $f=1$.
The diffusion time of CRs in this case is substantially larger than a
Hubble time, implying that CRp can be
accumulated in the volume of galaxy clusters and that their energy
budget increases with time, as first realized by 
\cite{voelk96, berezinsky97, ensslin97}.
Figure 8 suggests that even a fairly small 
level of magnetic field fluctuations in the ICM, $f <<1$, 
should be sufficient to confine 
most of the CRs in the gigantic volume of galaxy clusters for
a time-period comparable to the age of clusters themselves.

\begin{figure}[ht!]
\centering
\includegraphics[width=130mm]{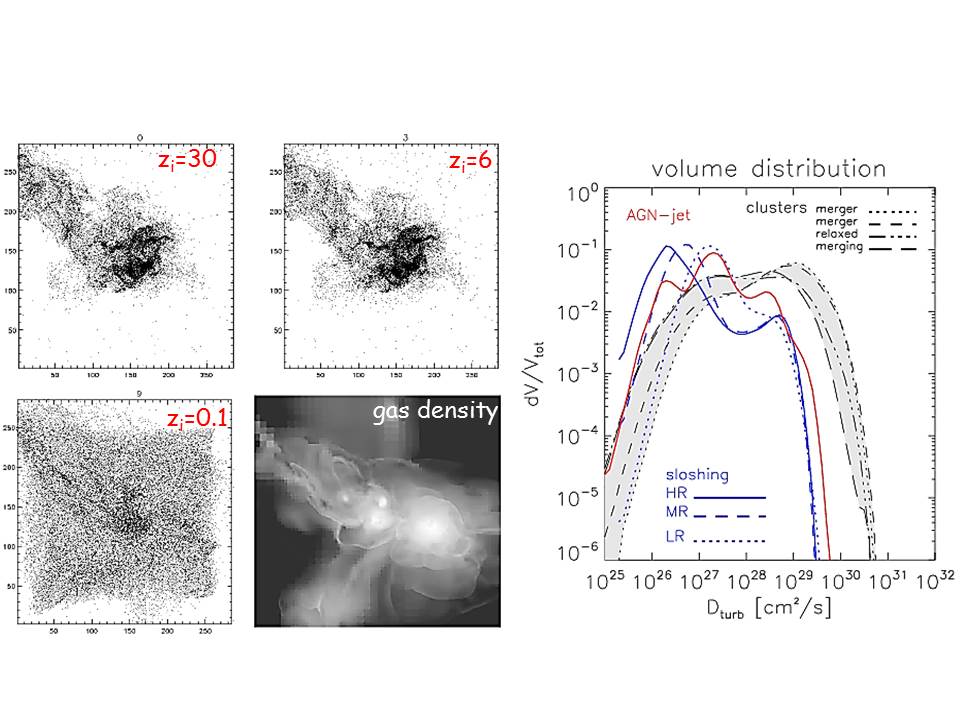}
\caption{
Left panel: spatial distribution of tracers particles in a galaxy
clusters at $z=0$ that have been uniformly generated in the cluster
region
in the simulated box at $z=$30, 6 and 0.1.
The bottom-right panel shows the gas density at $z=0$ (adapted
from \cite{vazzaghellerbrunetti10}).
Right panel:
volume-filling distribution of
the turbulent spatial-transport coefficient, $D \sim V_{l_o} l_o$,
that is estimated from the analysis of numerical simulations
of galaxy clusters. Curves refer to cases where turbulent motions are
driven by cluster mergers, AGN-jets and sloshing (from
\cite{vazza12}).}
\label{fig:1}       % Give a unique label
\end{figure}

Formally eq.\ref{eq:d}, which has been originally adopted to support 
CRs confinement in clusters, assumes an isotropic distribution of
Alfv\'enic fluctuations at resonant scales.
This is not true in the case of the strong\footnote{We note,
from the MHD turbulence literature, that 
the notion ``strong,''
refers to the strength of the non-linear interaction between turbulent 
waves, not to the amplitude of the turbulent fluctuations.}
incompressible turbulence, because its cascading process
is anisotropic with respect to the mean field \cite{GS95,
cholazarianvishniac03}.
Consequently, particle scattering is strongly reduced \cite{chandran00,
yanlazarian02, yanlazarian04}.
However, in general this does not imply that CRs scattering is
inefficient because, even limiting to the particular case of strong
incompressible turbulence\footnote{If a fraction of the turbulence is
compressible, scattering is efficient and dominated by fast modes.}, 
we can think that 
the non-resonant mirror interactions with the slow-mode
perturbations provide a lower limit to the rate of
scattering that is still orders of magnitude more 
efficient than that due to the gyroresonant scatter calculated according
to QLT \cite{beresnyaketal11}
We also note that Alfv\'en waves can be excited directly 
at resonant scales, for example, due to the streaming instability 
that is driven by CRs streaming along the field lines, eg.
\cite{wentzel74, schlickeiser02} (Fig. 6).
In this case the streaming speed (the effective ``drift speed'') 
of CRs gets limited to the
Alfv\'en speed. If this process is efficient in the ICM, CRs
drift at the Alfv\'en speed
and the time-scale necessary for CRs to cover Mpc distances is
$\sim$ Hubble time using a reference value, 
$v_A \sim 10^7$cm/s.
In the presence of background turbulence, the streaming instability 
can be partially suppressed \cite{yanlazarian04, wieneretal13}.
This is because turbulence
suppresses the waves responsible for self-confinement of CRs,
since they cascade to smaller scales before they have the
opportunity to scatter CRs. Such a background turbulence, however,
also limits the ``free flight'' of CRs, through the scattering
with the fluctuations of
the magnetic field induced by turbulence itself.
In the most ``favourable'' case, where we necglect
small scale fluctuations, CRs can ``fly'' along the magnetic 
field lines over maximum distances that are of the order of the smallest
scale on which the magnetic field is effectively advected by turbulent motions. 
That is the MHD, Alfv\'en scale, $l_A \simeq l_o/M_A^3$, where
$l_o$ is the turbulent-injection scale and
$M_A$ is the Alfv\'enic Mach number of the turbulence on that scale,
eg. \cite{lazarian06, brunettilazarian07} (Sect. 2.2.2, Fig. 6).
The resulting {\it maximum} diffusion
coefficient is $D \sim 1/3 c l_A \approx 10^{31} ({{l_A}\over{0.3 kpc}})$
cm$^2$ s$^{-1}$, still implying a diffusion time over Mpc-scales of
several Gyrs.
In conclusion, we believe that the wealth of waves that can be naturally 
generated in the ICM, on both small and large scales, supports the
paradigm of confinement of CRs
in galaxy clusters; namely, that most of the energy
budget of CRp is accumulated in the cluster volume over the
cluster life-time.

\noindent
Assuming, in the light of this discussion, that parallel diffusion of 
CRs of some energy $E$ 
is strongly suppressed by fluctuations on relatively small scales, the cluster-scale
dynamics of these
CRs is controlled by advection via gas flows accompanied 
by a process of turbulent-transport (Figure 9, left). 
This process is analogous to the transport of passive scalars by a
turbulent flow, and it induces the CR particles
to exibit a random walk behaviour, eg. \cite{cholazarian04JKAS}, within the bulk flow.
This regime is known as {\it Richardson} diffusion in hydro-turbulence
\cite{richardson1926}.
In this regime the transport is super-diffusive,
$L^2 \propto \tau_{diff}^{3/2}$, on scales
smaller than the injection scale of turbulent eddies and diffusive,
$L^2 \sim 4 D \tau_{diff}$, on larger scales.
Because it is controlled by the fluid motions, the CR transport-coefficient is 
energy-independent and  
can be estimated as $D \sim V_{l_o} l_o$, where
$V_{l_o}$ and $l_{o}$ are the velocity and scales of the largest
turbulent eddies.
According to numerical simulations of galaxy clusters the largest
values, 
$D \sim 10^{30-31}$cm$^2$s$^{-1}$, are derived in the case of merging 
systems \cite{vazza12} (Figure 9, right) as a result of large scale motions
and mixing generated during these events in the ICM. 
This has the potential implication that turbulence might
transport CRs on a scale that could be of the order of
cluster-cores, thus, potentially inducing a spatial distribution of
CRs that is broader than that of the thermal plasma.
Similarly, we note that the same mechanism could also spread metals through the
ICM leading to the formation of the flat spatial distributions of metals that are 
observed in non-cool core clusters \cite{rebusco06, vazzaghellerbrunetti10}.

To avoid confusion from the complexity of the above discussion, we clarify 
that the relative importance of particle scattering-based
diffusion parallel to the local mean magnetic field and bulk fluid 
turbulence-based transport of CRs in the ICM depend
on particle energies and the fluid turbulence properties, with 
the dynamics of very high energy CRs being dominated 
by scattering-based diffusion.

\subsection{Limits on CRp}

The expected confinement and accumulation of CRp generated during cluster 
formation (Sect. 3.2) motivate the quest for CRp in galaxy clusters.
Most of the thermal energy in the ICM is generated at shocks, as previously noted.
Estimates of CRp acceleration efficiency suggest as much as $\sim$10\% of the kinetic
energy flux at cosmological shocks may be converted into CRs\cite{kang11}. Then, one might
claim that the resulting energy budget of CRs should be a
substantial fraction of the ICM thermal energy. If true, the presence 
of the CRp could influence many
aspects of ICM dynamics, including, for examples, 
contribution to ICM pressure support and a partial quenching of
radiative cooling in core regions.

The most direct approach to constraining the energy content of
CRp in ICMs consists in the searches for
$\gamma$-ray emission from the decay of the neutral pions
due to CRp-p collisions in the ICM.
Early space-based $\gamma$-ray upper limits from EGRET observations
provided limits $E_{CR}/E_{ICM} < 0.3$ in several nearby
galaxy clusters \cite{reimer03}.
Subsequently, more stringent limits have been derived from
deep, pointed observations at energies $E_{\gamma}>$100 GeV with ground-based
Cherenkov telescopes
\cite{perkinsetal06, aharonian09a,aharonian09b,aleksic10,
aleksic12, veritas12}.
These limits, unfortunately, depend on the unknown spectral shape of
the CRp-energy distribution and the spatial distribution
of CRp in the clusters.
The most stringent limits are obtained assuming 
$\delta = 2.1$ ($N_{CR}(p)\propto p^{-\delta}$)
and a linear scaling between CRp and thermal energy densities
(Fig. 10); under these assumptions a
particularly deep limit $E_{CR}/E_{ICM} < 0.016$ is derived for the
Perseus cluster\cite{aleksic12}.
Constraints
are significantly less stringent for steeper spectra and for
flatter spatial distributions of the CRp component in the cluster.
The recent advent of the orbiting Fermi-LAT observatory has greatly improved
the detection prospects thanks to its unprecedented sensitivity
at MeV/GeV energies.
However, after almost 5 years of operations, no firm detection of any
ICM has been obtained. Only upper limits to the $\gamma$-ray
emission have been obtained for both individual nearby clusters and
from the stacking of samples of
clusters\cite{ackermann10, veritas12, Fermilat13, huberetal13,
prokhorov13, zandanelando13}.
Under the assumption 
that the spatial distribution of CRp roughly follows that
of the thermal ICM, in the most stringent cases (including stacking
procedures) these limits constrain $E_{CR}/E_{ICM} <$ about 1\% (Fig. 10),
with only a weak dependence on $\delta$. Similar limits are derived by
assuming a spatial distribution of CRp that is slightly broader than
that of the ICM\cite{Fermilat13}, such as that expected from simulations
by\cite{pinzke10}.
Limits on the CRp total energy become gradually less 
stringent if one assumes a 
flatter spatial
distribution compared to the ICM, since then more CRp reside in regions where
the number density of thermal-targets protons is lower and where they
consequently produce fewer $\pi^o$ and $\gamma$-rays.

\begin{figure}[ht!]
\centering
\includegraphics[width=130mm]{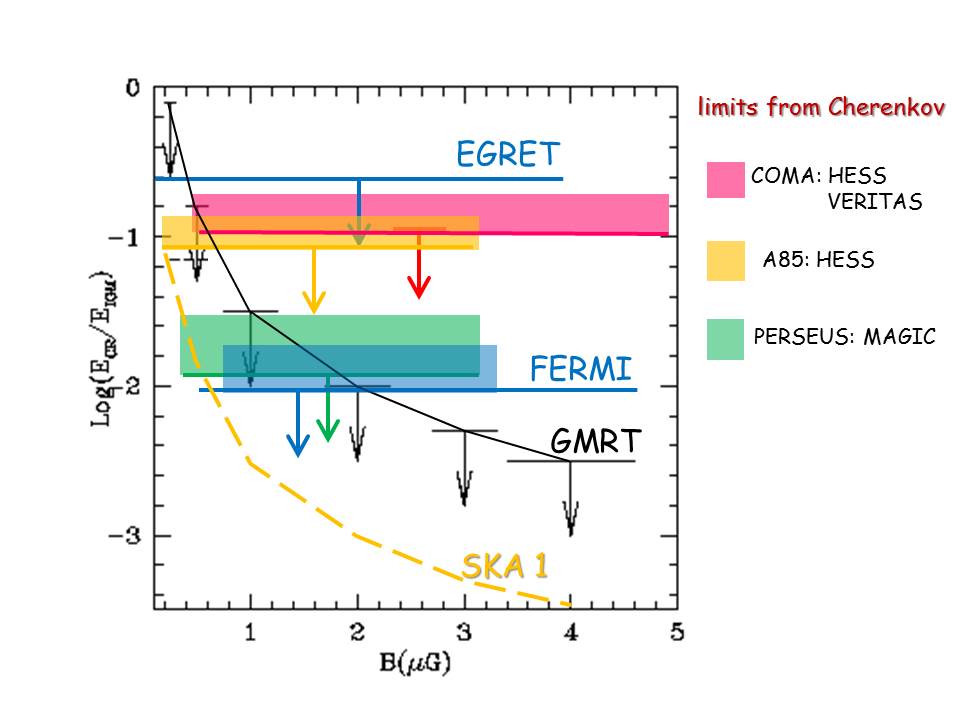}
\caption{
A collection of representative upper limits to the ratio of the energy
in CRp
and thermal ICM as derived from $\gamma$-ray and radio observations.
Radio-based upper limits depend on the magnetic field strength in a
Mpc$^3$
volume. Limits from observations with Cherenkov telescopes include
the case of Coma \cite{aharonian09b, veritas12}, A85 \cite{aharonian09a}
and Perseus \cite{aleksic12}. In \cite{aharonian09b, aharonian09a}
$\delta=2.1$ and a spatially constant ratio of CRp and ICM energy
densities are assumed. 
In the case of Coma\cite{veritas12} we report the limits obtained by 
these authors by adopting a spatial distribution and spectrum of CRp 
from numerical simulations.
In the case of Perseus\cite{aleksic12} the reported region (green) 
encompasses limits obtained using $\delta=2.1$ and a spatially constant
ratio of CRp and ICM energy densities and limits obtained
by adopting a spatial distribution and spectrum of CRp from numerical
simulations.
Limits from EGRET and Fermi-LAT are taken from \cite{reimer03}
and \cite{Fermilat13, huberetal13}, respectively. In \cite{reimer03,
huberetal13} 
limits are obtained assuming a spatially constant ratio of CRp and ICM
energy 
densities, whereas we report the limits obtained by \cite{Fermilat13}
by assuming
the spatial distribution of CRp from simulations.
Upper limits derived from radio observations are taken from
\cite{brunetti07}. The
sensitivity level to CRp from future SKA 1 observations is
also reported for clusters at $z=0.25$.
}
\label{fig:1}       % Give a unique label
\end{figure}

Radio observations of galaxy clusters also provide 
limits on $E_{CR}/E_{ICM}$, since these observations allow one to
constrain the generation rate of secondary CRe in the ICM
\cite{reimer04,brunetti07,brown11}.
Only a fraction of clusters show diffuse, cluster scale
synchrotron emission at the sensitivity level of present observations
(Sect. 4).  Most do not.
Radio upper limits to the cluster-scale emission
limit the combination of the energy densities of the
magnetic field and secondary CRe, and consequently 
the energy budget in the form of primary CRp as a function of the
magnetic field strength.
Faraday rotation measurements provide an indication that 
the central, Mpc$^3$-volume, regions of galaxy clusters
are magnetized at $\approx \mu$G 
level\cite{clarke01, carilli02, bonafede10}.
This information allows one to break the degeneracy between CRp and 
magnetic field
energy densities, resulting in limits $E_{CR}/E_{ICM} \leq$ few$\times
0.01$ (Fig. 10).

Current upper limits violate optimistic expectations for the
CRp energy content and $\gamma$--ray emission from galaxy clusters
derived in the last decade \cite{voelk96,miniati01gamma,
miniati03, blasi07, pfrommer08, colafrancesco08, pinzke10,
pinzkeetal11}.
Consequently the available limits now suggest that
the efficiency of CRp
acceleration that has been previously assumed for the most important
mechanisms operating
in galaxy clusters was too optimistic, or that eventually 
CRp diffusion and turbulent-transport (Sect. 3.2) play an important role. 
In this respect we notice 
that current observational constraints 
refer mainly to the innermost
($\sim$Mpc) regions of clusters where both the number density
of thermal protons (targets for $\pi^o$ production) and the magnetic
field are largest. 
In fact no tight constraints are available for the clusters outskirts
where the CRp contribution might be relatively larger.

\noindent
As a final remark on this point we note that future radio telescopes, 
including the phase 1 of the SKA, will have a chance to obtain
constraints one order of magnitude deeper.
These constraints will be better than current constraints 
from $\gamma$-ray observations even if we assume that the 
magnetic fields in the ICM are significantly weaker than those
estimated from Faraday rotation measurements (Fig. 10).
We note however that current limits on the hard X-ray 
and $\gamma$-ray emission\cite{ackermann10}
from galaxy clusters exclude the possibility that the magnetic fields
in the ICM are smaller than 0.1-0.2 $\mu$G.
Weaker fields than these in clusters hosting
giant radio halos would, as we explain in Sect. 5, necessarily 
require inverse Compton  hard X-rays above current upper 
limits from the
same CRe population responsible for the radio synchrotron emission.

\section{Diffuse synchrotron radio sources in galaxy clusters:
radio halos and relics}

Steep spectrum ($\alpha \geq 1$, with $F(\nu)\propto
\nu^{-\alpha}$), diffuse radio emission extended on cluster scales 
is observed in a number of galaxy clusters.
The emission is clearly associated with the ICM and not 
individual sources, implying the existence of
relativistic electrons and magnetic fields
mixed with the ICM.
Without entering in the details of the morphological
zoology that is observed, see e.g.\cite{ferrari08, feretti12},
in this Section we focus on the two main classes of
diffuse radio sources in galaxy clusters: radio halos and
giant radio relics.

\noindent
Halos and relics have different properties and presumably also
a different origin.
Radio halos are classified in {\it giant} (Figure 11) and {\it mini 
radio halos} (Figure 15),
peaking in intensity near the centre of galaxy clusters and having
good spatial coincidence with the distribution of the hot X--ray
emitting gas.
Radio relics (Figure 16) are typically
elongated and located at the cluster
periphery. Some clusters include both.
The two classes of radio
sources differ also in their polarization properties.
Halos are generally unpolarised, while relics are strongly polarised.
Synchrotron polarisation in the relics is a signature of significant
anisotropy in the magnetic field on large scales, probably due either to
compression (e.g., shocks) or possibly to shear (e.g., tangential 
discontinuities).
The absence of observed polarization in radio halos and their morphological
connection with the thermal X-ray emission suggest that the relativistic plasma
that generates that synchrotron radiation occupies a large fraction
of the volume filled by the hot X-ray emitting ICM.

\noindent
Faraday rotation measurements and limits to ICS X-rays (and $\gamma$-rays) 
indicate that the ICM is magnetised at $\mu$G levels\cite{carilli02}
(Sects. 4.1, 5), 
in which case the relativistic CRe emitting in the radio band
have energies of a few GeV (Lorentz factor $\gamma \sim 10^4$)
that have life-times $\approx$0.1 Gyr in the ICM (Sect. 3, Figure 8).
This short life-time,
combined with the excessively long time that is needed by these
CRe to diffuse across a sizable fraction of
the Mpc-scale of the observed emissions, 
requires that the emitting particles in halos and relics are 
continuously accelerated or generated {\it in situ} in the
emitting regions \cite{jaffe77}; this is known as the {\it diffusion
problem}.

\noindent
The hierarchical sequence of mergers and accretion of matter that
leads to the formation of clusters and filaments dissipates enormous
quantities of gravitational energy in the ICM through processes on
microphysical scales. Even if a small fraction of this energy is
converted into CRs acceleration we may expect non-thermal
emission from galaxy clusters and from the
Cosmic Web more generally. Cluster-scale radio sources may probe exactly 
this process
and consequently they are extremely important crossroads of cosmology,
astrophysics and plasma physics.
However, the very low radio surface brightness of the Mpc-scale 
radio sources in galaxy clusters\footnote{$\sim 1 \mu$ Jy arcsec$^{-2}$
at 1 GHz}, combined with their 
steep radio spectra (average values reported for
$\alpha$ are in the range 1.2--1.4), make their detection 
difficult\cite{venturi11, feretti12, farnsworthetal13}.
Presumably this implies that we are currently
detecting only the {\it tip of the
iceberg} of the non-thermal radio emission from the
Cosmic Web \cite{rudnicketal09}, which also implies that current
observational classification is possibly
subject to a revision in the next years
(for observational hints of diffuse emission on possibly very large scales, 
such as
radio bridges and complex emissions, 
see also \cite{feretti12} and ref. therein).

\noindent
Theoretically several mechanisms that are directly or indirectly
connected with the formation of galaxy clusters 
may contribute to the origin of the
observed radio emission. In this respect it is commonly accepted
that radio halos and relics are due to 
different mechanisms and consequently probe different
pieces of the complex physics in these environments.
Specifically giant radio halos probably trace turbulent regions in the
ICM where particles are trapped and accelerated by some
mechanism, while, on the other hand, radio relics are associated 
with cosmological
shock waves where particles can be accelerated.
As discussed in Sect 2.2 shocks and turbulent motions in the ICM are
tightly connected, and this ultimately may provide a possible physical
link between relics and halos \cite{markevitch10}.

\subsection{A brief note on B estimates in the ICM}

In the following Sections we shall use cluster-scale radio sources
as probes of the physics of CRs in the ICM.
The origin and distributions of magnetic fields in galaxy clusters
are not the main focus of our review.
However, for clarity we briefly comment on the current ideas for the
origin of magnetic fields in these systems. 
In particular, they rely on the possibility
that seed magnetic fields, of cosmological origin or
injected in the volume of galaxy clusters by galaxies and AGNs, are
mixed and amplified by compression and stretching through accretion and
turbulent-motions induced by shocks and
clusters dynamics\cite{roettiger99, dolag02, dolagjcap05,
brueggenetal05, subramanian06, ryuetal08, DuboisTeyssier08,
donnertetal09, xuetal10, xuetal11}.

\noindent
However, given the importance of ICM magnetic field strength estimates 
from Faraday rotation in constraining CR populations, it is appropriate 
here to outline briefly how they are currently obtained.
They depend, of course, on the circular birefringence at radio frequencies 
of a magnetized plasma.
Propagating through such a medium, the plane of linear polarization 
(generally in synchrotron emission) rotates through an
angle $\chi = RM\times \lambda^2$, 
where $RM = 812 \int_0^L n_e B_{\parallel} dl~\rm{rad/m}^2$, 
with $B_{\parallel}$ the magnetic field component along the line of
sight in $\mu$G, $n_e$ is the electron number density in units
of cm$^{-3}$, and $dl$ the differential path in kpc.  
Examination of the wavelength variation in polarization 
seen through the ICM 
is used to measure distributions of RM across the observed sources. 
The RM distribution carries essential information 
about the ICM magnetic field strength and structure
when combined with X-ray data to establish the electron density distribution. 
As already noted (Sects. 2-3), the ICM magnetic field is expected to 
be turbulent. 
In that case $B_{\parallel}$ will fluctuate (around zero 
if the turbulence is isotropic), reducing the
mean rotation measure, $<RM>$, towards zero if the mean aligned field vanishes,
but adding a nonzero dispersion, 
$\sigma_{RM} = 812 \bar{n}_e \sigma_{B\parallel} \Lambda \sqrt{L/\Lambda}$, 
where $\bar{n}$ the mean number density, $\sigma_{B\parallel}$ the rms
strength of the line-of-sight magnetic field, $L$ the path length,
and $\Lambda = (3/2) L_B$ , with $L_B$
the 3D correlation length of the magnetic field \cite{ensslinvogt03,
choryu09}. 
The units are the same as before.
In practice the magnetic field estimates depend on estimations 
for $\sigma_{RM}$ (or sometimes $RM_{RMS}$) and $\Lambda$. 
Accuracies in these
are controlled by limited statistics in the RM distribution 
(since available coverage of the ICM is generally rather sparse),
by the indirect connection between the 2D RM coherence scale 
and the 3D magnetic field correlation scale, $L_B$ (but see \cite{vogtensslin05}), 
and often by uncertainties in the total ICM path to a polarized source, 
if it is embedded. 
Typically, RM samples are taken from several moderately extended background 
and embedded synchrotron sources 
distributed across an area comparable to or
larger than the cluster core. By adopting simple models for the systematic 
radial scaling between the magnetic field and the ICM density, along 
with simple parameterization of the magnetic field turbulence spectrum 
several such analyses have been conducted, with typical resulting 
central ICM magnetic field values $B_0 \sim 1 - \rm{few}~\mu$G 
\cite{carilli02, murgiaetal04, vogtensslin05, govoni06,
guidettietal08, bonafede10, kucharensslin11, vaccaetal12,
bonafedecoma13}.
While uncertainties in individual cluster results are probably at least 
a factor $\sim 2$, the general pattern of results seems fairly robust.

\begin{figure}[ht!]
\centering
\includegraphics[width=130mm]{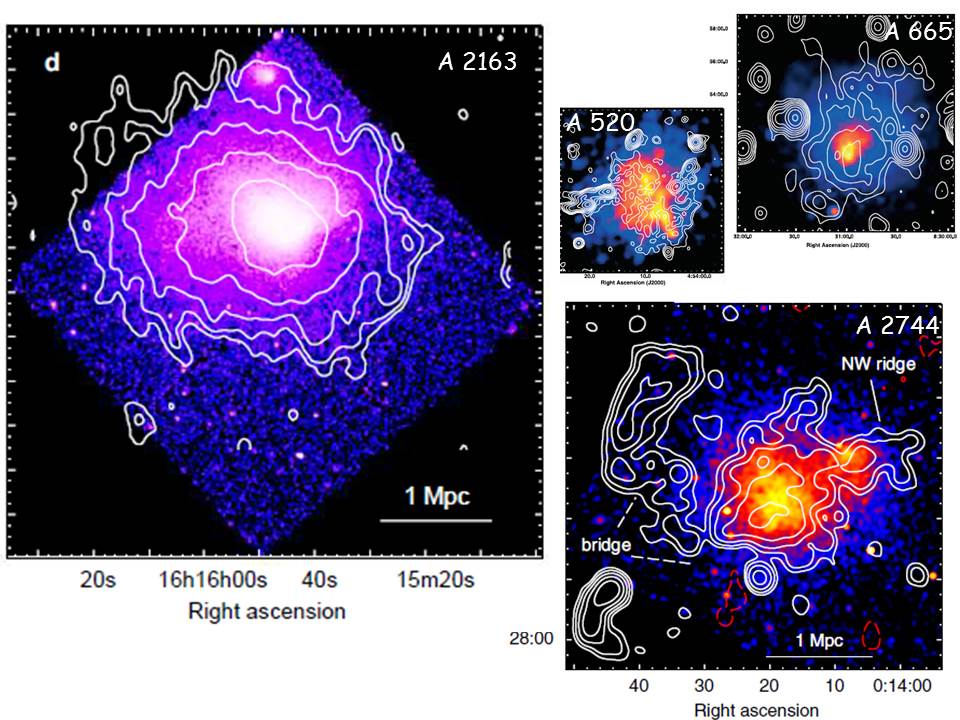}
\caption{
Radio images of giant radio halos (contours) overlaid on the thermal
X-ray emission of the hosting clusters.
Images are reported with the same physical scale (credits : Giacintucci
in prep. for A2163, \cite{govoni04} for A520 and A665, 
\cite{venturietal13} for A2744).
}
\label{fig:1}       % Give a unique label
\end{figure}

\subsection{Giant Radio Halos}

Returning to CRs, we note that 
two principal mechanism proposals are presently advocated to explain the
origin of CRe emitting in giant radio halos: i) (re)-acceleration 
of relativistic particles by MHD turbulence in the ICM 
\cite{brunetti01, petrosian01, ohno02, fujita03, brunettietal04,
cassanobrunetti05,
brunettilazarian07, petrosian08, brunetti08, 
donnertetal13, beresnyaketal13}
and, ii) continuous production of secondary
electron-positron pairs by inelastic hadronic
collisions between accumulated CRp and
thermal protons in the ICM \cite{dennison80, blasicolafrancesco99,
miniati01, pfrommerensslin04,  
keshetloeb10, ensslin11}, and their combination \cite{brunettiblasi05,
brunettilazarian11b}.

\noindent
The hadronic scenario is based on the physics described in Sect. 2.3
and allows one to resolve the {\it slow diffusion problem}\footnote{As
mentioned before, slow CR
diffusion in the ICM is incompatible with unavoidably rapid electron
energy loss rates unless observed electrons are injected or 
accelerated throughout radio halo volumes.}, because CRe
are continuously injected {\it in situ} throughout the ICM.
Also the morphological connection between radio and X-ray emission
in galaxy clusters can be explained by this scenario because the X-rays
trace the thermal matter that provides the targets for the
hadronic collisions.
An unavoidable consequence of this scenario is the emission of
$\gamma$-rays due to the decay of $\pi^0$ that are produced by the
same decay chain that is responsible for the injection of secondary
CRe (Sect. 2.3).

\noindent
The turbulent acceleration model is based on the physics described
in Sect. 2.2.2 and assumes that turbulence is generated strongly during cluster
mergers and that a fraction of its energy is dissipated into
(re)acceleration of CRe via Fermi II --type mechanisms.
In this case the slow diffusion problem is also solved, because the
emitting CRe are (re)accelerated {\it in situ}, for a fairly
long period (of about 1 Gyr) by merger--induced turbulence that is assumed
to fill a substantial fraction of the volume of clusters.
The most obvious expectation of this model is a tight connection
between giant radio halos and cluster mergers, because of the finite
decay time for merger-generated turbulence. 
According to this model radio halos could 
have complex, spatially
varying (and potentially very steep) spectra due to the breaks and cut-offs 
that are produced in the 
spectrum of the emitting CRe 
as a result of the balance between (spatially varying)
acceleration and cooling (Sect. 2.2.2, see also Fig. 7).

\subsubsection{Observational milestones and origin of giant radio halos}

Pioneering studies using Arecibo and the
NVSS and WENSS radio surveys suggested that radio halos 
are not common in galaxy clusters \cite{hanisch82,
giovannini99,kempnersarazin01}.
Current observations show that 
not all clusters have diffuse radio
emission, with only $\sim 1/3$ of X-ray luminous systems hosting giant radio
halos. This provides one of the most relevant constraints
for understanding the origin of CRe in radio halos.
In this respect an important step has been achieved in the last few years,
thanks to meter wavelength radio observational campaigns at the 
GMRT\footnote{http://gmrt.ncra.tifr.res.in/} 
combined with observations at higher radio
frequencies and in the X-ray band
\cite{venturi07,venturi08, cassano08, cassano10a, cassanoetal13, 
kaleetal13}.
The high sensitivity surveys with the GMRT found 
that clusters with similar thermal
X-ray luminosity, and presumably similar mass, branch into two
populations, one hosting radio halos and a second one
with no evidence for halo-type cluster-scale radio emission at the sensitivity
level of current observations \cite{brunetti07, brunettietal09}
(Figure 12, left). Related to this finding, 
another observational milestone that has been achieved in the last decade
is the connection between giant radio halos and the 
dynamics of the hosting clusters, with halos always found only
in merging systems \cite{buote01,feretti12}.
Firm statistical evidence of that has been recently
obtained from combined radio -- X-ray studies of galaxy clusters
in the GMRT surveys \cite{cassano10a, cassanoetal13} (Figure 12, right).
These studies have shown that the generation of giant radio halos 
occurs during mergers between galaxy clusters.  This leads to a
number of possible physical interpretations. The most obvious are that 
turbulence generated during cluster mergers 
may rapidly accelerate CRs (Sect. 2.2.2)\cite{brunetti07, brunettietal09,
donnertetal13}
or that the cluster magnetic field can be amplified by the turbulence during these
mergers \cite{kushniretal09, keshetloeb10}. 
Less obvious hypotheses to connect halos and mergers 
have been proposed, including 
the possibility that spatial diffusion or streaming of CRs plays 
a role in modifying
the level of synchrotron emission in galaxy clusters, provided that the diffusion or streaming
depends strongly on the dynamical state of the cluster\cite{ensslin11,
wieneretal13}.
More recent studies attempt to better constrain the occurrence of radio
halos in galaxy clusters 
by selecting clusters using the SZ-effect \cite{basu12}, 
that is a better indicator of cluster 
mass with respect to the clusters X-ray emission \cite{carlstrom02}.
The combination of the GMRT surveys, mentioned above, with the 
Planck SZ catalog \cite{plancksz13} shows that clusters branch into
two populations (radio halos and limits) also in 
a radio--SZ diagram (although this bimodal behaviour appears weaker
than that in X-ray diagrams), and that the two populations correlate with the
cluster dynamics \cite{cassanoetal13} as in previous X-ray
based studies. However the fraction of clusters hosting radio halos
using SZ selected samples appears larger than
that measured using X-ray clusters samples \cite{sommerbasu13}.
That may be due to the combination of a distinct time evolution
of the SZ ad X-ray signals from the
ICM during cluster mergers and a bias toward cool-core systems in
X-ray selected samples \cite{sommerbasu13}, and 
opens to complementary ways to study the connection of thermal and
non-thermal components in galaxy clusters.

\begin{figure}[ht!]
\centering
\includegraphics[width=130mm]{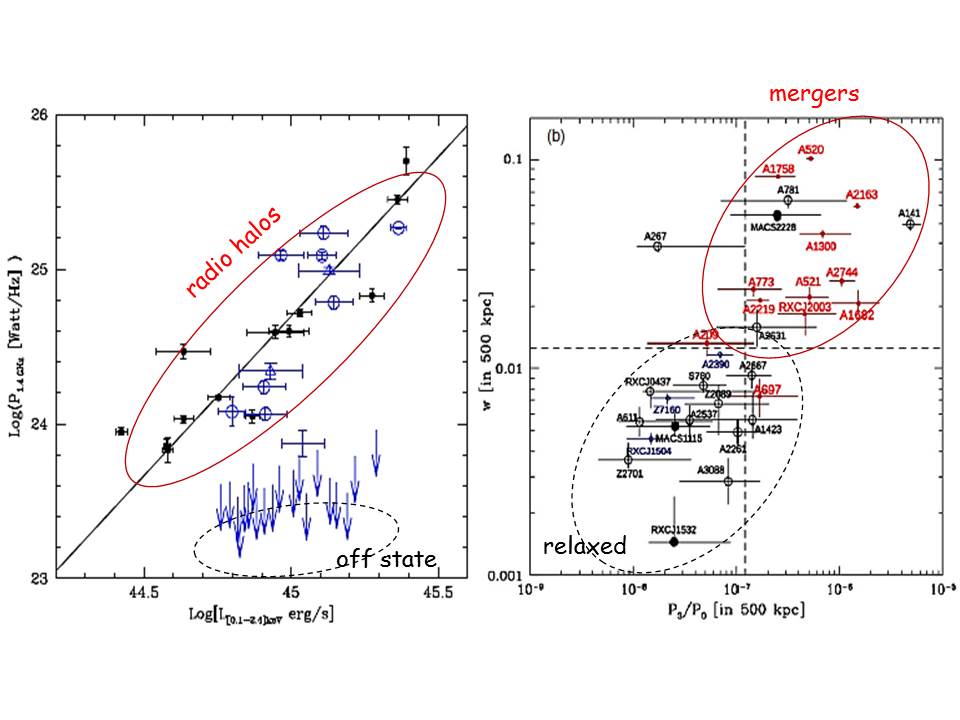}
\caption{
Left panel: distribution of galaxy clusters of the GMRT sample''
\cite{venturi08} in the radio power -- X-ray luminosity diagram,
showing that clusters branch into two :
giant radio halos (the merging systems in the left panel of the
figure) and
off state'', undetected, systems (the relaxed systems in the left panel
of the figure) (adapted from \cite{brunettietal09}).
Right panel: distribution of galaxy clusters
in the centroid-shift variance $w$ vs power ratio $P_3/P_0$
diagram. Mergers are expected in the top-right panel, relaxed systems
in the bottom-left panel. Clusters hosting giant radio halos are
reported in red (adapted from \cite{cassano10a}).
}
\label{fig:1}       % Give a unique label
\end{figure}

\noindent
The existence of a connection between thermal and non-thermal ICM components 
is also highlighted by point-to-point correlations
discovered between the synchrotron brightness of giant radio halos
and the X-ray brightness of the hosting clusters \cite{govoni01, feretti12}
and, in the case of the
Coma cluster, between the Compton $y$-parameter and the radio
brightness \cite{planck13}. 
These correlations have been used to claim that the spatial
distribution of CRs in galaxy clusters is generally broader than that
of the thermal ICM, with implications on the physics discussed in
Sect. 3.2. In addition a number of authors have identified the very
broad (flat) spatial profiles of giant radio halos as a potential
challenge for a purely hadronic origin of these sources, because in
some cases a large, or radially increasing, energy density of
CRp and/or magnetic field is required to generate the observed radio
emission from radially distant, low target- ICM density, regions
\cite{brunetti03, brunetti04, marchegianietal07, donnertetal10,
planck13, zandanel13}.
Potential constraints on the thermal -- non-thermal connection and on
the physics of radio halos also come from 
possible trends between the local
synchrotron spectral indices and ICM temperatures \cite{feretti12}.
Still, these relationships are only hints at the full picture.  
Only future radio observations that combine
extreme brightness sensitivity and high angular resolution, for
example using the SKA and its precursors, will provide the
opportunity to exploit adequately the physical information encoded
in the above correlations, including the interplay between particle 
acceleration and transport in the ICM and their connections with the
magnetic field properties.

Important constraints on the origin of giant radio halos already come from the
combination of radio and $\gamma$--ray observations.
Current upper limits to the $\gamma$-ray emission from galaxy
clusters, including clusters hosting radio halos, (Sect. 3.3, Figure 10) 
put especially stringent constraints on the
role of hadron-hadron collisions and their secondary products for
the origin of these sources.
Specifically, this information can strongly constrain 
cluster magnetic field requirements. 
From eq.(17), (19) and (21) the ratio of synchrotron luminosity
from secondary CRe and 
$\gamma$-ray luminosity due to $\pi^0$ decay
depends on the magnetic field strength in the ICM as,

\begin{equation}
{{P_{syn}}\over{P_{\gamma}}} \propto
\big\langle {{B^{1+\alpha}}\over
{B^2 + B_{IC}^2}} \big\rangle,
\end{equation}

where $<..>$ indicates an emission-weighted quantity and $\alpha$ is the
synchrotron spectral index.
Consequently, $\gamma$-ray upper limits derived for clusters
hosting giant radio halos provide corresponding lower 
limits to the strength
of the magnetic field in the volume occupied by radio halos. 
For a few giant (and nearby) radio halos, including their prototype in the
Coma cluster, current limits challenge a pure hadronic origin, 
because the values allowed for the magnetic fields are inconsistent
(or in tension) with 
those estimated from studies based on Faraday rotation 
measures \cite{brunetti09, jeltemaprofumo11,
brunetti12}.
Figure 13 (left) shows the case of the Coma cluster
radio halo, where the allowed region for the magnetic field is derived
within a pure hadronic model 
by combining the Fermi-LAT limits with both the spectrum of the halo and
its brightness distribution at 330 MHz \cite{brunetti12}.
In other words, assuming that rotation measures
give a relatively fair view of the magnetic fields in
clusters (Sect. 4.1, Fig. 13) and more explicitly do not seriously underestimate their strengths, 
the energy budget required for the CRp component to explain radio halos 
in the context of hadronic models is larger than that 
allowed by the existing limits to the cluster $\gamma$-ray emission.

\begin{figure}[ht!]
\centering
\includegraphics[width=130mm]{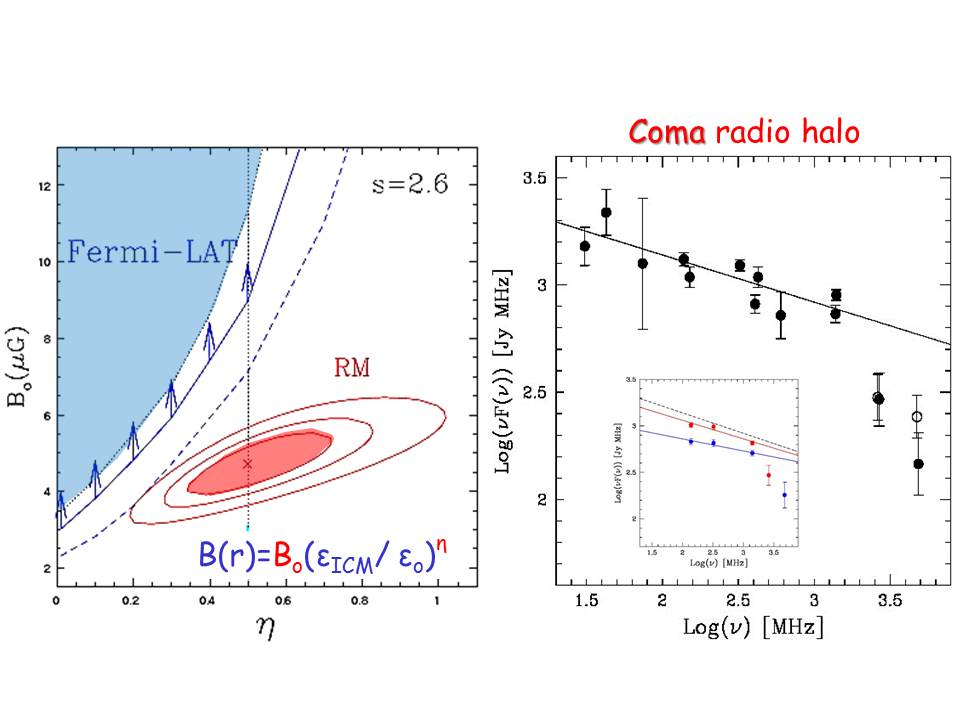}
\caption{
Left panel: the allowed region for the central magnetic field
$B_0$ -- $\eta$ parameter (defined in the bottom of the panel) 
derived from the combination of Fermi-LAT limits
and the properties of the radio halo at 330 MHz (blue region),
is compared with that derived 
from the analysis of RM (red region)
(adapted from \cite{brunetti12}).
Right panel: the spectrum of the Coma halo (empty points at
2.7 and 4.8 GHz are corrected for the SZ-decrement).
The inset is the spectrum of the Coma halo (corrected for the SZ
decrement) extracted using an aperture of 13 
(lower-blue) and 17.5 arcmin radius (upper-red). The spectra
show the steepening at higher frequencies (adapted from 
\cite{brunettietal13}).
}
\label{fig:1}       % Give a unique label
\end{figure}

The spectra of radio halos provide crucial information for the 
origin of the emitting CRe in the central Mpc-region of the hosting
clusters.
The accurate measurement of the integrated spectra of radio halos is a
difficult task. Radio halos usually embed a number of individual sources
and, sometimes, projected foreground and background sources,
whose flux density needs to be carefully subtracted from the total diffuse
emission. This requires high quality imaging over a range of resolutions.
Moreover, because of their steep spectra, diffuse cluster sources 
are best imaged at low frequencies.  
High quality, high resolution imaging at frequencies below 1.4 GHz 
has become available
only very recently. For this reason, a  meaningful spectrum
with data points at three or more frequencies spread over 
$\sim$ 1 order of
magnitude is currently available only for a few objects, e.g. \cite{venturi11}.

\noindent
The radio halo in the Coma cluster is the prototype of this
class of radio sources \cite{wilson70, venturi90, brownrudnick11}.
It is a unique case, with
spectral measurements spanning almost 2 orders of magnitude in frequency
\cite{thierbach03}(Figure 13, right).
The measured spectrum significantly steepens at frequencies above about 
1 GHz.
A power-law that fits the data at lower frequencies overestimates the
(SZ-corrected) flux measured at 2.7 and 5 GHz by factors of 2 and 3,
respectively \cite{brunettietal13}.
This suggests a break (or cut-off) in the spectrum of the emitting CRe
at a maximum energy, $E_{max}$, around a few GeV. 
Establishment of the steepening of the spectrum of the Coma halo is 
an observational
milestone, because it allows one to estimate the acceleration time-scale
(or equivalently the efficiency) of the mechanism responsible for the origin 
of the CRe responsible for the radio halo.
The maximum energy of accelerated CRe is given by the competition between the
acceleration rate and (for observable, GeV electrons) radiative losses. 
From eq. (22) the lifetime of radio-emitting CRe using 
viable magnetic fields in the Coma cluster is 
$\tau_e \sim~\rm{few}\times 10^8$ yrs, which is then also the time-scale of
the acceleration mechanism\footnote{This match applies, provided the acceleration 
process operates for a time-period
that is not short compared to the required acceleration time-scale (energy doubling time). 
Otherwise, a balance 
between acceleration and losses obviously cannot be achieved.}.
The existence of this spectral break implies that the mechanisms responsible for CRe
acceleration
must be of relatively low efficiency (long acceleration time-scale), or alternatively
that they
must intrinsically produce a break in the distribution of CRe at
$E_{max}\sim$few GeV.

\noindent
In the case of turbulent acceleration (Sect. 2.2.2) we
can roughly estimate an acceleration time-scale (see eq. (15) with
$\delta B^2/B_0^2 \sim {\rm V}_t^2/c_s^2$)

\begin{equation}
\tau_{acc} \sim  {{p^2}\over{ {\cal D}_{pp} }} \sim
{{c\, l}\over{{\rm v}_t^2}},
\end{equation}

\noindent
where ${\rm V}_t$ is the ICM turbulent velocity. With $l \sim$ kpc 
and ${\rm V}_t \sim 100~\rm{km~s}^{-1}$
the acceleration times, $\tau_{acc} \sim 10^8$ yrs, providing a relative
match at energies $E_{max} \sim$few GeV between acceleration and cooling.
The spectrum of higher energy electrons will be steepened by
the increasing relevance of cooling (see Figure 7 for a representative
re-accelerated spectrum), 
thus leading to the observed break in the
radio synchrotron spectrum at a frequency

\begin{equation}
\nu_{max} \approx C_{SYN} E_{max}^2 B 
\propto
{{ {\cal D}_{pp}^2 B }\over{(B^2 + B_{IC}^2)^2}}.
\end{equation}

\noindent
where $C_{SYN} \sim 2 \times 10^7$ if frequency, energy and 
magnetic field are measured in Hz, GeV and $\mu$G, respectively.

\noindent
Schlickeiser et al (1987) \cite{schlickeiser87}
first realized the importance of the
measurements of the spectrum of the Coma radio halo and indeed suggested
an origin of the CRe based on stochastic acceleration due to turbulence,
disfavouring other mechanisms that included the generation of secondary
electrons by hadronic collisions. In this latter case the 
CRe spectrum extends in principle, to very high energies, if
CRp extends to very high energies (say $E_{CRp} > 100$ GeV), as
indeed expected from the discussion in Sect. 2.
Therefore, no intrinsic cut-off would be expected in the radio spectrum
\cite{blasi01, blasi07}.

\noindent
Spectral behaviours qualitatively similar to that of the Coma halo are expected 
in other radio halos if they are powered by turbulent acceleration.
However, the generality of this behaviour is still unclear from the observational side,
because of the observational difficulties in obtaining reliable
measurements of radio halos spanning an adequately large frequency range.
As a matter of a fact, the spectra of most giant radio halos 
are constrained by only a few data-points (typically 2-3)
spanning less than one order
of magnitude in frequency range, and are consistent with a simple 
power law\footnote{see however e.g. \cite{kale10, vanweerenlofar12} for A2256}.
Most importantly, however, in the last decade it
has been found that
the observed values of the slopes of the integrated spectrum 
of radio halos span a broad range of values, $\alpha \sim 1 - 2$
($F(\nu)\propto \nu^{-\alpha}$, e.g. \cite{venturi11,
feretti12, venturietal13}).
This implies (at least) that the synchrotron spectrum of radio 
halos is not
a {\it universal} power law, and puts constraints
on the mechanisms responsible for the acceleration of the CRe.
Particularly stringent constraints derive from 
halos with extreme spectral properties, $\alpha \sim 1.5-2$. 
In these cases 
the energy budget of CRs that would follow from the assumption
that the energy distribution of CRe is a power law extending to lower 
energies, is unacceptably large\cite{brunetti04,
pfrommerensslin04, brunetti08, macarioetal10, macarioetal13}.
One possibility is that the spectrum of the 
emitting CRe breaks towards high energies,
producing a very steep radio spectral shape within the observed
frequency range. This requires, of course, that such a
break in the CRe spectrum is formed at suitable energies (i.e., in the GeV
range responsible for the radio 
emissions)\cite{cassanobrunettisetti06, brunetti08}. 
Under this hypothesis the constraints on the 
acceleration mechanisms are similar to those derived above for
the Coma radio halo.

\subsubsection{Open problems and Future Efforts}

Despite the discussion above (and in Sect. 2.2.2) that provides arguments 
that are consistent with the hypothesis that giant radio halos trace turbulent
regions in merging clusters and possibly originate due to  
(some type of) turbulent-related mechanism of CRs acceleration, many 
important points are still open.
First of all, the role of secondary CRe is still unclear.
It is true in the case of several radio halos that  $\gamma$-ray limits 
and arguments connected with the observed radio spectra (and brightness
distributions) 
seriously challenge a purely hadronic, 
simply injected $e^{\pm}$ population for the observed emission.
However, it is also fair to say that in general the existing 
data are still insufficient to discriminate clearly among 
different more complex scenarios.
Most important, secondaries will exist at some level, since
CRp are virtually certain to exist in clusters and will undergo inelastic 
collisions with the thermal ICM (Sects. 2.3 and 3).
Consequently Mpc-scale radio emission should
be generated at some level in ``all'' clusters with luminosities 
that should vary depending on the cluster 
dynamical histories  (e.g., accretion shock histories) and/or 
magnetic properties (e.g., histories of magnetic flux injection 
and amplification)\cite{miniati01, keshetloeb10, brown11, cassano12,
zandanel13}.
In this respect future radio surveys with 
sensitivities to diffuse cluster-scale emission much better than current
surveys, will provide unique and critical constraints to the role played
by secondaries. For reference, {\it Hybrid models}, that assume 
radio halos to be generated by
the (re)acceleration of secondary particles by turbulence
in cluster mergers \cite{brunettiblasi05, brunettilazarian11b}, 
predict that secondaries generated
in more relaxed systems 
produce radio emission with luminosity marginally smaller
than current upper limits in Figure 12 \cite{brunettilazarian11b,
brown11}.

\begin{figure}[ht!]
\centering
\includegraphics[width=130mm]{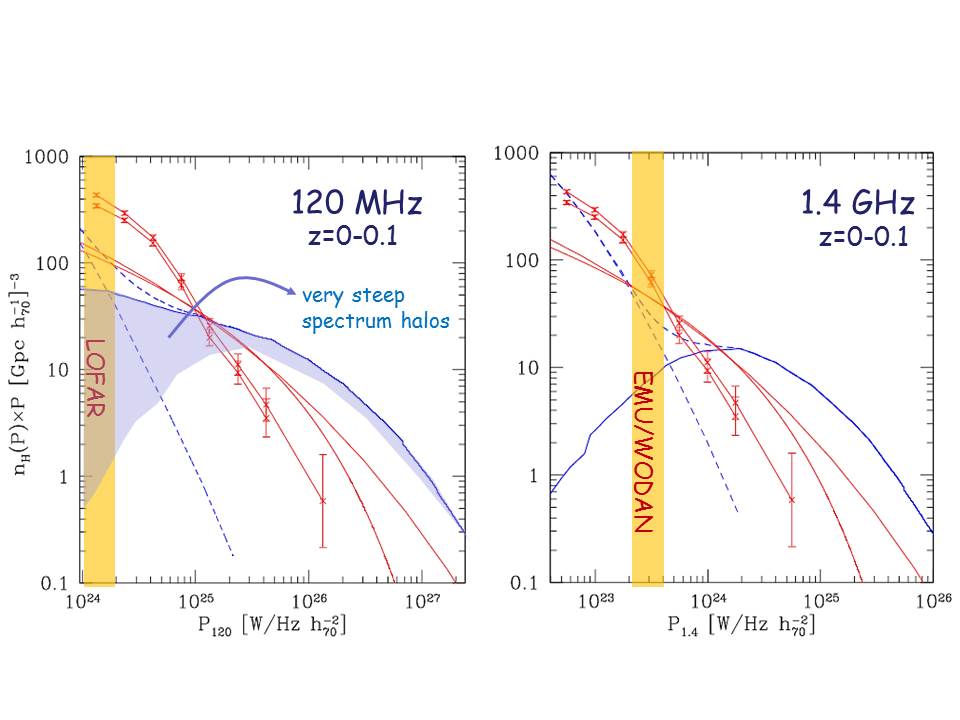}
\caption{
Radio luminosity functions of giant radio halos at $z=0-0.1$
predicted by different models at 1.4 GHz (right) and 120 MHz (left).
Blue lines are based on turbulent re-acceleration models (solid line)
and secondary emission (thin dashed lines) from `off state'' systems (i.e.
undetected halos in Fig. 12); the thick dashed-lines represent the
sum of the two contributions (adapted from \cite{cassano12}).
Red-solid lines are from \cite{ensslinrottgering02}, while the
red-points are from numerical simulations by \cite{zandanel13} that
are based on secondary models and assume that 10\% of systems host
radio halos (note that those simulations include both giant and mini
halos).
The blue region in the left panel shows the contribution
to the luminosity function at 120 MHz due to very steep spectrum halos
that are predicted in the turbulent re-acceleration model and do not
contribute at higher frequency.
The nominal sensitivity level of LOFAR and EMU/WODAN surveys is also
reported.}
\label{fig:1}       % Give a unique label
\end{figure}

Future radio surveys are likely to produce a big step forward in our
understanding of the formation and evolution of radio halos.
The measure of the formation rate of radio halos in galaxy 
clusters and its dependence on the 
cluster thermal properties (dynamics, mass, temperature etc) 
provides unique information on the physics of these 
sources. One of the main limitations of current studies is  
that only X-ray luminous clusters have been 
scrutinized with adequate radio
observations, whereas only very poor information is available for less
luminous systems \cite{feretti12, farnsworthetal13}
that are expected to host fainter 
halos according 
to the correlation in Fig. 12. A similar limitation applies also to the
most recent studies that
focus on mass-selected cluster samples, that are based on SZ-catalogs,
that are indeed limited to
cluster masses $M_{500} \geq 6 \times 10^{14}M_{\odot}$\cite{cassanoetal13}.
These limitations are due to the limited sensitivity of present radio
observations, but soon will be eliminated thanks to a
new generation of
radio telescopes, such as LOFAR, ASKAP, and in the 
longer period, the SKA, that will survey the sky with 
unprecedented sensitivities to the cluster scale diffuse emission.

LOFAR and, to a somewhat lesser extent, MWA and LWA,
will be particularly important
because they will observe at lower frequencies, between 10 MHz
and 200 MHz, stepping into a basically unexplored frequency range. 
LOFAR observations and their combination with
JVLA\footnote{https://science.nrao.edu/facilities/vla} observations and,
for nearby radio halos, observations with single dish radio telescopes 
(such as GBT and Effelsberg), will allow derivation of 
spectral measurements of radio halos over unprecedented, wide 
frequency coverage, thus 
providing crucial constraints on theoretical models. In particular, one 
of the most intriguing theoretical hypotheses is the existence of
radio halos with very steep spectra that are undetected by
radio surveys at higher frequencies and that should be discovered 
by observations at lower frequencies\cite{cassanobrunettisetti06,
brunetti08, cassano10b, cassano12}.
Such a hypothesis is based on turbulent acceleration models that 
predict that less energetic merger events (less turbulent)
generate radio halos with
very steep spectra. Since these less energetic mergers are more common,
one can speculate that steep spectrum halos constitute a large
population of halos that is presently invisible. 
According to present calculations that use the crude assumption
that a {\it fixed} fraction of the cluster-merger energy goes 
into MHD turbulence available for particle acceleration on Mpc-scale, 
the LOFAR surveys should detect about 500 new radio halos.
About half of these halos should have very steep spectra 
\cite{cassano10b, cassano12}.
Despite their still-early development and need for greater sophistication,
current models have been able to make some notable predictions
\footnote{We note that, independent of the theoretical
scenario that is adopted for the origin of radio halos, 
predictions of the formation rate of
radio halos and their number counts are severely limited by
uncertainties in the properties of cluster magnetic fields - including
their evolution with cluster mass and cosmic epoch - and CR dynamics.}. 
For instance, re-acceleration models predict that the shape of the Luminosity
Functions of radio halos change with observing frequency due to the
contributions of the expected population of steeper-spectrum halos
at lower frequencies (Figure 14).
This differs from predictions based on other scenarios, including the
hadronic model, where all halos should have to a first approximation
similar spectra (independent of the mass of the hosting 
clusters and of the energy
released during mergers), and provides a valuable way to put constraints on
the origin of these sources with future radio surveys.
In particular, as shown in Figure 14, the most powerful way will be probably 
the combination of LOFAR surveys, at low frequencies, with future surveys with ASKAP or 
Apertif\footnote{http://www.astron.nl/general/apertif/apertif}, at higher frequencies.

\begin{figure}[ht!]
\centering
\includegraphics[width=130mm]{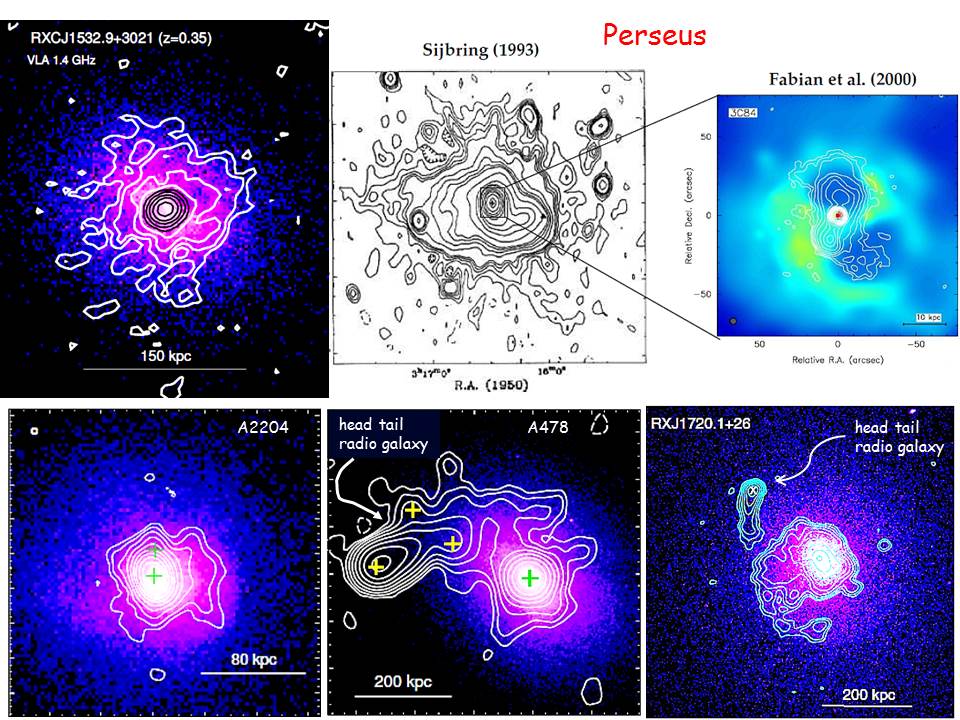}
\caption{Images of radio mini-halos (contours) and the X-ray emission
from the hosting clusters (colors). Credits: \cite{giacintuccietal13}
for RXCJ1532, A2204, A478, Giacintucci in prep. for 
RXCJ1720, and \cite{gittietal04} for Perseus.
}
\label{fig:1}       % Give a unique label
\end{figure}

\subsection{Mini Halos}

In addition to the disturbed, merging clusters that have been found 
to host giant radio halos, 
a number of relaxed, cool-core clusters host faint,
diffuse radio emission with a steep spectrum and a 
size comparable to that of the cool-core region \cite{feretti12} 
(Figure 15), so substantially smaller than the giant halos.
As a first approximation one might have thought these sources as 
a version of giant radio halos scaled on smaller scales of the 
order of few 100 kpc in size rather than Mpc size. 
However, the two 
classes of radio sources show prominent differences.
First of all, in clear contrast to giant radio halos,
mini halos are always found in dynamically relaxed
systems suggesting that cluster mergers do not
play a major role for their origin.
Also the synchrotron volume emissivity of mini halos differs from that
of giant halos, being typically larger
\cite{cassanogittibrunetti08,
murgiaetal09}.
Whether the underlying physical mechanisms that accelerates the 
CRe in mini halos differ substantially from the analogous mechanisms 
in giant radio halos is still unclear.
Clusters hosting mini halos always have
central radio-loud AGNs exhibiting outflows
in the form of radio lobes and bubbles inject CRe into the central regions
of galaxy clusters (Sect. 2.1).  In principle these AGNs could
represent the primary source of the CRs 
in mini halos, however
they are not sufficient by themselves, at least without some dynamical
assistance, to explain the diffuse
radio emission.  In particular, a {\it  slow diffusion problem} once again exists for
mini halos; that is, the energy loss time scale of the radio emitting 
CRe is still much shorter
than the time needed by these particles to diffuse efficiently across the emitting
volume. 

Similarly to giant halos, two physical mechanisms 
have been identified as possibly responsible for the
radio emission in mini halos: i) re-acceleration of CRe (leptonic models or
re-acceleration models) and ii) generation of secondary CRe (hadronic
or secondary models).

\noindent
According to leptonic models mini halos originate due
to the re-acceleration of pre-existing, relativistic CRe
in the ICM by turbulence in the core
region, e.g. \cite{gittibrunettisetti02}. 
So, this model is similar in character to the turbulence 
model for giant halos, except
that the responsible turbulence is concentrated in the cluster core region. In this case 
obvious sources of the seed CRe are, for example, 
the buoyant bubbles that are inflated by the central AGN 
and disrupted by gas motions in the core 
\cite{heinz06,bruggen09,cassanogittibrunetti08}.
A key question in this model
is the origin of the turbulence responsible
for re-accelerating the electrons. Unlike giant halos, merger shocks 
and Mpc-scale motions/flows would not be candidates.  
Gitti et al.(2002)\cite{gittibrunettisetti02} originally
proposed that the cooling flow of gas inward within the core
could generate the needed turbulence. More recent X-ray observations however 
have shown that these ``classical cooling flows'' 
do not materialize, e.g. \cite{petersonfabian06}.
On the other hand, 
observations suggest that even relatively
relaxed clusters have large-scale gas motions in 
their cores at significant fractions of the local sound speed.  
The clearest observational signatures of these gas 
motions are spiral-shaped “cold fronts” seen in the majority of
cool-core clusters \cite{markevitchvikhlinin07}.
These cold fronts are believed to be produced by the cold gas of the core
“sloshing” in the cluster’s deep potential well, 
in response to passing dark matter 
subhalo motions, for example.
Those sloshing motions, can advect ICM across the cluster core and can also
produce turbulence there \cite{fujita04,
askasibarmarkevitch06, zuhone10, zuhone11}.
Remarkably, a correlation between radio mini halos and
cold fronts has been discovered in a few clusters and has been used to
suggest a connection between radio mini halos and turbulence generated by the
sloshing motions\cite{mazzottagiacintucci08}.
Recent simulations\cite{zuhone13} support this hypothesis 
and have shown that the motions and turbulence generated by
core sloshing in galaxy clusters 
can re-accelerate and spatially redistribute seed CRe. 
Those CRe and magnetic fields amplified at the same time produce 
radio emission that resembles mini halos, being 
diffuse, steep spectrum, and connected with the spatial features of 
the X-ray emitting gas. As noted early in this review, 
AGN activity also can dump considerable kinetic
energy into the ICM and drive cluster-core scaled ICM motions (Sect. 2.2.2).  
While these motions can mix CRe into ICM relatively near the AGN outflows, 
larger scale flows, such as sloshing
are probably necessary for those CRe to become broadly 
distributed across the full cluster cores.
Current observational information on turbulent motions 
in cool cluster cores coming from X-ray emission lines is limited 
by relatively poor
available spectral resolution. 
In the few cases of compact cool cores the turbulent 
velocities have been constrained 
and the turbulent pressures (due to fluctuations on scales much smaller
than the size of these compact cores) are found to be less 
than $\sim$ 10\% of the thermal 
pressures \cite{sandersetal10, bulbuletal12}.
Future observations with ASTRO-H will provide much better constraints 
as they will be able to reveal the presence of small-scale turbulence
(with velocities down to about 100 km/s) from the study of X-ray
lines \cite{takahashi12, zhuravleva12}.

The second mechanism proposed for the origin of mini halos
is, just as for giant halos, based on the generation of secondary
particles via inelastic collisions between CRp and thermal protons
\cite{pfrommerensslin04, keshetloeb10, zandanel13}.
In this case the primary CRp could be provided by the central
AGN and then advected, streamed or diffused across the cluster core.
In this case an interesting issue is the possible connection
between these CRp, which ultimately generate the 
observed mini halos in this picture,
and the heating of the ICM in the region of the cool core.
Colafrancesco \& Marchegiani (2008)\cite{colafrancescomarchegiani08} 
proposed a scenario where
the cooling of the gas in the cool cores is balanced by the
heating due to the CRp via Coulomb and hadronic interactions.
In order to balance the cooling of the gas, these models assume that CRs in 
cool cores carry a substantial
fraction of the energy of the ICM, a condition that however is
ruled out by current $\gamma$-ray observations that put severe
limits to the energy budget of CRs (Sect. 3.3) in core
regions\cite{huberetal13}.
More recent approaches that model the CRp-driven heating mechanism
rely on the possibility that CRp streaming and the 
related generation of MHD waves can heat the ICM in the cool cores
with an efficiency that may be much larger than that due to Coulomb
and hadronic interactions\cite{guooh08, fujitaohira12}.
Under this hypothesis several calculations have derived consistent
pictures of cool-core heating and generation of radio 
mini halos\cite{fujitaohira13}.

Discrimination between a leptonic and hadronic origin of radio mini halos
is very challenging due to severe limitations of current 
observational constraints.
Although a correspondence between radio mini halos and cool core
clusters is well established,
current radio studies do not provide an exhaustive view of the
occurrence of radio mini halos in galaxy clusters more generally. For example,
it is still not clear whether these radio sources 
are common or rare in cool cores\cite{giacintuccietal13}. 
In addition, $\gamma$-ray upper limits from  the Fermi satellite and ground
based Cherenkov
telescopes are not yet deep enough to put strong constraints on the origin
of these radio sources. For example, these limits do not put significant
tension on a purely hadronic origin of mini halos \cite{perkinsetal06,
aleksic12}. Generally, in comparison with the case
of giant halos, the combination of radio and $\gamma$-rays constraints are 
less restrictive for mini halos, because the ratio of $\gamma$-ray luminosity 
to radio luminosity 
is smaller.  This is due to the fact that the likely value of the
average magnetic field in the much smaller and more central volume occupied 
by mini halos is larger
than that in the much bigger volume of giant halos (see eq. 26).
In addition, we note that for the best test case, the mini halo in
the nearby Perseus cluster, the presence  
of strong $\gamma$-ray emission from the central
radio source 3C 84 (NGC 1275)
does not allow one to put stringent limits on the diffuse emission from the
cluster core in the critical 0.2-10 GeV band \cite{ackermann10}.

One possibility to discriminate between different models is to look for 
differences in the integrated spectrum and
spectral index maps. For example, compelling evidence
for very steep spectra and/or for the existence 
of spectral breaks at high frequencies would favour
a re-acceleration scenario rather than a hadronic origin in the mini halos.
Similarly to the case of giant halos, current  mini halo radio observations 
cover too short a frequency range to constrain spectral curvatures.
Furthermore, the fractional contribution to the observed flux 
that is provided by discrete radio sources embedded in the halo volume is
larger for mini halos, making good spectral measurements more
challenging than in the case of giant halos \cite{feretti12,
giacintuccietal13}.
Better spectral constraints 
might come from future observations at low (e.g. with LOFAR) and 
higher (JVLA) radio frequencies. These radio telescopes,
and their combination, should allow deeper
imaging, with unprecedented dynamic range, and measurement of the diffuse 
synchrotron spectrum over a fairly wide frequency range. 

\subsubsection{A connection between giant and mini radio halos ?}

An obvious, interesting question is whether there is any connection
between mini halos and giant radio halos or do they have
independent physical origins? 
Despite the apparently distinct population distributions, there are, in principle, 
several arguments in support of a possible connection.
Indeed, one could argue that the driven flows and
turbulent motions that destroy the ICM cores of clusters during merger 
could also transport 
and re-accelerate CRs on larger, Mpc, scales. This, in fact,
is a ``generic'' candidate-scenario that we propose 
to switch off bright mini halos and to begin powering giant
radio halos in dynamically active systems.
In this scenario complex situations where a central mini-halo
is embedded in a lower brightness radio emission on larger scales
should exist in dynamically ``intermediate'' systems; Abell 2142
could be one example\cite{rossettietal13, farnsworthetal13}.

\noindent
In a recent work Zandanel et al.(2013)\cite{zandanel13} proposed that
mini-halos are primarily of hadronic origin, while giant radio halos
experience a transition from central hadronic emission to leptonic
emission component in the external regions due to CRe re-acceleration.
Potentially, also diffusion and advective transport of CRs in the ICM can  
induce an evolution from mini to giant halos (and vice-versa), assuming 
that this process is sufficiently fast (i.e., $\leq$ Gyr) on spatial scales 
of clusters cores\footnote{Streaming has also been proposed to explain the 
bimodality of giant radio halo populations
\cite{ensslin11,wieneretal13}}.
A similar evolution could be driven by the evolution (amplification
and dissipation) of the magnetic field in the cluster core and periphery 
in response to mergers \cite{kushniretal09, keshetloeb10}.

Observations with the next generation of observational facilities 
and similar advances in numerical simulations will hopefully clarify 
these issues in the near future.

\subsection{Giant Radio Relics}

Some merging clusters host peripheral, giant radio relics.
As in the case of giant radio halos,
a few tens of giant relics have been discovered so far.
Radio relics differ from radio halos in the morphologies, peripheral
locations and polarization 
(typically being up to 30\% level in integrated
linear polarization)\cite{feretti12} (Figure 16).
These properties provide a clear starting point in establishing
their origin. Specifically, there is broad consensus that the giant 
radio relics trace shocks outside cluster cores, 
probably relatively strong merger shocks,
where the emitting CRe can be accelerated 
or re-accelerated\cite{ensslin98,
roettiger99b,miniati01, ensslingopal01, markevitchetal05, bagchi06,hoeft07,
hoeft08,vanweeren10,skillman11,bruggen12, kang11,
kang12, pinzkeetal13}. This model is based on
the physics described in Sect. 2.2.1.

\noindent
In addition to probing the origin of CRs in galaxy clusters, relics are also
important probes of the magnetic field properties in the ICM
periphery, as they are found at distances up to a large fraction 
of the cluster virial radius, e.g. \cite{bruggen12}.

\begin{figure}[ht!]
\centering
\includegraphics[width=130mm]{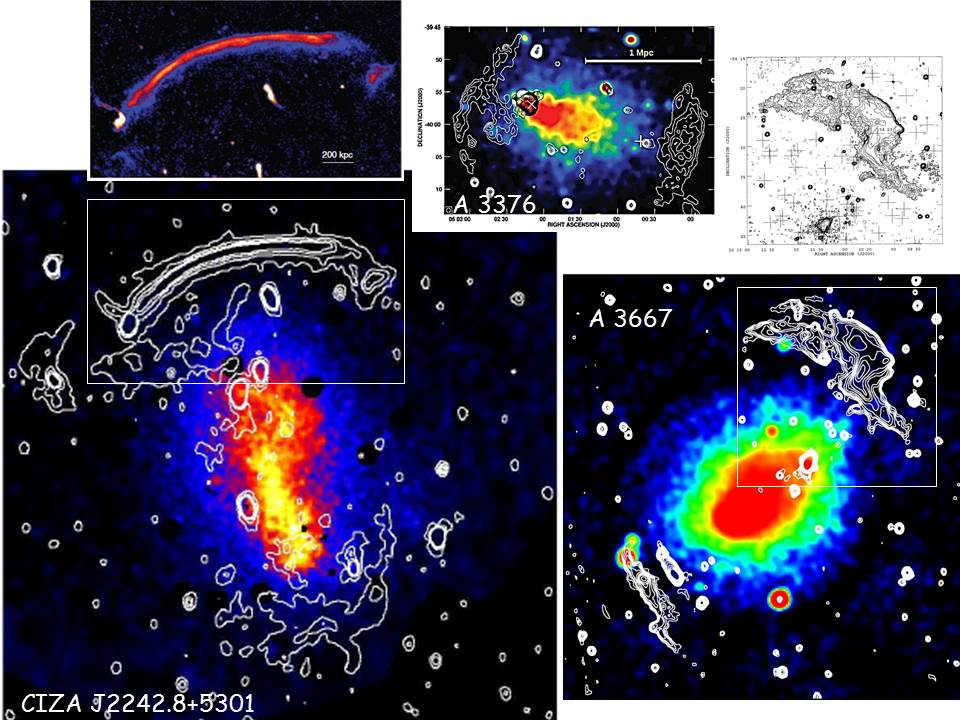}
\caption{
Images of giant radio relics (contours) overlaid on the X-ray emission
from the hosting systems (colors). The three radio relics are reported
with the same physical scale. Upper-left and upper-right panels
highlight the high-resolution radio images of the northern
relics in CIZA2242 and A3667, respectively (credits: \cite{vanweeren10,
ogrean13} for CIZA2242, \cite{rottgering97} for 
A3667, \cite{kaleetal12} for A3376).}
\label{fig:1}       % Give a unique label
\end{figure}

\subsubsection{Shock--relics connection}

The association between giant radio
relics and shocks is based partly on the usually 
elongated morphologies of the relics, consistent with a shock seen relatively 
edge-on. In addition, the
fact that in some cases the relics occur in pairs on opposite 
sides of the cluster 
core \cite{rottgering97, bagchi06, vanweeren10,
bonafede12} is telling, since emerging merger shocks should form in such pairs. 
In those paired relic cases a
line between the relics is typically consistent with an apparent merger
axis, as established by other observations, such as in the X-rays.
In addition, the fact that relics are strongly polarized with an
orientation that generally implies the magnetic field is aligned with 
the long axis of the relic, suggests that they
originate in regions where the magnetic field is compressed in the
shock plane \cite{clarke06, vanweeren10, bruggen12}.
Most important, in a number of cases merger shocks that have been
detected using X-ray observations coincide closely with radio relics
(or with sharp edges of radio halos) reinforcing the idea of a 
direct connection between shocks and the acceleration of the radio
emitting CRe in those regions \cite{giacintucci08, markevitch10, 
fino10, macario11, akamatsuetal12, akamatsukavahara13,
bourdin13, ogreanshock13, owersetal13}.

\begin{figure}[ht!]
\centering
\includegraphics[width=130mm]{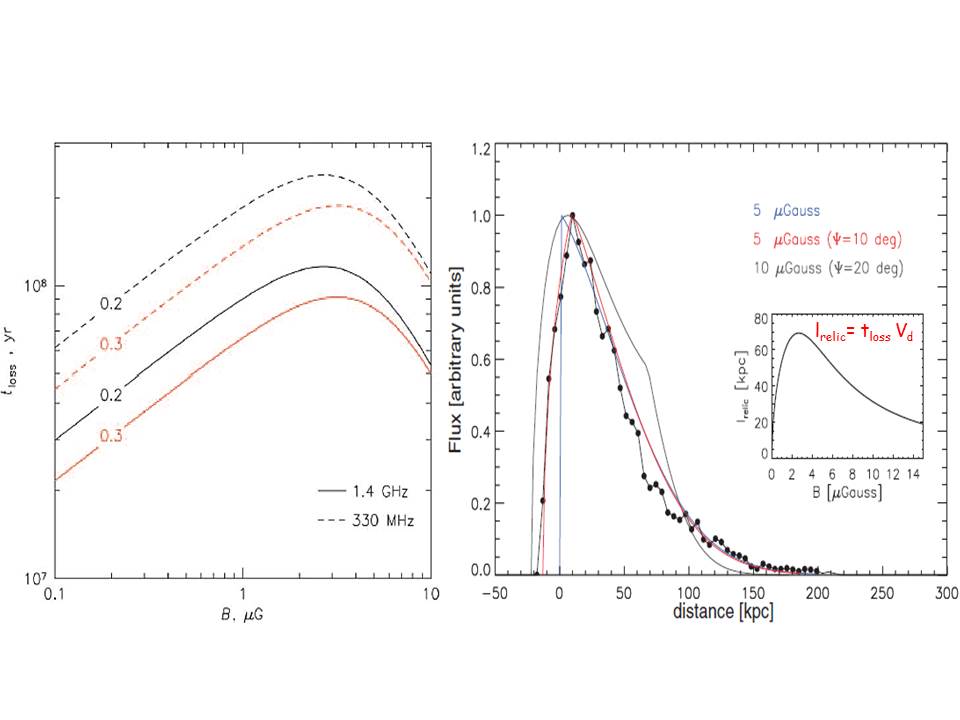}
\caption{
Left panel: the
life-time due to synchrotron and inverse Compton losses of CRe
emitting at
330 and 1400 MHz (dashed and solid lines, respectively) as a function
of the magnetic field strength in the downstream region.
Calculations are shown for redshift
$z=$0.2 and 0.3 (black (upper) and red (lower) lines, respectively)
(adapted from
\cite{markevitchetal05}).
Right panel: the
brightness profile of the northern radio relic in CIZA2242 (data-points)
compared to scenarios that assume different magnetic fields
in the downstream regions and inclination angles. The inset shows
the life time of the emitting particles as a function of the magnetic
field and the relationship between the thickness of the relic, 
$l_{relic}$ (distance),
the life-time, $\tau_{loss}$, and the downstream velocity, $V_d$
(adapted from \cite{vanweeren10}).
}
\label{fig:1}       % Give a unique label
\end{figure}

Presumably the observed CRe are accelerated or re-accelerated at these shocks.
Those CRe are then advected into the downstream
region at the velocity of the downstream flow, $V_d$ as they cool from
ICS and synchrotron losses.
This process sets the thickness of radio relics as seen at different
frequencies; the emitting CRe can travel a maximum distance
from the shock $= V_d \tau$ , where $\tau$ is the radiative life-time
of CRe emitting at the observed frequency $\nu_o$,
$\tau \propto \nu_o^{-1/2}$. Typically the life-time 
of CRe radiating in the radio band 
is of the order of 100 Myr implying a 
transverse size of radio relics $\sim 100$ kpc for a reference
downstream velocity $V_d = 1000$ km $s^{-1}$ and ICS dominated losses 
(with $z \sim 0-0.3$). This is consistent with
observations, once projection effects are properly taken into account. 
In this case radio
synchrotron spectral steepening with distance from the shock front 
is expected in radio relics with spatially resolved cross sections 
as a consequence of the fact that the oldest
population of CRe is also the most distant from the shock. 
This expectation 
is in agreement with several observations that provide evidence 
for steepening in radio relics along their transverse direction,
from their front to the back\cite{clarke06, giacintucci08, vanweeren10}.
Furthermore, recent analyses have shown evidences for 
synchrotron spectral curvatures along the transverse dimension
of a few particularly favorable relics and 
also that the curvature increases with distance in the downstream
area \cite{vanweeren12, stroe13}.
The life-time of radio-emitting CRe in the downstream region 
depends also on the magnetic field, $\tau \propto E_e^{-1} B^{1/2}/(B^2 +
B_{IC}^2)$, and the energy of electrons emitting at frequency
$\nu_o$ is $E_e \propto (\nu_o / B)^{1/2}$. 
For this reason the measure of the transverse size of relics at a given
frequency $\nu_o$, that is $= V_d \tau$,  
provides also constraints on the magnetic field strength in these sources, 
provided the downstream velocity is known \cite{markevitchetal05};
this is shown in Figure 17.
The downstream velocity, $V_d = c_s (M^2 + 3)/(4M)$ ($c_s$ is the
sound speed in the upstream region and $M$ is the shock Mach-number), 
can be derived from the 
density/temperature jumps at the shock 
measured by X-ray observations, 
or alternatively from the synchrotron injection spectrum measured at the
leading edge of the relic (assuming this is the location of the 
physical shock), using $\alpha_{inj}= (\delta_{inj}-1)/2$ (eq.3).

\subsubsection{Open problems \& particle re-acceleration at shocks}

The above observational facts support a connection 
between giant radio relics and merger shocks, suggesting that 
CRe are energized locally at these shocks and then age 
in the downstream region. However these observations do not
tell us more about the physics of the acceleration mechanisms.
Shocks discovered in the X-rays, including those that
coincide with radio relics, are apparently weak, 
with Mach numbers determined from the X-rays estimates for density
and/or temperature jumps across the shock region, 
$M \approx 1.5-3$ \cite{markevitch01, markevitchvikhlinin07}. 
High resolution cosmological simulations are consistent with 
the observational data, suggesting
that shocks with $M>3$ are rare well inside cluster virial radii 
\cite{ryu03, pfrommer06, skillman08, ryu09, vazza09b, vazza10}.
This, however, poses a problem for the origin of giant radio relics,
because, as discussed in Sect. 2.2.1, weak shocks are expected to 
be relatively ineffective as particle accelerators, especially 
if they are accelerating
CR from nonrelativistic energies.
This is partly because the
equilibrium DSA spectrum of locally injected CRs in such shocks 
is quite steep (eq. 3),
so most particles injected at the shock carry away relatively little
energy. In addition, thermal leakage injection is likely to be
less effective in weak shocks than strong shocks \cite{kang02}. 
Using reasonable parameters for shocks with $M\leq 3$, less
than $\sim 0.01$\% of the thermal ion flux through 
such shocks should be injected
into the CRp population and less
than a few percent of the shock energy flux should be transferred
to freshly injected CRp \cite{kang05, kang11}.
It is important to keep in mind that all the above efficiencies
refer to CRp, while CRe are the particles observed in radio relics. Again,
as pointed out above, primary CRe are much more difficult to inject from
the thermal pool into the CR
population than CRp (Sect.2.2.1). 
In the galactic CRs $N_{CRe}/N_{CRp} \sim 1/100$,
presumably reflecting that constraint. An electron fraction even that
large, and likely representing DSA at strong shocks, is already
a theoretical challenge.
Thus, merger shocks seem
very unlikely to inject thermal electrons and then to accelerate them
to relativistic energies with efficiencies high enough to extract more than
a tiny fraction of a
percent of the energy flux through the shock \cite{spitk11}. 
Thus, the question is whether radio relics can be powered by
the acceleration of the thermal particles of the ICM
at weak shocks. This may point to an alternate source for the CRe that 
are available to be re-accelerated in the shocks
by the DSA process.

\noindent
The synchrotron luminosity that is emitted in the radio band by
assuming
that a fraction $\eta_{CRe}$ of the kinetic energy flux through the 
shock-surface, $S$, is transferred to supra-thermal (and relativistic)
CRe is (see also \cite{hoeft07}): 

\begin{equation}
\nu_o P_{Syn}(\nu_o) \sim 1/2 \rho_u V_{sh}^3 \eta_{CRe} S
(1 + ( {{B_{IC} }\over{B}} )^2 )^{-1} {\cal F}^{-1}\left(\alpha (M) \right)
\label{eq:relicpower}
\end{equation}

\noindent
where the dimensionless
function ${\cal F}$ is a few for $\alpha_{inj}=0.5$ and rapidly 
decreases for steeper spectra (weaker shocks, eq. 3). 
That drop results from the
fact that a progressively 
smaller fraction of the energy of the accelerated 
CRe is associated with GeV radio-emitting 
electrons at these shocks.
Figure 18 shows the radio luminosity (left panel) 
and luminosity per unit surface area (right panel) of 25 giant
radio relics as a function of their projected distance from cluster 
centers (see Figure caption for details).
Figure 18 highlights that a large scatter exists 
in the radio luminosity of relics
at a given distance. This may be due to projection effects (namely the
relics with lower radio power are seen at large angles) or it
may imply large variations of either the acceleration 
efficiency, $\eta_{CRe}$, or the magnetic field strength in radio relics.
The most luminous radio relics in Fig. 18 put tension on the standard
scenario where the radio-emitting 
CRe are accelerated at shocks from the thermal pool.
Indeed if we adopt typical parameters of the ICM at the location of 
peripheral radio relics, $V_{sh} \sim 3000$ km s$^{-1}$ and ${\cal F}
\sim$ few, Eq. \ref{eq:relicpower} implies that 
$\eta_{CRe} \geq 10^{-4}$ 
is necessary to generate relics with synchrotron
luminosity above $\sim 10^{41}$erg/s. 
\begin{figure}[ht!]
\centering
\includegraphics[width=130mm]{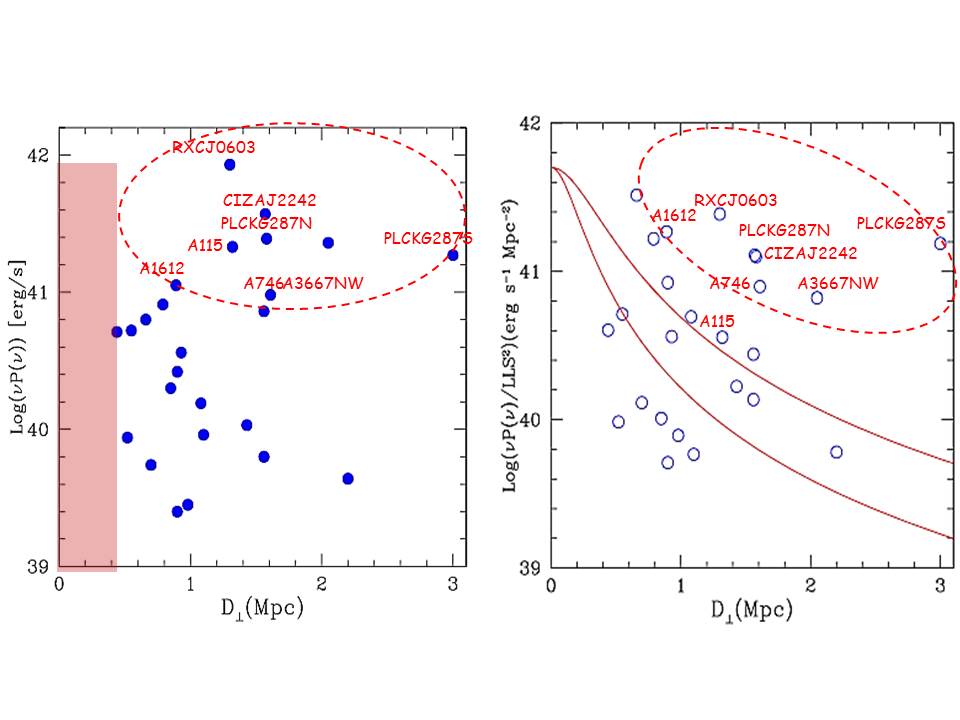}
\caption{
Left panel :
the synchrotron radio luminosity ($\nu P(\nu)$ evaluated at 1400 MHz)
as a function of projected distance from cluster center for 25 giant
radio relics with unambiguous classification. Data are taken from 
the compilations in\cite{feretti12, nuzaetal12}, although we only use 
the 25 sources that can be ``unambiguously'' classified
as giant radio relics.
Right panel : the synchrotron radio luminosity ($\nu P(\nu)$ evaluated at
1400 MHz) per unit relic-surface area
as a function of projected distance for the same relics in the left
panel. 
Solid lines mark reference radial behaviours of the kinetic energy
flux per unit area through shocks with equal velocity (in arbitrary
units); they re obtained using typical beta-models for
the gas-density distribution in the hosting clusters.
The pink vertical band in the left panel marks the typical minimum
distance where giant radio relics are found \cite{vazzarelics12}.}
\label{fig:1}       % Give a unique label
\end{figure}
For some of the most powerful relics in Fig. 18, such as RXCJ0603, A3667 and 
CIZAJ2242, a Mach number 2.4-3 shock is seen by X-ray observations
\cite{akamatsukavahara13, ogreanshock13}.
In these cases if we assume a reference
acceleration efficiency of CRp at
weak, $M \sim 2-3$, shocks of about 1\%, e.g. \cite{kangetal07}, 
an unusually large CRe/CRp ratio $\sim 0.1$ is required to match the
observed synchrotron luminosity.

\noindent
Another problem faced by the standard scenario where relics originate
from shock acceleration of thermal particles is that 
in several cases the observed radio spectrum is flatter than that
expected from DSA by assuming the Mach number of shocks derived from  
X-ray observations (eq. 3)\cite{akamatsukavahara13, ogreanshock13}.
For example, the well studied relic in CIZA J2242.8+55301 has a 
radio spectra at the putative shock location 
$\alpha_{inj} = (\delta_{inj} - 1)/2 = 0.6$ (so, $\delta_{inj} = 2.2$),
implying from eq. 3
a rather large shock Mach number $M = 4.6$ \cite{vanweeren10},
if the radiating electrons
represent a locally injected population. On the other hand, X-ray
observations with Suzaku derived a Mach
number $\sim 3$ from the temperature jump at the relic 
\cite{akamatsukavahara13}, which  
would imply $\alpha_{inj} \simeq 0.75$.
In this respect it is worth mentioning that 
numerical simulations show variations of the Mach number 
over shocks\cite{skillman13} that potentially might produce 
some differences
in the shock Mach numbers derived from X-rays (temperature
or density jumps) and from radio. In this case the radio Mach number
would be biased high because synchrotron emission is likely to be 
locally stronger in regions with higher Mach number.

In principle, all the challenges discussed above can be naturally circumvented 
by adopting a scenario where radio relics are generated by the
re-acceleration of pre-existing (seed) CRe at merger shocks
\cite{markevitchetal05, kang12, pinzkeetal13}.
If a pre-existing population of CRs with a hard (flat) spectrum passes
through such weak shocks, DSA can enhance the energy content of those
CRs by a factor of a few. This makes DSA re-acceleration at merger
shocks a relatively more efficient process\cite{kang11}. 
Additionally, the
re-accelerated particle spectrum is similar to the spectrum of the pre-existing
population if the latter is flat, so not the classic  DSA 
spectrum for shock injected CRs given in eq. 3 \cite{kang05, kang11}. 
The physical acceleration is still DSA, however; the change reflects only the 
assumed source of particles being accelerated.
The potentially higher efficiency of re-acceleration in these shocks is possible 
because the
acceleration time needed to double the energy of each CR can still be relatively
short (see eq. 4). 
In this case a model that can
account for the observed properties of CIZA J2242.8+55301 could
involve a Mach 2 shock re-accelerating an upstream CRe population with
an energy spectral index, $\delta = 2.2$ and a plausible upstream
CRe pressure $\sim 10^{-4}$ compared
to the upstream thermal pressure \cite{kang12}.
Then the challenge becomes defining acceleration processes 
just upstream of the shock (since radiative lifetimes are still very short), 
such as, e.g., turbulence, 
that can account for the CRe population entering the shock (Sects. 2.2.2, 4.2-4.3).

$\gamma$-ray upper limits (Sect. 3.3) may provide a complementary way to
constrain the origin of radio relics, because the ratio of $\gamma$ and
radio-relic luminosities essentially resembles the ratio of the 
acceleration-efficiency of CRp and CRe at clusters shocks.
In principle this also applies to the case of re-acceleration, 
as the same shocks would re-accelerate both CRe and CRp. Their relative
abundances are then the principal uncertainty.
According to Vazza \& Bruggen (2013)\cite{vazzabruggen13}, 
by assuming the standard paradigm for
the origin of giant radio relics, based on shock acceleration or
re-acceleration, and a ``canonical'' ratio CRp/CRe,
too many CRp should be produced in galaxy clusters violating current
$\gamma$-ray limits.
One possibility to reconcile a shock-acceleration
origin for radio relics with the current lack of $\gamma$-rays 
from clusters, would be to assume that cluster shocks accelerate 
CRe and CRp with an ``unusually'' large CRe to CRp energy ratio $\geq 0.1$.
Apparently, this is in contrast with constraints based on SNR (which,
however, refer to very different environments and strong shocks) and with
several theoretical arguments (Sect. 2.2.1). However 
recent PIC simulations (that are focused however on strong
shocks) suggest that the configuration of the magnetic field may play 
an important role, with quasi-perpendicular shocks being efficient CRe
accelerators\cite{spitk11}. 
This explanation for relics generally, would seem, 
then, to require magnetically
perpendicular shocks over Mpc scales, or at least perpendicular shock 
geometry being exceedingly common.
As we will discuss in Section 5.2 
an ``unusually'' large ratio CRe/CRp 
at clusters shocks would also have consequences for the $\gamma$-ray
emission from galaxy clusters produced via ICS.

\noindent
However, we also mention that another possibility to reconcile
a shock-re-acceleration origin of relics with $\gamma$-ray limits is
that the spectra of the pre-existing CRe differs from a power law
and breaks at a maximum Lorentz factor $\gamma \sim 1000$; this does not
apply to shock acceleration.
In this 
case re-acceleration from weak shocks could provide high-energy CRe
without accelerating CRp too much, e.g. \cite{vazzabruggen13}.
Rather the pre-existing electron population is ``bumped-up'' 
in energy enough at the shock to take it from an invisible
to a visible state.
We note that this scenario for the pre-existing CRe
is somewhat similar to that assumed in
several calculations of turbulent re-acceleration for giant radio halos
(Sect. 4.2 and also Fig. 7) and could be motivated by the
fact that the life-time of CRe in the ICM is maximum at energies in
the range 0.1-0.3 GeV (Fig. 8).

\section{High energy emissions from galaxy clusters}

Non-thermal high energy emission from galaxy clusters is an unavoidable
consequence of the presence of energetic CRe in the ICM as shown by 
radio observations (Sect. 4). High energy emission is also expected from the
theoretical scenario of CR acceleration and confinement discussed in
Sects. 2 \& 3.
Standard mechanisms for the production of high energy photons from the
ICM are supra-thermal bremsstrahlung from supra-thermal and primary CRe,
ICS of seed photons from primary and
secondary CRe, and the decay of $\pi^0$ generated by CRp-p collisions.
In addition, if ultra-high-energy CRs are present in the ICM, the CRe
produced through the interaction of these CRs with ambient photons will
produce both synchrotron and ICS emission at high energies.
In this Section we will discuss the relevant processes, current
observational constraints and future prospects for hard-X-ray emission
and $\gamma$-ray emission from galaxy clusters.

\subsection{Hard x-ray emission from nonthermal electrons}

While thermal emission dominates ICMs in the keV energy range, 
it declines rapidly at higher energies, allowing the possibility 
of detection of 
non-thermal excess emission at energies $>$ 10-20 keV. 
ICS of CMB photons (from both primaries and secondary CRe) and supra-thermal
bremsstrahlung are expected to be the most important mechanisms responsible
for hard X-rays in the ICM \cite{sarazin99,
ensslin99, blasi00, dogiel00, petrosianbykovrephaeli08}.
Their relative contributions 
depend on the spectral energy distributions of supra-thermal
electrons and more energetic CRe in the ICM.
The formation of prominent hard, supra-thermal tails of electrons in the ICM 
is difficult, because Coulomb collisions with the thermal ICM 
induce a relatively efficient thermalization of the supra-thermal
particles \cite{petrosian01, petrosianeast08, chernyshov12}. 
At the same time, however, a population of 
ultra-relativistic CRe exists in the ICM, as demonstrated by the
diffuse radio sources (halos and relics) discussed in Sects. 4.2-4.4.
Consequently, ICS of CMB
photons is an unavoidable process that produces some level of high 
energy emission in galaxy clusters \cite{rephaeli79}. 
The typical energy of ultra-relativistic
electrons emitting photons observed in the hard X-ray
band (with energy $E_{ph}$) via ICS of the seed CMB photons is:

\begin{equation}
E_e({\rm GeV}) \approx 3 \big( {{ E_{ph} }\over {30 {\rm keV}}} \big)^{{1 \over
2}},
\end{equation}

\noindent
while that of CRe emitting synchrotron radiation, emitted at redshift
$z$ and observed at frequency $\nu_0$ is 

\begin{equation}
E_e({\rm GeV}) \approx 7 \big( {{ \mu G}\over{B}} {{\nu_0}\over{{\rm GHz}}}
\big)^{{1 \over 2}} (1 + z)^{{1 \over 2}}
\end{equation}
\noindent
which means that the two processes sample pretty much the same 
population of CRe. More specifically, by assuming a power law energy
distribution of the emitting
CRe, in the form $N(E_e) = K_e E_e^{-\delta}$, both the
synchrotron and ICS spectrum are power-laws with slope $\alpha
= (\delta -1)/2$ and the ICS flux received by the observer
at frequency $\nu_X$ 
can be calculated from the synchrotron flux received at the frequency
$\nu_R$ :

\begin{equation}
F_{ICS}(\nu_X) = 1.38 \times 10^{-34} \big( {{F_{Syn}(\nu_R)}\over{
{\rm Jy}}}\big) {{ (1+z)^{\alpha +3} }\over{
\langle B_{\mu G}^{1+\alpha} \rangle }}
\left( {{\nu_X/{\rm keV} }\over{\nu_R/ {\rm GHz}}} \right)^{-\alpha}
{\cal C}(\alpha)
\label{ICSSYN}
\end{equation}
where $F_{ICS}$ is in c.g.s. units,
$\langle {\ldots}  \rangle$ denotes the emission-weighted quantity
in the emitting volume and the dimensionless
function ${\cal C}$ is given in Table 1
for different values of $\alpha$.

\begin{table}[ph]
\tbl{Numerical values of the function ${\cal C}$ in eq. 32}
{\begin{tabular}{@{}ccccccccccccccc@{}} \toprule
\colrule
$\alpha$ & 0.7 & 0.8 & 0.9 & 1.0 & 1.1 & 1.2 & 1.3 \\
${\cal C}$ & 4.78$\times 10^2$ & 9.09$\times 10^2$ &
1.70$\times 10^3$ & 3.16$\times 10^3$ & 5.83$\times 10^3$ &
1.07$\times 10^4$ & 1.95$\times 10^4$ \\
\botrule
$\alpha$ & 1.4 & 1.5 & 1.6 & 1.7 & 1.8 & 1.9 & 2.0 \\
${\cal C}$ & 3.55$\times 10^4$ &
6.45$\times 10^4$ & 1.17$\times 10^5$ & 2.11$\times 10^5$ &
3.80$\times 10^5$ & 6.83$\times 10^5$ & 1.23$\times 10^6$ \\ \botrule
\end{tabular}}
\end{table}

\noindent
Eq. 32 shows the well known fact that measuring ICS from galaxy
clusters in the hard X-ray band constrains (essentially provides 
a measure for, if the emission is detected) 
the average value of the magnetic field in the emitting
region.

\noindent
The search for diffuse ICS emission 
from galaxy clusters at hard X-ray energies has been underway 
for many years.
One of the most persistent discussions of possible ICS detections 
involve the Coma cluster that hosts the prototype of giant radio halos
(Fig. 1, and Sect. 4.2).  
Non-thermal hard X-ray excess emission has been claimed 
from Coma by Rephaeli \& Gruber (1999, 2002)\cite{rephaeli99,
rephaeli02} with the Rossi X-ray Timing Explorer (RXTE) and by Fusco-Femiano
et al. (1999, 2004)\cite{fusco99, fusco04} with BeppoSAX.
However, the detection has remained controversial\cite{rossetti04,
fusco07, eckert07, lutovinov08, wik09}.
We note that a $\simeq$4$\sigma$ hard X-ray excess emission,
at a level consistent with that in\cite{fusco04}, was also found 
by\cite{wik09} with the Suzaku HXD-PIN assuming a best fit value of 
the cluster temperature $= 8.2$ keV as derived from XMM-Newton data,
that was consistent with the temperature adopted in the analysis 
of\cite{fusco04}. 
However, \cite{wik09} concluded that there was no statistical evidence 
for ICS emission within the field of view of Suzaku HXD-PIN because 
these authors used an higher temperature, $= 8.45 \pm 0.06$ keV, 
that was obtained from a more complex fit combining XMM-Newton and 
Suzaku HXD-PIN data.
Recently Wik et al.\cite{wik11} performed a joint analysis
of 58-month Swift Burst Alert Telescope (BAT) and of the
XMM-Newton EPIC-pn spectrum derived from mosaic
observations of the Coma cluster
and found no evidence for ICS emission at the
level expected from previous detections even if the
such emission is assumed to be very extended.
More recently Fusco-Femiano et al.(2011)\cite{fusco11} attempted to
reconcile the limits and detections obtained from different instruments by
carrying out a complex joint model-analysis based on the 
temperature and density profiles, derived from XMM-Newton and ROSAT
data, and on the BeppoSAX data.
So far, additional candidate clusters for ICS emission 
are the Ophiucus cluster \cite{eckert08} (but see \cite{ajello09})
and the Bullet cluster \cite{petrosian06, ajello10}.
Other few cases where a detection of hard X-ray excess emission 
has been claimed are at low significance level\cite{fuscofemiano00, 
rephaeli03, nevalainen04}.
Based on that we would reach the conservative conclusion that 
there is no compelling evidence 
for a detection of ICS emission from any ICM with current
instruments and that limits imply (averaged) fields
$B \geq 0.1-0.3 \mu$G. Thus future observations are necessary to
solve current controversies.

In the next years the detectors onboard of
NuSTAR\footnote{http://www.nustar.caltech.edu/ , was launched in 2012
and first results on the Bullet cluster are in preparation\cite{nustar13}}
and ASTRO-H will
improve the sensitivity 
in the hard X-ray band by more than one order of
magnitude with respect to the current facilities.

\noindent
From eq. 32, clusters hosting 
bright radio halos, with fluxes at Jy level at 1 GHz
(such as Coma, Perseus, A2319), should have ICS fluxes
integrated in the band 20-80 keV of several $\times
10^{-13}$erg/s/cm$^{2}$, if we assume 
an average magnetic field $\simeq 1 \mu$G in the
halo region. Coma is a relevant example. 
Current analysis of the Coma cluster
using RM \cite{bonafede10} favor a value $\simeq 2 \mu$G for
the magnetic field averaged on Mpc$^3$-volume.
This implies an expected ICS flux $\sim 2 \times
10^{-13}$erg/s/cm$^2$ in the 20-80 keV band measured on
an aperture radius $= 35$ arcmin, if we adopt the radio flux of
the halo measured on the same aperture radius \cite{brunettietal13}
and a synchrotron spectral index $\alpha =1.25$.
Such a flux distributed on a 35 arcmin aperture-radius is probably
difficult to detect with ASTRO-H. However, even a relatively small
departure from that
magnetic field value, say a factor 2 lower, may be sufficient to
lead to a detection. 
In fact, the important point here is
that even the non detection of nearby
clusters hosting bright radio halos
in the hard X-rays at the sensitivity level of the ASTRO-H telescope
would be a result, providing very meaningful constraints on the
magnetic field strength in these systems.

\noindent
Similar considerations apply for clusters with giant radio relics.
Some clusters host bright relics, with fluxes at Jy level
at 1 GHz (A2256, A3367), and should produce ICS fluxes in the
hard X-rays of several $\times 10^{-13}$erg/s/cm$^{2}$, provided that 
the average magnetic fied in the relic region is $\simeq 1 \mu$G.
On one hand radio relics cover an aperture angle that is smaller than
giant radio halos and have peripheral locations, 
making their detection in the hard X-rays less
problematic. On the other hand, however, relics are 
associated with shock compression
regions (Sect. 4.4) where the average magnetic fields is 
presumebly larger than that
(averaged) in the region of giant halos, thus reducing the expected ICS flux.
A textbook example is the relic A3667-NW, the brightest giant radio relic.
Suzaku observations put a limit to the cluster ICS hard X-ray emission of about
$5 \times 10^{-12}$erg/s/cm$^{2}$ \cite{nakazawaetal09, akamatsuetal12}.
Deeper limits have been obtained in the X-rays with Suzaku XIS and
XMM-Newton, taking advantage of the peripheral location of the relic
where thermal cluster's emission is fainter\cite{nakazawaetal09, finoguenovetal10,
akamatsuetal12}. These constraints imply $B \geq 2-3 \mu$G if the power law
spectrum measured in radio (corresponding to the energies of ICS
emitting CRe in the hard X-rays, eq. 30-31) is extrapolated at energies
about 10 times smaller (producing ICS at keVs).
Constraints become significantly shallower in the case that the
spectrum of CRe in the relic-downstream 
flattens at lower energies, e.g. \cite{nakazawaetal09}.
We note that such a flattening might result from the fact that the
life-time of CRe producing ICS photons at keV energies
becomes comparable to the shock life-time, implying that
their spectrum gradually approaches the injection spectrum (with 
slope $\sim \delta_{inj}$, see Sect. 2.2.1).
The future ASTRO-H or NuSTAR observations will overcome these uncertainties.
From eq. 32 and assuming the spectrum and flux of the
radio relic, $\alpha =1.1$ and 3.7 mJy at 1.4 GHz\cite{melanie04}, we
obtain an expected ICS flux in the 10-40 keV band $\simeq 2.5 \times
10^{-13}/ (B_{\mu G}/3)^{2.1}$erg/s/cm$^2$. We note that $B \sim 3-5
\mu$G is derived from Faraday RM in the relic region\cite{melanie04}.

As a general conclusion, however, it is worth stressing that, unless 
the magnetic field 
strength in galaxy clusters is much smaller than that constrained from Faraday 
rotation measurements, 
future telescopes with 10 times more effective area than ASTRO-H 
are needed to explore non-thermal electron emission from galaxy clusters 
in the hard X-rays.

\subsection{Gamma ray emission}

The initial motivation for the interest in the $\gamma$-ray emission from
clusters arose from the possibility of significant CRp
confinement in the ICM (Sect. 2)\cite{berezinsky97, 
blasicolafrancesco99, pfrommerensslin04G}.
As already discussed, a natural byproduct of CR confinement is the
emission of $\gamma$ radiation from both the decay of
secondary $\pi^0$ and the ICS of CMB photons by high energy
secondary CRe (Sect. 2.3).
The latter channel becomes sub-dominant when integrated over a cluster 
at energies $> $100 MeV \cite{blasi01, miniati03} (Figure 19, left,  
for an example of the expected $\gamma$-ray spectrum).

\begin{figure}[ht!]
\centering
\includegraphics[width=130mm]{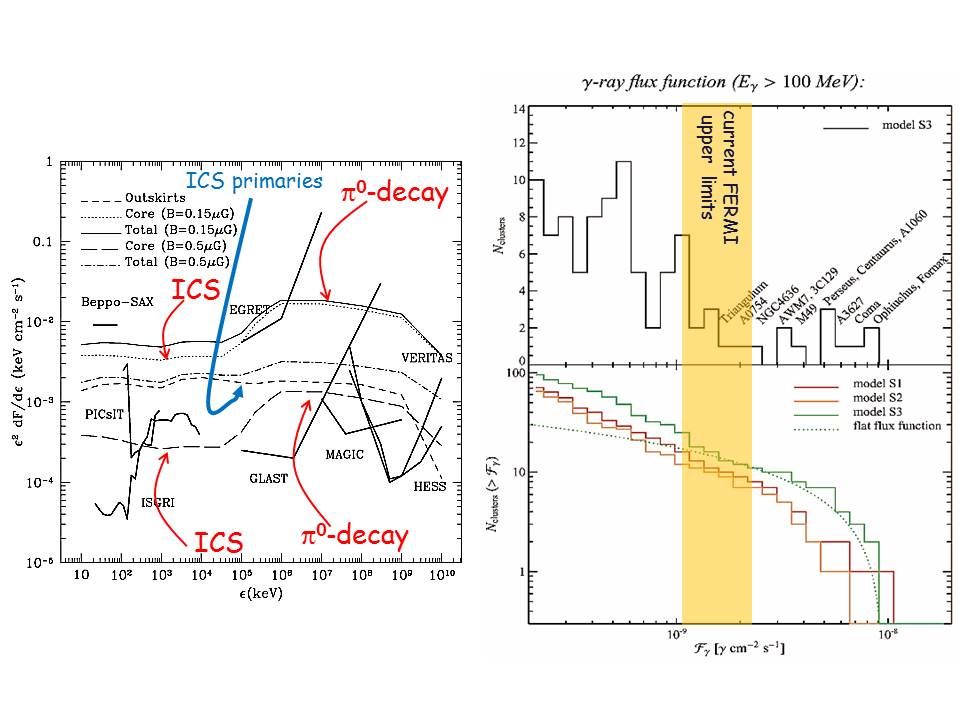}
\caption{Left panel: the $\gamma$-ray spectrum of the Coma cluster
obtained by assuming that the radio halo is due to secondary CRe 
generated by CRp-p collisions in the ICM (adapted from \cite{miniati03}).
Calculations account for the $\pi^0$-decay and ICS of CMB photons by
secondary pairs and 
are obtained assuming an average magnetic field $=0.15$ and
$=0.5 \mu$G. The dashed line is the contribution from ICS of primary
CRe accelerated at strong shocks in the cluster outskirts.
Expectations are compared to the sensitivities of several high energy
instruments.
Right panel: the expected number counts of galaxy clusters
in the $\gamma$-rays from simulations by \cite{pfrommer08} (adapted from
\cite{pfrommer08}). 
The reference range of sensitivity of Fermi-LAT observations is 
reported as a yellow area.}
\label{fig:1}       % Give a unique label
\end{figure}

After the first pionereing theoretical attempts \cite{voelk96, 
berezinsky97,
sarazin99, voelkatoyan99, loebwaxman00, blasi01, gabiciblasi04, 
pfrommerensslin04G} in the 
last decade numerical simulations allowed useful estimates of the expected 
$\gamma$--ray emission from galaxy clusters, under different assumptions.
These simulations, that include some models for CR physics
and the acceleration of CRs at cosmological shocks provided a picture
of the radio to $\gamma$--ray properties of galaxy clusters.
The first simulations of this kind
predicted that clusters would be potentially
detectable in $\gamma$-rays with the Fermi-LAT 
telescope \cite{miniati01gamma,
miniati03, pfrommer08} (Figure 19).
The most important assumption in these simulations is in the
efficiency of particle acceleration at weak shocks that, as explained
in Sect. 2.2, is, however, poorly known. 
Subsequent simulation studies have attempted to reconcile expectations 
with the limits from 18 months observations with
Fermi-LAT and from 
Cherenkov telescopes observations of the Perseus cluster\cite{pinzke10, 
aleksic10, pinzkeetal11}.
However the most recent limits, that are derived using almost 5 years of 
Fermi-LAT data and stacking procedures, are 
definitely inconsistent with the predictions of 
these simulations \cite{Fermilat13, huberetal13}.
That outcome may imply that the models of particle 
acceleration adopted in
these simulations were too optimistic, in which case
current limits can be used to constrain the acceleration efficiency.
For example,
by assuming the CRp spectral slope and spatial distribution that are
predicted by simulations in\cite{pinzke10}, Ackermann et al. (2013)\cite{Fermilat13} 
and Zandanel \& Ando (2013)\cite{zandanelando13} derived limits on the maximum
efficiency of shock acceleration in the ICM
which refers to the case of strong shocks, $\eta_{CRp} \leq 0.15$.
Alternatively one may speculate that the spatial distribution of CRp
is much broader than that expected from these simulations
and also than that of the thermal ICM. That would also 
indicate that the diffusion/transport of CRp with energy $<$TeV is 
fairly efficient, putting meaningful constraints on the
physics discussed in Sect. 3.2.
\begin{figure}[ht!]
\centering
\includegraphics[width=130mm]{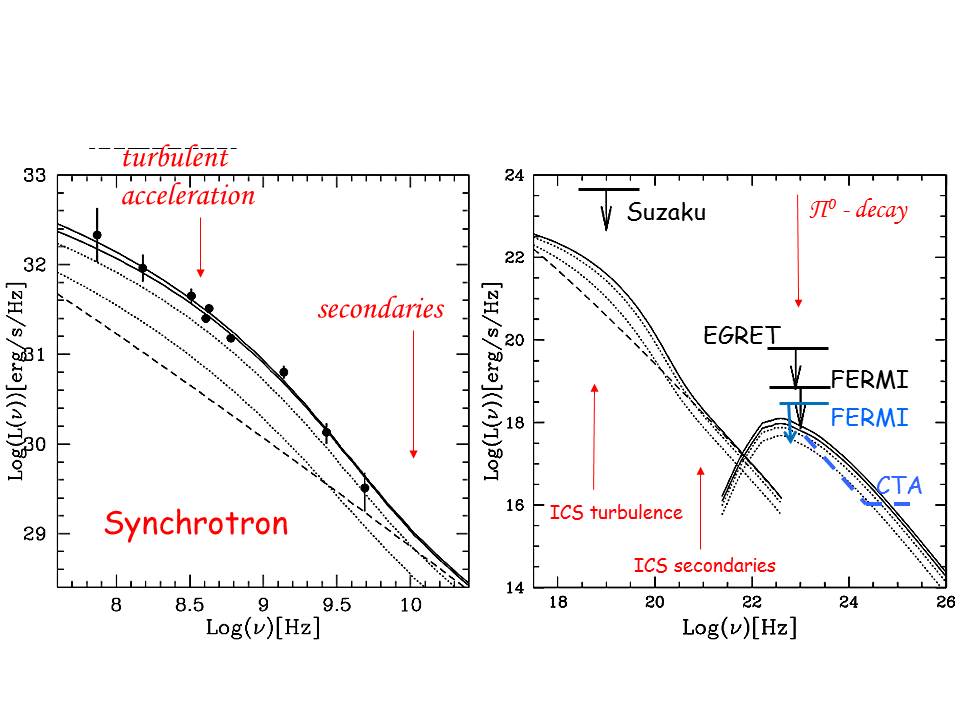}
\caption{
Left panel: the synchrotron spectrum of the Coma radio halo
(data-points) is compared with expectations based on turbulent reacceleration of
secondary particles (tick-solid lines) (adapted from\cite{brunettilazarian11b}). 
Dotted-lines show the time-evolution
of the emission that is produced during different stages of the mergers.
The dashed line shows the expected emission for a non-turbulent relaxed system that
is powered via pure-hadronic cascades (see \cite{brunettilazarian11b}
for details).
Right panel: the high-energy (ICS and $\pi^0$ decay) emission of the
Coma cluster compared with current limits (blue is the Fermi-LAT limit
based on about 4 years of data\cite{Fermilat13}). 
The style of the lines is the same of the 
left panel (adapted from \cite{brunettilazarian11b}).
In all models/calculations the magnetic field is fixed to 
the value and radial profile that are favoured by current RM analysis\cite{bonafede10}.
The bump in both the radio and hard X-ray spectra is produced by the
combined effect of turbulent reacceleration and radiative losses on
the CRe spectrum.}
\label{fig:1}       % Give a unique label
\end{figure}

In reality, the microphysics of the ICM and of CRs in galaxy clusters
is very complicated and, unfortunately, well 
beyond the capabilities of current simulations (Sects. 2.2, 3.2).
A way to circumvent these difficulties for now is to derive expectations 
by anchoring model parameters to the properties of the
diffuse synchrotron emission in the form of radio halos 
in nearby clusters \cite{reimer04, pfrommerensslin04G,
brunetti09, brunetti12} (Sect. 4.2, 4.3).
Potentially this also allows one to put combined constraints
on both the origin of diffuse cluster-scale emission (such as radio halos)
and on the expected high energy emissions from the hosting clusters.
In Sect. 4.2 we have already discussed that, at least in a few observed cases,
the combination of 
radio-halo properties and of the Fermi-LAT upper limits disfavor a purely
 hadronic origin of these radio sources. That  limits
the role played by secondaries for the origin of the observed radio
emission and, consequently, also the expected $\gamma$-ray luminosity generated
by CRp-p collisions.
In particular for the Coma cluster, where the most
stringent constraints have been derived using the properties of
the radio halo at lower frequencies, Fermi-LAT upper limits imply
a magnetic energy density in the cluster that is much 
larger than that constrained by Faraday rotation measures
\cite{brunetti12}.

\noindent
Based on that, one way to obtain meaningful expectations of the level
of $\gamma$-rays from galaxy 
clusters expected because of the existence of radio halos is provided 
by joint modelings of the production of secondary particles and of the 
re-acceleration of these secondaries (and of the primary CRp)
due to MHD turbulence in the ICM \cite{brunettiblasi05}.
Our ability to apply turbulent
re-acceleration to refine the picture based on CRp confinement and
injection of secondaries in the ICM is limited primarily
by our limited understanding of ICM turbulence, its
origins, evolution distribution and spectral properties,
especially on scales where resonant interaction with
CR take place (Sect. 2.2.2).
Pioneering calculations in this direction have been recently developed in
\cite{brunettilazarian11b} assuming that compressible MHD turbulence, generated
in galaxy clusters at large scales during cluster-cluster mergers, cascades
to smaller scales and also assuming that a pre-existing population of long-living
CRp is mixed with the ICM.
These calculations allow one to model the temporal evolution of the
non-thermal emission from galaxy clusters, from radio (radio halos)
to $\gamma$-rays,
and to connect the evolution with cluster merger histories (Figure 20). 
These calculations 
show that radio halos and cluster-scale ICS 
emission are generated in a turbulent ICM (presumably in merging systems),
while a fainter long-living radio (and ICS) emission sustained by the process
of continuous injection of high energy secondary CRe is expected to be
common in clusters more generally.
The strength of this latter, persistent component is proportional to the energy density
of the primary CRp in the ICM and, in the context of these modelings, it
can be constrained by the upper limits to the cluster-scale radio emission 
in galaxy clusters without radio halos (Fig. 12, Sect. 4.2).
Contrary to the {\it transient} nature of giant radio halos
and hard X-ray emission
in galaxy clusters, $\gamma$-ray emission 
is expected to be common at a moderate level in all clusters
and not directly correlated with the presence of the currently
observed giant radio halos.
Since the CRp population of a cluster reflects its integrated
history (assuming CRp do not escape) there should be a moderate
range in $\gamma$-ray luminosities, e.g. \cite{miniati01}.
We note that expectations of $\gamma$-ray emission
(from $\pi^0$ decay and ICS from secondaries) according to
these calculations are
optimistic, because they are still based on a secondary origin of the
seed electrons to re-accelerate.
They can be used to constrain the
minimum level of $\gamma$-rays from galaxy clusters {\it under the hypothesis
that secondaries play an important role} for the origin of halos, being the only seed
electrons to re-accelerate in the ICM.
Figure 20 shows that 
the expected level of $\gamma$--ray emission from a Coma--like cluster
is at a level few times below present Fermi-LAT
limits. Note that expectations in Fig. 20 are obtained by assuming
the magnetic field in the Coma cluster that is favored by current 
analysis of Faraday RM, namely a central magnetic field $B = 4.6
\mu$G and a radial decline of the magnetic energy with the thermal
energy density\cite{bonafede10}. A lower value of the field would 
imply a larger $\gamma$-ray luminosity (eq. 26).

\noindent
In general the population of CRe in the ICM is a mix of re-accelerated
primaries and of locally injected secondaries. The $\gamma$-ray luminosity should 
roughly scale with the ratio of secondaries to primary CRe,  
because primary CRe (re)accelerated in
Mpc-scale turbulent regions cannot produce $\gamma$-rays 
(as explained in Sect. 2.2.2, 
their maximum energy does not exceed significantly about a few times 10 GeV).
Consequently if the secondary particles play a minor role and radio 
halos are powered by primary CRe the expected $\gamma$-ray luminosity
is much smaller than that in Figure 20.
For this reason future $\gamma$-ray observations with a sensitivity
level few times better than current ones, using 10 years of Fermi-LAT
data or with the Cerenkov Telescope Array (CTA), are 
extremely important as they will definitely clarify the role of
CRp and their secondaries for the origin 
of non-thermal emission in galaxy clusters.

\begin{figure}[ht!]
\centering
\includegraphics[width=130mm]{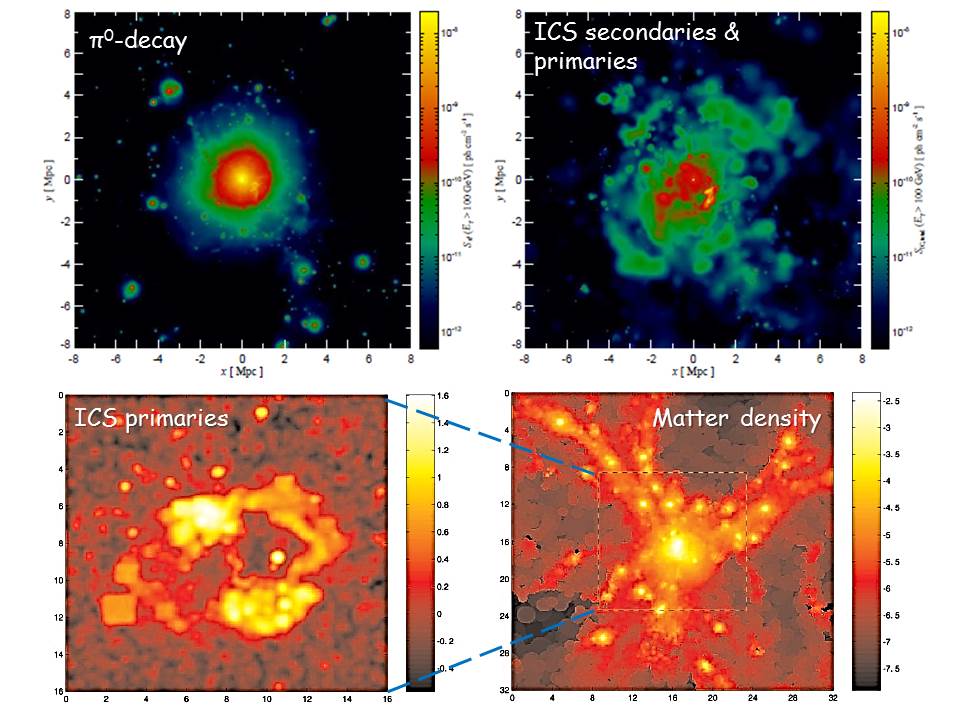}
\caption{
Upper panels (adapted from \cite{pinzke10}): $\gamma$-ray emission
from a simulated cluster due to $\pi^0$-decay (left panel) and
due to the combined ICS of CMB photons from shock-accelerated
(primary) CRe and secondary CRe (right panel).
Bottom panels (adapted from \cite{keshet03}): $\gamma$-ray emission
from a simulated cluster due to shock-accelerated
(primary) CRe (left panel), and projected matter density distribution in
a simulated box containing that cluster (right panel); the
square-region in
the right-panel highlights the region shown in the left panel.}
\label{fig:1}       % Give a unique label
\end{figure}

Primary CRe can be accelerated at cosmological shocks
up to energies of tens of TeV (Sect 2.2.1), and the resulting
ICS emission extends to multi-TeV 
energies \cite{sarazin99, loebwaxman00, miniati01gamma,
blasi01, keshet03, blasi07, pinzke10}.
There is agreement that the expected $\gamma$-ray spectrum
of galaxy clusters should be dominated by the decay of the neutral
pions in the central cluster regions and by ICS of primary CRe accelerated by
strong shocks in cluster outskirts \cite{miniati01gamma, blasi07,
kushnirwaxman09, pinzke10}(Figs. 19, 21).
The relative importance of the two mechanisms
depends on the ratio of the efficiency of acceleration of CRp 
and CRe at shocks (Sect. 2.2.1)
and on the dynamics/transport of CRp in the ICM (Sect. 3.2)
\cite{miniati03, pfrommer08, kushnirwaxman09}.
Under the assumption of CRp confinement and of ``canonical'' CRp/CRe
ratios at shocks, the $\pi^0$ decay is the
dominant channel for the production of $\gamma$-rays from clusters.

\noindent
Instead of the rather centrally concentrated $\gamma$-ray emission
from $\pi^0$-decay, the
ICS from primary CRe accelerated at strong shocks should
appear very different, giving rise 
to a spatially extended (rather flat) emission, potentially edge
brightened,
on scales comparable to the cluster virial region (Figure 21).
The ICS luminosity from CRe accelerated at the strong shocks that
surround clusters outskirts (assuming $B_{IC} >> B$) is :

\begin{equation}
\nu_o P_{ICS}(\nu_o) \sim 1/2 \rho_u V_{sh}^3 \eta_{CRe} S_{cl}
{\cal F}^{-1}\left(\alpha (M) \right)
\label{eq:ICSpower}
\end{equation}

\noindent
where ${\cal F}\left(\alpha (M) \right)$ is of the order of few for
strong (external) shocks that are of interest for this discussion. 
Numerical simulations allow estimates of the 
shock-kinetic energy flux across the cluster surface,  
$1/2 \rho_u V_{sh}^3 S_{cl} \sim 10^{41-42}$erg/s (e.g., Fig. 3), 
implying that a luminosity $\geq 10^{40}$erg/s, that might become
detectable in nearby clusters with future $\gamma$-ray observations, can be 
generated only provided that the acceleration of 
CRe at these shocks is ``unusually''
efficient, about $\eta_{CRe} \geq 0.01$.
Although this is theoretically challenging and also not supported
by current observations of SNR (Sect. 2.2.1), we note that the
possibility of an efficient CRe (re-)acceleration at clusters shocks
is also under discussion for the origin of
radio relics (see Sect. 4.4).
Consequently data from 10 years of Fermi-LAT observations will provide very
important constraints.

Other mechanisms for the production of $\gamma$-rays in the ICM include
non thermal bremsstrahlung \cite{blasi01, miniati03, blasi07} and ICS
from pairs generated by Bethe-Heitler processes (photo-pair 
and photo-pion 
production) due to the interaction between Ultra High
Energy protons ($E \geq 10^{18}$ eV) and 
photons of the CMB\cite{aharonian02, rordford04}. 
As discussed in Sect. 2.2.1, the acceleration of these 
ultra high energy CRp in galaxy clusters is a theoretical challenge (Figure 5 
for an ``optimistic'' view). However if EeV CRp are present, 
the high energy electron-positron pairs that are produced 
should radiate synchrotron and ICS emission
peaking in the hard X-ray and TeV energy bands, respectively
\cite{inoue05, vannoni11}, where future observations with ASTRO-H and 
CTA may obtain interesting constraints.

\section{Concluding Remarks}
\label{sec:5}

The importance of non-thermal components in galaxy clusters is twofold. 
One point is that they are complementary probes of the mechanisms
that dissipate energy at micro-physical scales; that is, especially 
the gravitational energy associated with the hierarchical sequence of mergers
and accretion of matter that lead to the formation of clusters and cosmic filaments.
The second is that non-thermal components, magnetic fields and
CRs, actually strongly affect the physics of the ICM and evolve somewhat distinctly 
from thermal plasma components. 
Thus, they could also have a role in the 
evolution of clusters themselves.

\noindent
In this review we have discussed the physics of CRs in galaxy clusters,
including their acceleration and transport in the ICM, and the most 
relevant observational milestones that have provided
relevant constraints on this physics.
Finally, looking forward, we have placed emphasis on what appear to be the most important 
prospects for the near future from radio and high-energy observations.

As we have discussed in Sect. 3, the bulk of CR protons deposited
within the cluster volume would be magnetically confined there for
cosmological times. Therefore, due to their long life-times
they should accumulate at increasing
levels over time as the cluster is assembled.
The amount of energy that is associated with these CRp depends on the 
efficiency of particle acceleration mechanisms in the ICM. 
This has been discussed in Sect. 2.
CRs, and CR protons in particular, should be accelerated at shocks associated
with cluster formation and in turbulent regions. Galaxies
within the cluster may also supply CRs by way of supernovae and
high energy outflows from supermassive black holes in their nuclei (that is, AGNs).
A notable and unavoidable consequence of CR proton acceleration and
confinement in the
ICM is $\gamma$-ray emission resulting from decay of neutral pions
produced by inelastic collisions involving CR protons
and thermal protons (Sect. 2.3). 
For this reason $\gamma$-ray observations are powerful tools to
constrain the physics of acceleration and transport of CRs in galaxy
clusters. This has been discussed in Sects. 3.3 and 5.2.
The non-detection of galaxy clusters after almost 5 years of
observations with the Fermi-LAT telescope is one of the most relevant
recent observational milestones in this field (Sect. 3.3).
It contradicts several optimistic expectations
that had been developed from attempts to model CR production in
these environments.
The failure, so far, to detect GeV - TeV $\gamma$-rays in any cluster
generally limits the energy budget of CR protons in the
central (Mpc-scale) regions of galaxy clusters to the percent level
of the thermal energy.
However, this observed limit does not yet imply that CRs are dynamically
unimportant everywhere in the
ICM, because CRs may still be able to
contribute significantly to the cluster energy
budget in the outermost regions of these systems\footnote{For a fixed
ratio of CRp to thermal
energy the luminosity of secondary $\gamma$-rays scales with the square of the 
thermal gas density (eq. \ref{q_pi}, \ref{gamma}).}.
Most important, current $\gamma$-ray limits do not constrain the most
significant aspect of having CRs and magnetic fields in the ICM; that
is, radical changes of the (micro-)physics of the ICM that are
potentially induced by these non-thermal components. Even weak
magnetic fields
will control important transport processes within the ICM, including
viscosity and thermal and electrical conductivities. CRs can also play
important roles in damping turbulence as well as driving
instabilities that influence the small-scale structure of the
cluster magnetic field.

Radio observations of synchrotron radiation from the ICM
provide the most important guide-lines 
for understanding the physics of CRs acceleration in galaxy clusters.
This has been discussed extensively in Sect. 4.
Observations detect diffuse, steep spectrum, cluster-scale
synchrotron emission clearly associated with the ICM in the form 
of giant radio halos and relics.
One of the most significant 
observational milestones for galaxy clusters in the last
decade was the establishment of a connection between these giant radio sources 
and cluster mergers. 
That connection provides a starting point to understand radio emission origins
and, most important, to identify a theoretical link between CR
acceleration/transport in the ICM and the effects induced by cluster
mergers on the physics of the ICM itself.
In this respect it is quite commonly accepted that giant radio halos trace 
turbulent regions 
that form during mergers and where particles are trapped and 
accelerated/generated by some mechanism (Sect. 4.2) and 
where magnetic fields are amplified. 
On the other hand,
giant radio relics are associated with cosmological shock waves where 
particles can be accelerated 
from the upstream thermal particle pool reaccelerated from nonthermal particle 
populations somehow resident in the inflowing plasma (Sect. 4.4).
In Sect. 4.3 we have also discussed the case
or radio-mini halos. These diffuse radio sources are found in relaxed,
cool-core clusters and have a size comparable to that of the cool-core region, so substantially 
smaller than the giant halos. The fact that mini-halos are always found
in dynamically relaxed systems suggests that situations other than cluster mergers 
can also drain energy ``in situ'' into CRs and, probably, magnetic fields. 
Yet, as we have discussed in Sect.
4.3, whether or not the underlying physical mechanisms that accelerate the
CRe in mini halos differ substantially from the analogous mechanisms
in giant radio halos is still unclear.

\subsection{Radio Emission}
%\centerline{\it Radio Emission}

The challenges to explain the origin of radio halos and relics triggered 
a lively debate in the literature. Although fundamental steps 
have been successfully obtained in the last years, 
several important aspects of the physics of these sources are
still unclear. The most relevant observational and theoretical
breakthroughs have been summarised in Sect. 4 along with the future
observational prospects in the field.

As motivated in Sect. 4.2, a 
promising model for giant radio halos is based on the hypothesis that CRs are re-accelerated
by turbulence generated in the ICM during cluster mergers. Turbulence
will not accelerate CR electrons directly out of the thermal
electron population; some other injection sources are needed.
Such seed electrons probably come in various proportions from
large scale shocks within the ICM, from energetic phenomena
associated with individual galaxies and as secondary $e^{\pm}$
also produced in CR$p-p$ collisions. The last of these
contributors has received much attention. In fact direct secondary CRe
generation without any re-acceleration was also proposed
by itself to explain giant radio halos.
Indeed one of the most relevant theoretical discussions for the origin of giant
radio halos is the role of these secondary particles.
One way to constrain the importance of these secondaries 
starts from the Fermi-LAT limits to $\gamma$-ray emission.
This is especially meaningful when combined with radio frequency
information in nearby clusters hosting radio halos.  
In particular deep radio observations
allow detection of the radio emission in these clusters up
to their largest spatial scales and provide constraints on the spectrum
of radio halos. These studies appear to challenge the classical, pure
secondary injection model for radio halos.

\noindent
A unique and clear expectation of radio halo models based on turbulent
acceleration is the existence of a large population of radio halos with very steep
spectra. This prediction reflects the intrinsic difficulties of
the mechanisms of stochastic acceleration by turbulence in
accelerating CRe to very high energies, and suggests that clusters with weaker turbulence
will produce steeper radio spectra.
In the next few years the LOFAR radio telescope will survey the
northern sky at (as yet) unexplored low radio frequencies,
thus allowing to unveil a large number of these sources if they exist.

Giant, peripheral radio relics have recently attracted much attention.
This has been discussed in Sect. 4.4.
There is general agreement on the fact that radio relics originate
at shocks waves that cross galaxy clusters during mergers.
A shock connection is motivated by strong arguments, including 
their morphology, polarization and spectral properties, but also by
the observed spatial coincidence between radio relics and shocks found
in some cases by X-ray observations. On the other hand,
theory and X-ray observations agree on the fact that merger shocks are
relatively weak, with Mach numbers less than a few. Significantly,  they are much weaker
than shocks in supernova remnants, where efficient
CRs acceleration is proved and strong magnetic field amplification is
strongly suggested. There are very good reasons to expect particle acceleration to
be much less efficient in weak shocks. So, these processes, that are so important in SNR
shocks, should be much less effective in cluster merger shocks and, so long
as the accelerated population is locally injected at low energies, lead to a 
rather steep spectrum compared to those seen in some
relics. In this respect radio relics are ``unique laboratories'' to study 
the acceleration of particles at large-scale, weak shocks in high-beta plasma
environments.
While first order, diffusive shock acceleration (DSA) at these
shocks is the apparent candidate to explain the CR electron population seen
in relics, there are fundamental concerns about the detailed, micro-physical mechanisms.
In fact the most relevant concern is
whether shocks in galaxy clusters are efficient enough
to accelerate the emitting CRe from the thermal pool to observed multi-GeV energies.
As discussed in Sect. 4.4 this appears not to be the case.
Although the classical model of DSA
for locally injected CR electrons could account for properties of 
some relics, such a model is challenged by energy arguments combined
with simulations of CR acceleration at weak shocks.
This is especially a concern in the case of the most powerful relics
and those with relatively flat spectra giant radio relics (we note that
very steep spectra of giant relics also challenge the classical DSA model due 
to associated energy arguments).
A natural candidate-mechanism is the re-acceleration of pre-existing
CRe at these shocks, provided there is a scenario that produces them in sufficient numbers and
maintains a sufficiently flat spectrum upstream of the shock. DSA is 
more efficient at extracting shock energy 
in this situation and does not constrain the
spectrum of the downstream CR population to the ``classical DSA'' value
if the CR population entering the shock has a flatter spectrum
than this.
Radio relics pin-point the (re)acceleration of CRe at shocks in the
ICM. However, what about CRp ?
As discussed in Sect. 2, 
shock acceleration is generally expected to be more efficient for CRp
acceleration rather than for CRe. 
In this respect a consequence of current models of radio relics is the likely
acceleration of a substantial population of CRp in galaxy clusters.
That, however, is strongly constrained by current luminosity limits in
$\gamma$-rays. 
Consequently we can expect in the coming years the combination of radio and $\gamma$-ray
observations to provide interesting constraints on the physics of
(re-)acceleration of CRs at clusters shocks and on the
ratio CRp/CRe at these shocks.

\subsection{High Energy Emission}
%\centerline{\it High Energy Emission}

High energy emission from galaxy clusters has been discussed in Sect. 5.
Galaxy clusters should be high energy sources. This is not only due to
theoretical arguments based on acceleration and confinement of CRs in
these systems, but it is unavoidably expected at som elevel
from the existence of the
high energy CRe that produce the synchrotron radiation observed in the
radio band. 

\noindent
In particular hard X-ray emission is expected from ICS of the CMB
photons by essentially the same CRe that produce synchrotron radiation in the
radio band. In this case the expected ICS luminosity depends on 
the magnetic field strength in the ICM. 
As a conservative summary in Sect. 5.1 we stated that 
no ICS emission is convincingly detected in the hard X-ray band, implying
a lower limit to the cluster magnetic field of about 0.1-0.3 $\mu$G in
the cases of several nearby clusters hosting radio halos. These limits
are substantially smaller than the current estimates of the magnetic
field strength based on Faraday RM. Consequently the expected ICS
emission could very well be much smaller and well below current observational
capabilities.

\noindent
Gamma-ray emission from galaxy clusters is discussed in Sect. 5.2 (and
Sect. 3.3). These emissions are mainly expected from the
combination of $\pi^0$-decay from CRp-p collisions and ICS from
primary CRe. The former mechanism should dominate in
cluster central regions, while the latter one could be the most
important mechanism in cluster outskirts. The $\gamma$-ray
luminosity of galaxy clusters and the expected ratio between
$\pi^0$-decay and
ICS components are uncertain, as a consequence of our poor understanding 
of several aspects of the CRs acceleration and transport processes in the ICM.
These mechanisms are, however, strongly connected with the physics of the
observed radio sources. For this reason multi-frequency modelings
and the combination of observations at both radio frequencies and
high-energies have allowed derivation of important constraints on the 
origin of halos and relics, and also meaningful expectations for the 
resulting high energy emission from galaxy clusters. This has been extensively
discussed in Sect. 5.2.

Another important point that has been discussed in this review is the
maximum energy of CRs that can be achieved in galaxy clusters. This has
been discussed to some extent in Sect. 2 where we focus on
CRe and CRp; we do not study the case of nuclei.
Of course, this is still very uncertain. Based on current understanding
of CRs acceleration and transport/confinement in galaxy clusters it is 
possible to conclude that CRe can be accelerated to multi-TeV energies,
while CRp might also reach EeV energies.
In particular the possibility to accelerate EeV protons in the ICM is
very appealing. As we mention in Sect. 5.2, 
one possible way to unveil these particles, if they exist in galaxy clusters, is 
to detect the synchrotron hard X-rays and ICS $\gamma$-rays (TeV) that are emitted 
by the secondary pairs that are produced through the interaction of
these EeV CRp with the CMB photons. 

\subsection{The Future}
%\centerline{\it The Future}

The next decade represents a ``golden age'' for 
studies of non-thermal components in galaxy clusters. 
This is one of the reasons that motivated us to write a review at this time.

\noindent
This expectation is especially true in radioastronomy, thanks to a new generation 
of radio telescopes, such as LOFAR and the 
SKA pathfinders/precursors (ASKAP, MeerKAT, MWA, LWA). They will enter into
unexplored territories, reaching 
unprecedented sensitivities to cluster scale emission over a
broad frequency range.
These telescopes will also allow polarization and Faraday Rotation studies of 
background and cluster radio sources with unprecedented statistics,
frequency and dynamic range, probing cluster scale magnetic fields.
As we briefly discussed in Sect. 4.1 (and Sect. 4.2-4.4), magnetic 
field properties and the
polarization of diffuse cluster-scale sources are indeed central 
ingredients also for 
our understanding of CRs acceleration and transport in the ICM.

\noindent
One of the most notable examples of new, expected information is the spectrum of halos and relics.
As discussed in Sect. 4, the spectrum of these sources provide crucial information 
for the origin of the emitting CRe. However, current observations allow
derivation of high quality spectra only for a few cases. Most important,
spectra are typically measured in a limited frequency range
$\nu_{max}/\nu_{min} \leq 10$ that probes the energy spectra 
of the emitting CRe over a very narrow energy range.
The combination of LOFAR with JVLA observations (potentially also in
combination with data from single
dish radio telescopes) will allow spatially resolved and 
accurate measurements of halo and relic spectra over a frequency
range 10 times bigger. That extension will open the possibility of deriving meaningful 
constraints of the CRe energy distributions and their spatial evolution/variation.

\noindent
As discussed in Sect. 4, the surveys planned with the next generation of radio 
telescopes have the potential to probe the formation and evolution
of cluster scale radio sources along with the evolution of the hosting
clusters themselves.
The apparent connection between halos and mergers suggests that these
sources are valuable probes of the clusters merging history throughout the
cosmic epochs. In this respect combined X-ray -- radio and SZ -- radio statistical
studies of the occurrence of radio halos in clusters will provide 
unique information on the origin of non-thermal components and on their
connection with the clusters thermal properties.
In this respect LOFAR surveys will be of great interest, as they will
probe an unexplored frequency range where current models predict the
existence of cluster-scale radio emission with very steep spectrum.

\noindent
In our review we also stressed that presumably {\it current observations}
detect only the {\it tip of the iceberg} of the non-thermal 
radio emission from the Cosmic Web. 
In this respect, what is still unseen is the most attractive science case of
the new generation of radio telescopes. In principle, according to the current
theoretical picture, one may speculate that halos and relics pin-point 
only the regions where 
CRe acceleration is most efficient. Fainter synchrotron emission
might be much more common and it may extend to scales even larger, thus
tracing the complex pattern of shocks and turbulence that embed 
clusters and filaments. Unfortunately, our current ignorance of plasma-physics in
these environments does not yet allow physically solid attempts 
to establish whether the SKA and its pathfinders will be able
to detect these signals.

The future is bright also at high photon energies.
First of all the hard X-ray telescopes NuSTAR and ASTRO-H will shortly
derive constraints on the presence of non-thermal excess emission
in galaxy clusters at a level that is about 10 times fainter than current studies.
As discussed in Sect. 5.1 the detection of ICS emission from 
the brightest halos and relics is only possible if 
the magnetic field is substantially smaller, at least a factor 2-3, than that estimated
from current RM analysis. However, even the non detection of nearby
clusters hosting bright radio halos and relics 
at the sensitivity level of these telescopes
would be a valuable result, providing very meaningful constraints on the
magnetic field strength in these systems.

\noindent
However, one of the most desired results from ASTRO-H is the first measure, 
from the study of metal lines, of turbulent motions in the ICM. 
This has been mentioned in Sects. 2 and 4.
Turbulence is a fundamental ingredient for current theoretical scenario,
not only because it is important for the acceleration and transport of
CRs, but also because it is expected to strongly affect the
micro-physics of the ICM. In this respect we also mention that 
a gigantic step forward represented by measured turbulent motions in galaxy clusters
is a central science case of X-ray telescopes with much better spectral resolution
and effective area, such as {\it Athena+}, that may be launched in
about 15 years.

\noindent
Current constraints from $\gamma$-ray observations allowed a gigantic
step in our understanding of the CRs content in galaxy clusters.
This has been discussed in Sects. 3.3, 4 and 5 also in the context of the
impact of these constraints on models of giant halos and relics, and mini-radio halos. 
Better constraints are expected from future and ongoing observations
with the Fermi-LAT and from observations with the next generation of
Cherenkov telescopes, such as the CTA.
As discussed in Sect. 5 these observations will allow derivation of
fundamental constraints on the physics of CRe/CRp acceleration 
and transport in the ICM,
and on the role played by secondary particles in the origin of
giant and mini-radio halos.

\section*{Acknowledgments}
This work was supported in part by the National Science Foundation under
Grant No. PHYS-1066293 and the hospitality of the Aspen Center for
Physics. GB acknowledges the Simons Foundation for support.
TWJ is supported by NASA grant NNX09AH78G, by US NSF grants AST0908668 and
AST1211595 and the Minnesota Supercomputing Institute.
We thank Kaustuv Basu, Marcus Bruggen, Reiner Beck, Damiano Caprioli, Rossella
Cassano, Julius Donnert, Yutaka Fujita, Roberto Fusco-Femiano, Myriam Gitti,
Joseph Lazio, Hui Li, Franco Vazza and Fabio Zandanel
for comments and discussions. We also thank help from
Franco Vazza for Fig. 2, from Paul Edmond for Fig. 5, from
Simona Giacintucci for Figs. 11 \& 15, and from
Rossella Cassano and Fabio Zandanel for Fig. 14.


\begin{thebibliography}{00}
\bibitem{sarazinbook} Sarazin, C. L., {\it X--ray emission from clusters of galaxies}, Cambridge: Cambridge University Press (1988). 
\bibitem{borgani} Kravtsov, A.~V., \& Borgani, S.\ 2012, \araa, 50, 353
\bibitem{briel01} Briel, U.~G., Henry, J.~P., Lumb, D.~H., et al.\ 2001, \aap, 365, L60 
\bibitem{planck11} Planck Collaboration, Ade, P.~A.~R., Aghanim, N., et al.\ 2011, \aap, 536, A8 
\bibitem{brownrudnick11} Brown, S., \& Rudnick, L.\ 2011, \mnras, 412, 2
\bibitem{carlstrom02} Carlstrom, J.~E., Holder, G.~P., \& Reese, E.~D.\ 2002, \araa, 40, 643
\bibitem{normanbryan99} Norman, M.~L., \& Bryan, G.~L.\ 1999, The
Radio Galaxy Messier 87, LNP, 530, 106 
\bibitem{ryuetal08} Ryu, D., Kang, H., Cho, J., \& Das, S.\ 2008, Science, 320, 909 
\bibitem{kaastrabook} Kaastra, J.~S., Bykov, A.~M., Schindler, S., et al.\ 2008, \ssr, 134, 1 
\bibitem{norman95} Norman, C.~A., Melrose, D.~B., \& Achterberg, A.\ 1995, \apj, 454, 60
\bibitem{sarazin99} Sarazin, C.~L.\ 1999, \apj, 520, 529
\bibitem{loebwaxman00} Loeb, A., \& Waxman, E.\ 2000, \nat, 405, 156
\bibitem{ryu03} Ryu, D., Kang, H., Hallman, E., \& Jones, T.~W.\ 2003, \apj, 593, 599
\bibitem{cassanobrunetti05} Cassano, R., \& Brunetti, G.\ 2005, \mnras, 357, 1313
\bibitem{brunetti08} Brunetti, G., Giacintucci, S., Cassano, R., et al.\ 2008, \nat, 455, 944
\bibitem{voelk96} V{\"o}lk, H.~J., Aharonian, F.~A., \& Breitschwerdt, D.\ 1996, \ssr, 75, 279 
\bibitem{berezinsky97} Berezinsky, V.~S., Blasi, P., \& Ptuskin, V.~S.\ 1997, \apj, 487, 529
\bibitem{ensslin97} Ensslin, T.~A., Biermann, P.~L., Kronberg, P.~P., \& Wu, X.-P.\ 1997, \apj, 477, 560 
\bibitem{blasicolafrancesco99} Blasi, P., \& Colafrancesco, S.\ 1999, Astroparticle Physics, 12, 169 
\bibitem{lazarianbrunetti11} Lazarian, A., \& Brunetti, G.\ 2011, \memsai, 82, 636 
\bibitem{colafrancescoblasi98} Colafrancesco, S., \& Blasi, P.\ 1998, Astroparticle Physics, 9, 227 
\bibitem{voelkatoyan99} V{\"o}lk, H.~J., \& Atoyan, A.~M.\ 1999, Astroparticle Physics, 11, 73
\bibitem{miniati01} Miniati, F., Jones, T.~W., Kang, H., \& Ryu, D.\ 2001, \apj, 562, 233 
\bibitem{pfrommerensslin04} Pfrommer, C., \& En{\ss}lin, T.~A.\ 2004, \mnras, 352, 76 
\bibitem{blasi07} Blasi, P., Gabici, S., \& Brunetti, G.\ 2007, International Journal of Modern Physics A, 22, 681 
\bibitem{reimer03} Reimer, O., Pohl, M., Sreekumar, P., \& Mattox, J.~R.\ 2003, \apj, 588, 155
\bibitem{perkinsetal06} Perkins, J.~S., Badran, H.~M., Blaylock, G., et al.\ 2006, \apj, 644, 148
\bibitem{kiuchietal09} Kiuchi, R., Mori, M., Bicknell, G.~V., et al.\ 2009, \apj, 704, 240
\bibitem{aharonian09a} Aharonian, F., Akhperjanian, A.~G., Anton, G., et al.\ 2009, \aap, 495, 27
\bibitem{ackermann10} Ackermann, M., Ajello, M., Allafort, A., et al.\ 2010, \apjl, 717, L71
\bibitem{aleksic12} Aleksi{\'c}, J., Alvarez, E.~A., Antonelli, L.~A., et al.\ 2012, \aap, 541, A99
\bibitem{Fermilat13} The Fermi-LAT Collaboration : Ackermann, M., et al.\ 2013, arXiv:1308.5654 
\bibitem{huberetal13} Huber, B., Tchernin, C., Eckert, D., et al.\ 2013, \aap, 560, A64
\bibitem{reimer04} Reimer, A., Reimer, O., Schlickeiser, R., \& Iyudin, A.\ 2004, \aap, 424, 773
\bibitem{brunetti07} Brunetti, G., Venturi, T., Dallacasa, D., et al.\ 2007, \apjl, 670, L5
\bibitem{brown11} Brown, S., Emerick, A., Rudnick, L., \& Brunetti, G.\ 2011, \apjl, 740, L28
\bibitem{ferrari08} Ferrari, C., Govoni, F., Schindler, S., Bykov, A.~M., \& Rephaeli, Y.\ 2008, \ssr, 134, 93
\bibitem{feretti12} Feretti, L., Giovannini, G., Govoni, F., \& Murgia, M.\ 2012, \aapr, 20, 54
\bibitem{bruggen12} Br{\"u}ggen, M., Bykov, A., Ryu, D., R\"ottgering, H.\ 2012, \ssr, 166, 187
\bibitem{brunetti11} Brunetti, G.\ 2011, \memsai, 82, 515
\bibitem{sarazin04} Sarazin, C.~L.\ 2004, Journal of Korean Astronomical Society, 37, 433 
\bibitem{nuzaetal12} Nuza, S.~E., Hoeft, M., van Weeren, R.~J., Gottl{\"o}ber, S., \& Yepes, G.\ 2012, \mnras, 420, 2006
\bibitem{kang96} Kang, H., Ryu, D., \& Jones, T.~W.\ 1996, \apj, 456, 422
\bibitem{fujita03} Fujita, Y., Takizawa, M., \& Sarazin, C.~L.\ 2003, \apj, 584, 190
\bibitem{brunettilazarian07} Brunetti, G., \& Lazarian, A.\ 2007, \mnras, 378, 245
\bibitem{ensslin11} Ensslin, T., Pfrommer, C., Miniati, F., \& Subramanian, K.\ 2011, \aap, 527, A99
\bibitem{jones11} Jones, T.~W.\ 2011, Journal of Astrophysics and Astronomy, 32, 427
\bibitem{morlinocaprioli12} Morlino, G., \& Caprioli, D.\ 2012, \aap, 538, A81 
\bibitem{miley80} Miley, G.\ 1980, \araa, 18, 165
\bibitem{burns90} Burns, J.~O.\ 1990, \aj, 99, 14 
\bibitem{bestetal07} Best, P.~N., von der Linden, A., Kauffmann, G., Heckman, T.~M.,  \& Kaiser, C.~R.\ 2007, \mnras, 379, 894 
\bibitem{mittaletal09} Mittal, R., Hudson, D.~S., Reiprich, T.~H., \& Clarke, T.\ 2009, \aap, 501, 835
\bibitem{macnamara07} McNamara, B.~R., \& Nulsen, P.~E.~J.\ 2007, \araa, 45, 117
\bibitem{gitti12} Gitti, M., Brighenti, F., \& McNamara, B.~R.\ 2012,
Advances in Astronomy, 2012, 950641
\bibitem{rafferty06} Rafferty, D.~A., McNamara, B.~R., Nulsen, P.~E.~J., \& Wise, M.~W.\ 2006, \apj, 652, 216
\bibitem{dunnfabian06} Dunn, R.~J.~H., \& Fabian, A.~C.\ 2006, \mnras, 373, 959 
\bibitem{oneill10} O'Neill, S.~M., \& Jones, T.~W.\ 2010, \apj, 710, 180
\bibitem{gitti11} Gitti, M., Nulsen, P.~E.~J., David, L.~P., McNamara, B.~R., \& Wise, M.~W.\ 2011, \apj, 732, 13
\bibitem{hardcastlecroston10} Hardcastle, M.~J., \& Croston, J.~H.\ 2010, \mnras, 404, 2018
\bibitem{crostonhardcastle13} Croston, J.~H., \& Hardcastle, M.~J.\ 2013, arXiv:1312.5183 
\bibitem{deyoung06} De Young, D.~S.\ 2006, \apj, 648, 200
\bibitem{dursipfrommer08} Dursi, L.~J., \& Pfrommer, C.\ 2008, \apj, 677, 993
\bibitem{dongstone09} Dong, R., \& Stone, J.~M.\ 2009, \apj, 704, 1309
\bibitem{oneilletal09} O'Neill, S.~M., De Young, D.~S., \& Jones, T.~W.\ 2009, \apj, 694, 1317
\bibitem{stockeetal09} Stocke, J.~T., Hart, Q.~N., \& Hallman, E.~J.\ 2009, American Institute of Physics Conference Series, 1201, 206 
\bibitem{burns98} Burns, J.~O.\ 1998, Science, 280, 400
\bibitem{hardcastlesakelliou04} Hardcastle. M. J., \& Sakelliou, I. 2004, \mnras, 349, 560
\bibitem{cassano10a} Cassano, R., Ettori, S., Giacintucci, S., et al.\ 2010, \apjl, 721, L82
\bibitem{takizawa00} Takizawa, M., \& Naito, T.\ 2000, \apj, 535, 586
\bibitem{gabiciblasi03} Gabici, S., \& Blasi, P.\ 2003, \apj, 583, 695
\bibitem{pfrommer06} Pfrommer, C., Springel, V., En{\ss}lin, T.~A., \& Jubelgas, M.\ 2006, \mnras, 367, 113
\bibitem{skillman08} Skillman, S.~W., O'Shea, B.~W., Hallman, E.~J., Burns, J.~O., \& Norman, M.~L.\ 2008, \apj, 689, 1063
\bibitem{vazza09b} Vazza, F., Brunetti, G., \& Gheller, C.\ 2009, \mnras, 395, 1333
\bibitem{vazza12CR} Vazza, F., Br{\"u}ggen, M., Gheller, C., \& Brunetti, G.\ 2012, \mnras, 421, 3375
\bibitem{roettiger99} Roettiger, K., Stone, J.~M., \& Burns, J.~O.\ 1999, \apj, 518, 594
\bibitem{dolag02} Dolag, K., Bartelmann, M., \& Lesch, H.\ 2002, \aap, 387, 383
\bibitem{markevitch01} Markevitch, M., \& Vikhlinin, A.\ 2001, \apj, 563, 95
\bibitem{markevitch10} Markevitch, M.\ 2010, arXiv:1010.3660
\bibitem{vazza10} Vazza, F., Brunetti, G., Gheller, C., \& Brunino, R.\ 2010, \na, 15, 695
\bibitem{berrington03} Berrington, R.~C., \& Dermer, C.~D.\ 2003, \apj, 594, 709
\bibitem{kangetal07} Kang, H., Ryu, D., Cen, R., \& Ostriker, J.~P.\ 2007, \apj, 669, 729 
\bibitem{vazzacomparison11} Vazza, F., Dolag, K., Ryu, D., et al.\ 2011, \mnras, 418, 960
\bibitem{miniati01gamma} Miniati, F., Ryu, D., Kang, H., \& Jones, T.~W.\ 2001, \apj, 559, 59
\bibitem{pfrommer07} Pfrommer, C., En{\ss}lin, T.~A., Springel, V., Jubelgas, M., \& Dolag, K.\ 2007, \mnras, 378, 385 
\bibitem{bell78} Bell, A.~R.\ 1978, \mnras, 182, 147
\bibitem{drury83} Drury, L.~O.\ 1983, Reports on Progress in Physics, 46, 973
\bibitem{blandford87} Blandford, R., \& Eichler, D.\ 1987, \physrep, 154, 1
\bibitem{jones91} Jones, F.~C., \& Ellison, D.~C.\ 1991, \ssr, 58, 259
\bibitem{malkov01} Malkov, M.~A., \& O'C Drury, L.\ 2001, Reports on Progress in Physics, 64, 429
\bibitem{blasi02} Blasi, P.\ 2002, Astroparticle Physics, 16, 429
\bibitem{kang02} Kang, H., Jones, T.~W., \& Gieseler, U.~D.~J.\ 2002, \apj, 579, 337
\bibitem{kang05} Kang, H., \& Jones, T.~W.\ 2005, \apj, 620, 44
\bibitem{kang09} Kang, H., Ryu, D., \& Jones, T.~W.\ 2009, \apj, 695, 1273
\bibitem{kang13} Kang, H., Jones, T.~W., \& Edmon, P.~P.\ 2013, \apj, 777, 25 
\bibitem{kangjones02} Kang, H., \& Jones, T.~W.\ 2002, Journal of Korean Astronomical Society, 35, 159 
\bibitem{blasi01} Blasi, P.\ 2001, Astroparticle Physics, 15, 223
\bibitem{kang12} Kang, H., Ryu, D., \& Jones, T.~W.\ 2012, \apj, 756, 97
\bibitem{ensslin98} Ensslin, T.~A., Biermann, P.~L., Klein, U., \& Kohle, S.\ 1998, \aap, 332, 395
\bibitem{hoeft07} Hoeft, M., \& Br{\"u}ggen, M.\ 2007, \mnras, 375, 77
\bibitem{berezinskygrigoreva88} Berezinskii, V.~S., \& Grigor'eva, S.~I.\ 1988, \aap, 199, 1
\bibitem{blasi04} Blasi, P.\ 2004, Journal of Korean Astronomical Society, 37, 483
\bibitem{vink14} Vink, J., \& Yamazaki, R.\ 2014, \apj, 780, 125 
\bibitem{amato06} Amato, E., \& Arons, J.\ 2006, \apj, 653, 325
\bibitem{burgess06} Burgess, D.\ 2006, \apj, 653, 316
\bibitem{amano09} Amano, T., \& Hoshino, M.\ 2009, \apj, 690, 244
\bibitem{spitk11} Riquelme, M.~A., \& Spitkovsky, A.\ 2011, \apj, 733, 63
\bibitem{gargate12} Gargat{\'e}, L., \& Spitkovsky, A.\ 2012, \apj, 744, 67
\bibitem{capriolispitkovsky13} Caprioli, D., \& Spitkovsky, A.\ 2013, arXiv:1310.2943
\bibitem{parketal12} Park, J., Workman, J. C., Blackman, E. G., Ren, C. \& Siller, R., Physics of Plasmas, 19, 2904, 2012
\bibitem{bell78b} Bell, A.~R.\ 1978, \mnras, 182, 443
\bibitem{bell00} Lucek, S.~G., \& Bell, A.~R.\ 2000, \mnras, 314, 65
\bibitem{bell04} Bell, A.~R.\ 2004, \mnras, 353, 550
\bibitem{amato09} Amato, E., \& Blasi, P.\ 2009, \mnras, 392, 1591
\bibitem{biermann50} Biermann, L.\ 1950, Zeitschrift Naturforschung Teil A, 5, 65 
\bibitem{weibel59} Weibel, E.~S.\ 1959, Physical Review Letters, 2, 83
\bibitem{gedalinetal10} Gedalin, M., Medvedev, M., Spitkovsky, A., Krasnosolskikh, V., Balikhin, M., Vaivads, A. \& Perri, S, Physics of Plasmas, 17, 2108, 2010
\bibitem{davieswidrow00} Davies, G., \& Widrow, L.~M.\ 2000, \apj, 540, 755
\bibitem{caprioli12} Caprioli, D.\ 2012, \jcap, 7, 38
\bibitem{jaffe77} Jaffe, W.~J.\ 1977, \apj, 212, 1
\bibitem{deiss96} Deiss, B.~M., \& Just, A.\ 1996, \aap, 305, 407
\bibitem{heinz06} Heinz, S., Br{\"u}ggen, M., Young, A., \& Levesque, E.\ 2006, \mnras, 373, L65
\bibitem{bruggen09} Br{\"u}ggen, M., \& Scannapieco, E.\ 2009, \mnras, 398, 548
\bibitem{parrish07} Parrish, I.~J., \& Stone, J.~M.\ 2007, \apj, 664, 135
\bibitem{ricker01} Ricker, P.~M., \& Sarazin, C.~L.\ 2001, \apj, 561, 621
\bibitem{subramanian06} Subramanian, K., Shukurov, A., \& Haugen, N.~E.~L.\ 2006, \mnras, 366, 1437
\bibitem{vazza09a} Vazza, F., Brunetti, G., Kritsuk, A., et al.\ 2009, \aap, 504, 33
\bibitem{dolag05} Dolag, K., Vazza, F., Brunetti, G., \& Tormen, G.\ 2005, \mnras, 364, 753
\bibitem{iapichinoniemeyer08} Iapichino, L., \& Niemeyer, J.~C.\ 2008, \mnras, 388, 1089
\bibitem{keshet10} Keshet, U., Markevitch, M., Birnboim, Y., \& Loeb, A.\ 2010, \apjl, 719, L74
\bibitem{iapichino11} Iapichino, L., Schmidt, W., Niemeyer, J.~C., \& Merklein, J.\ 2011, \mnras, 414, 2297
\bibitem{paul11} Paul, S., Iapichino, L., Miniati, F., Bagchi, J., \& Mannheim, K.\ 2011, \apj, 726, 17
\bibitem{zuhone11} ZuHone, J.~A., Markevitch, M., \& Lee, D.\ 2011, \apj, 743, 16
\bibitem{hallman11} Hallman, E.~J., \& Jeltema, T.~E.\ 2011, \mnras, 418, 2467
\bibitem{vazzaetal11} Vazza, F., Brunetti, G., Gheller, C., Brunino, R., \& Br{\"u}ggen, M.\ 2011, \aap, 529, A17 
\bibitem{vazza12} Vazza, F., Roediger, E., \& Br{\"u}ggen, M.\ 2012, \aap, 544, A103
\bibitem{nagaietal13} Nagai, D., Lau, E.~T., Avestruz, C., Nelson, K., \& Rudd, D.~H.\ 2013, \apj, 777, 137 
\bibitem{miniati13} Miniati, F.\ 2013, arXiv:1310.2951
\bibitem{beresnyaketal13} Beresnyak, A., Xu, H., Li, H., \& Schlickeiser, R.\ 2013, \apj, 771, 131 
\bibitem{lazarian06} Lazarian, A.\ 2006, \apjl, 645, L25
\bibitem{schekochihin05} Schekochihin, A.~A., Cowley, S.~C., Kulsrud, R.~M., Hammett, G.~W., \& Sharma, P.\ 2005, \apj, 629, 139
\bibitem{schekochihin10} Schekochihin, A.~A., Cowley, S.~C., Rincon, F., \& Rosin, M.~S.\ 2010, \mnras, 405, 291
\bibitem{levinson92} Levinson, A., \& Eichler, D.\ 1992, \apj, 387, 212
\bibitem{pistinner96} Pistinner, S., Levinson, A., \& Eichler, D.\ 1996, \apj, 467, 162
\bibitem{brunettilazarian11a} Brunetti, G., \& Lazarian, A.\ 2011, \mnras, 412, 817
\bibitem{yanlazarian11} Yan, H., \& Lazarian, A.\ 2011, \apj, 731, 35
\bibitem{wentzel74} Wentzel, D.~G.\ 1974, \araa, 12, 71
\bibitem{pistinner97} Pistinner, S.~L.\ 1997, in Galactic Cluster Cooling Flows, ASP Conf. Series, 115, 165 
\bibitem{watheralleilek99} Weatherall, J.~C., \& Eilek, J.\ 1999, in Diffuse Thermal and Relativistic Plasma in Galaxy Clusters, 255
\bibitem{churazov12} Schuecker, P., Finoguenov, A., Miniati, F., B{\"o}hringer, H., \& Briel, U.~G.\ 2004, \aap, 426, 387 
\bibitem{sandersscience13} Sanders, J.~S., Fabian, A.~C., Churazov, E., et al.\ 2013, Science, 341, 1365
\bibitem{shuecker04} Schuecker, P., Finoguenov, A., Miniati, F., B{\"o}hringer, H., \& Briel, U.~G.\ 2004, \aap, 426, 387 
\bibitem{sunyaev03} Sunyaev, R.~A., Norman, M.~L., \& Bryan, G.~L.\ 2003, Astronomy Letters, 29, 783 
\bibitem{vazzaghellerbrunetti10} Vazza, F., Gheller, C., \& Brunetti, G.\ 2010, \aap, 513, A32
\bibitem{takahashi12} Takahashi, T., Mitsuda, K., Kelley, R., et al.\ 2010, Proceedings of the SPIE, 7732, 77320Z-77320Z-18
\bibitem{zhuravleva12} Zhuravleva, I., Churazov, E., Kravtsov, A., \& Sunyaev, R.\ 2012, \mnras, 422, 2712
\bibitem{zhuravleva13} Zhuravleva, I., Churazov, E., Sunyaev, R., et al.\ 2013, \mnras, 435, 3111
\bibitem{melrose80} Melrose, D. B., {\it Plasma astrohysics. Nonthermal processes in diffuse magnetized plasmas}, New York: Gordon and Breach, (1980)
\bibitem{schlickeiser02} Schlickeiser, R., {\it Cosmic ray astrophysics}, Springer (2002)
\bibitem{lazarian99} Lazarian, A., \& Vishniac, E.~T.\ 1999, \apj, 517, 700
\bibitem{petrosianeast08} Petrosian, V., \& East, W.~E.\ 2008, \apj, 682, 175
\bibitem{chernyshov12} Chernyshov, D.~O., Dogiel, V.~A., \& Ko, C.~M.\ 2012, \apj, 759, 113 
\bibitem{brunetti01} Brunetti, G., Setti, G., Feretti, L., \& Giovannini, G.\ 2001, \mnras, 320, 365
\bibitem{petrosian01} Petrosian, V.\ 2001, \apj, 557, 560
\bibitem{jokipii66} Jokipii, J.~R.\ 1966, \apj, 146, 480
\bibitem{schlickeiser93} Schlickeiser, R., \& Achatz, U.\ 1993, Journal of Plasma Physics, 49, 63
\bibitem{miller95} Miller, J.~A., \& Roberts, D.~A.\ 1995, \apj, 452, 912
\bibitem{footekulsrud79} Foote, E.~A., \& Kulsrud, R.~M.\ 1979, \apj, 233, 302
\bibitem{yanlazarian04} Yan, H., \& Lazarian, A.\ 2004, \apj, 614, 757
\bibitem{dung94} Dung, R., \& Petrosian, V.\ 1994, \apj, 421, 550
\bibitem{kirk88} Kirk, J.~G., Schneider, P., \& Schlickeiser, R.\ 1988, \apj, 328, 269
\bibitem{skilling75} Skilling, J.\ 1975, \mnras, 172, 557
\bibitem{brunettilazarian11b} Brunetti, G., \& Lazarian, A.\ 2011, \mnras, 410, 127
\bibitem{ohno02} Ohno, H., Takizawa, M., \& Shibata, S.\ 2002, \apj, 577, 658
\bibitem{brunettietal04} Brunetti, G., Blasi, P., Cassano, R., \& Gabici, S.\ 2004, \mnras, 350, 1174
\bibitem{chandran00} Chandran, B.~D.~G.\ 2000, Physical Review Letters, 85, 4656
\bibitem{yanlazarian02} Yan, H., \& Lazarian, A.\ 2002, Physical Review Letters, 89, 1102
\bibitem{brunettiblasi05} Brunetti, G., \& Blasi, P.\ 2005, \mnras, 363, 1173
\bibitem{moskalenko98} Moskalenko, I.~V., \& Strong, A.~W.\ 1998, \apj, 493, 694
\bibitem{stecker70} Stecker, F.~W.\ 1970, \apss, 6, 377
\bibitem{badhwar77} Badhwar, G.~D., Golden, R.~L., \& Stephens, S.~A.\ 1977, \prd, 15, 820
\bibitem{stephens81} Stephens, S.~A., \& Badhwar, G.~D.\ 1981, \apss, 76, 213
\bibitem{dermer86a} Dermer, C.~D.\ 1986, \apj, 307, 47
\bibitem{dermer86b} Dermer, C.~D.\ 1986, \aap, 157, 223
\bibitem{kamae05} Kamae, T., Abe, T., \& Koi, T.\ 2005, \apj, 620, 244
\bibitem{kelner06} Kelner, S.~R., Aharonian, F.~A., \& Bugayov, V.~V.\ 2006, \prd, 74, 034018
\bibitem{dolagensslin00} Dolag, K., \& Ensslin, T.~A.\ 2000, \aap, 362, 151
\bibitem{kamae06} Kamae, T., Karlsson, N., Mizuno, T., Abe, T., \& Koi, T.\ 2006, \apj, 647, 692
\bibitem{bhattacharjee00} Bhattacharjee, P.\ 2000, \physrep, 327, 109
\bibitem{GS95} Goldreich, P., \& Sridhar, S.\ 1995, \apj, 438, 763
\bibitem{cholazarianvishniac03} Cho, J., Lazarian, A., \& Vishniac, E.~T.\ 2003, in Turbulence and Magnetic Fields in Astrophysics, ed. E. Falgarone, and T. Passot., Lecture Notes in Physics, 614, 56
\bibitem{beresnyaketal11} Beresnyak, A., Yan, H., \& Lazarian, A.\ 2011, \apj, 728, 60
\bibitem{wieneretal13} Wiener, J., Oh, S.~P., \& Guo, F.\ 2013, \mnras, 434, 2209 
\bibitem{cholazarian04JKAS} Cho, J., \& Lazarian, A.\ 2004, Journal of Korean Astronomical Society, 37, 557 
\bibitem{richardson1926} Richardson, L.~F.\ 1926, Royal Society of London Proceedings Series A, 110, 709
\bibitem{rebusco06} Rebusco, P., Churazov, E., B\"ohringer, H., \& Forman, W.\ 2005, \mnras, 372, 1840  
\bibitem{kang11} Kang, H., \& Ryu, D.\ 2011, \apj, 734, 18
\bibitem{aharonian09b} Aharonian, F., Akhperjanian, A.~G., Anton, G., et al.\ 2009, \aap, 502, 437
\bibitem{aleksic10} Aleksi{\'c}, J., Antonelli, L.~A., Antoranz, P., et al.\ 2010, \apj, 710, 634
\bibitem{veritas12} Arlen, T., Aune, T., Beilicke, M., et al.\ 2012, \apj, 757, 123
\bibitem{prokhorov13} Prokhorov, D.~A., \& Churazov, E.~M.\ 2013, arXiv:1309.0197
\bibitem{zandanelando13} Zandanel, F., \& Ando, S.\ 2013, arXiv:1312.1493
\bibitem{pinzke10} Pinzke, A., \& Pfrommer, C.\ 2010, \mnras, 409, 449
\bibitem{clarke01} Clarke, T.~E., Kronberg, P.~P.,B\"ohringer, H.\ 2001, \apjl, 547, L111
\bibitem{carilli02} Carilli, C.~L., \& Taylor, G.~B.\ 2002, \araa, 40, 319
\bibitem{bonafede10} Bonafede, A., Feretti, L., Murgia, M., et al.\ 2010, \aap, 513, A30
\bibitem{miniati03} Miniati, F.\ 2003, \mnras, 342, 1009
\bibitem{pfrommer08} Pfrommer, C.\ 2008, \mnras, 385, 1242
\bibitem{colafrancesco08} Colafrancesco, S., \& Marchegiani, P.\ 2008, \aap, 484, 51
\bibitem{pinzkeetal11} Pinzke, A., Pfrommer, C., \& Bergstr{\"o}m, L.\ 2011, \prd, 84, 123509
\bibitem{govoni04} Govoni, F., Markevitch, M., Vikhlinin, A., et al.\ 2004, \apj, 605, 695
\bibitem{venturietal13} Venturi, T., Giacintucci, S., Dallacasa, D., et al.\ 2013, \aap, 551, A24
\bibitem{giacintuccietal13} Giacintucci, S., Markevitch, M., Venturi, T., et al.\ 2014, \apj, 781, 9
\bibitem{gittietal04} Gitti, M., Brunetti, G., Feretti, L., \& Setti, G.\ 2004, \aap, 417, 1
\bibitem{vanweeren10} van Weeren, R.~J., R{\"o}ttgering, H.~J.~A., Br{\"u}ggen, M., \& Hoeft, M.\ 2010, Science, 330, 347 
\bibitem{ogrean13} Ogrean, G.~A., Br{\"u}ggen, M., R{\"o}ttgering, H., et al.\ 2013, \mnras, 429, 2617
\bibitem{rottgering97} Rottgering, H.~J.~A., Wieringa, M.~H., Hunstead, R.~W., \& Ekers, R.~D.\ 1997, \mnras, 290, 577
\bibitem{kaleetal12} Kale, R., Dwarakanath, K.~S., Bagchi, J., \& Paul, S.\ 2012, \mnras, 426, 1204
\bibitem{venturi11} Venturi, T.\ 2011, \memsai, 82, 499
\bibitem{farnsworthetal13} Farnsworth, D., Rudnick, L., Brown, S., \& Brunetti, G.\ 2013, \apj, 779, 189
\bibitem{rudnicketal09} Rudnick, L., Alexander, P., Andernach, H., et al.\ 2009, astro2010: The Astronomy and Astrophysics Decadal Survey, 2010, 253 
\bibitem{dolagjcap05} Dolag, K., Grasso, D., Springel, V., \& Tkachev, I.\ 2005, \jcap, 1, 9 
\bibitem{brueggenetal05} Br{\"u}ggen, M., Ruszkowski, M., Simionescu, A., Hoeft, M., \& Dalla Vecchia, C.\ 2005, \apjl, 631, L21 
\bibitem{DuboisTeyssier08} Dubois, Y., \& Teyssier, R.\ 2008, \aap, 482, L13 
\bibitem{donnertetal09} Donnert, J., Dolag, K., Lesch, H., M\"uller, E.\ 2009, \mnras, 392, 1008 
\bibitem{xuetal10} Xu, H., Li, H., Collins, D.~C., Li, S., \& Norman, M.~L.\ 2010, \apj, 725, 2152
\bibitem{xuetal11} Xu, H., Li, H., Collins, D.~C., Li, S., \& Norman, M.~L.\ 2011, \apj, 739, 77 
\bibitem{ensslinvogt03} Ensslin, T. A. \& Vogt. C., 2003, A\&A, 401, 835
\bibitem{choryu09} Cho, J. \& Ryu, D. 2009, ApJ, 705, L90
\bibitem{vogtensslin05} Vogt, C. \& Ensslin, T. A. 2005, A\&A, 434, 67
\bibitem{murgiaetal04} Murgia, M., Govoni, Feretti, L., Giovannini, Dallacasa, G., Fanti, R., Taylor, G. B. \& Dolag, K., 2004, A\&A, 424, 429
\bibitem{govoni06} Govoni, F.\ 2006, Astronomische Nachrichten, 327, 539
\bibitem{guidettietal08} Guidetti, D., Murgia, M., Govoni, F., et al.\ 2008, \aap, 483, 699
\bibitem{kucharensslin11} Kuchar, P., \& En{\ss}lin, T.~A.\ 2011, \aap, 529, A13
\bibitem{vaccaetal12} Vacca, V., Murgia, M., Govoni, F., et al.\ 2012, \aap, 540, A38
\bibitem{bonafedecoma13} Bonafede, A., Vazza, F., Br{\"u}ggen, M., et al.\ 2013, \mnras, 433, 3208
\bibitem{donnertetal13} Donnert, J., Dolag, K., Brunetti, G., \& Cassano, R.\ 2013, \mnras, 429, 3564
\bibitem{petrosian08} Petrosian, V., \& Bykov, A.~M.\ 2008, \ssr, 134, 207
\bibitem{dennison80} Dennison, B.\ 1980, \apjl, 239, L93
\bibitem{keshetloeb10} Keshet, U., \& Loeb, A.\ 2010, \apj, 722, 737
\bibitem{hanisch82} Hanisch, R.~J.\ 1982, \aap, 111, 97
\bibitem{giovannini99} Giovannini, G., Tordi, M., \& Feretti, L.\ 1999, \na, 4, 141
\bibitem{kempnersarazin01} Kempner, J.~C., \& Sarazin, C.~L.\ 2001, \apj, 548, 639
\bibitem{venturi07} Venturi, T., Giacintucci, S., Brunetti, G., et al.\ 2007, \aap, 463, 937
\bibitem{venturi08} Venturi, T., Giacintucci, S., Dallacasa, D., et al.\ 2008, \aap, 484, 327
\bibitem{cassano08} Cassano, R., Brunetti, G., Venturi, T., et al.\ 2008, \aap, 480, 687
\bibitem{cassanoetal13} Cassano, R., Ettori, S., Brunetti, G., et al.\ 2013, \apj, 777, 141
\bibitem{kaleetal13} Kale, R., Venturi, T., Giacintucci, S., et al.\ 2013, \aap, 557, A99
\bibitem{brunettietal09} Brunetti, G., Cassano, R., Dolag, K., \& Setti, G.\ 2009, \aap, 507, 661
\bibitem{buote01} Buote, D.~A.\ 2001, \apjl, 553, L15
\bibitem{kushniretal09} Kushnir, D., Katz, B., \& Waxman, E.\ 2009, \jcap, 9, 24
\bibitem{basu12} Basu, K.\ 2012, \mnras, 421, L112
\bibitem{plancksz13} Planck Collaboration, Ade, P.~A.~R., Aghanim, N., et al.\ 2013, arXiv:1303.5089 
\bibitem{sommerbasu13} Sommer, M.~W., \& Basu, K.\ 2014, \mnras, 437, 2163
\bibitem{govoni01} Govoni, F., En{\ss}lin, T.~A., Feretti, L., \& Giovannini, G.\ 2001, \aap, 369, 441
\bibitem{planck13} Planck Collaboration, Ade, P.~A.~R., Aghanim, N., et al.\ 2013, \aap, 554, A140
\bibitem{brunetti03} Brunetti, G.\ 2003, in Matter and Energy in Clusters of Galaxies, ASP Conf. Series, 301, 349
\bibitem{brunetti04} Brunetti, G.\ 2004, Journal of Korean Astronomical Society, 37, 493
\bibitem{marchegianietal07} Marchegiani, P., Perola, G.~C., \& Colafrancesco, S.\ 2007, \aap, 465, 41
\bibitem{donnertetal10} Donnert, J., Dolag, K., Cassano, R., \& Brunetti, G.\ 2010, \mnras, 407, 1565
\bibitem{brunetti09} Brunetti, G.\ 2009, \aap, 508, 599
\bibitem{jeltemaprofumo11} Jeltema, T.~E., \& Profumo, S.\ 2011, \apj, 728, 53
\bibitem{brunetti12} Brunetti, G., Blasi, P., Reimer, O., et al.\ 2012, \mnras, 426, 956
\bibitem{wilson70} Willson, M.~A.~G.\ 1970, \mnras, 151, 1
\bibitem{venturi90} Venturi, T., Giovannini, G., \& Feretti, L.\ 1990, \aj, 99, 1381
\bibitem{thierbach03} Thierbach, M., Klein, U., \& Wielebinski, R.\ 2003, \aap, 397, 53
\bibitem{brunettietal13} Brunetti, G., Rudnick, L., Cassano, R., et al.\ 2013, \aap, 558, A52
\bibitem{schlickeiser87} Schlickeiser, R., Sievers, A., \& Thiemann, H.\ 1987, \aap, 182, 21
\bibitem{kale10} Kale, R., \& Dwarakanath, K. S. \ 2010, \apj, 718, 939
\bibitem{vanweerenlofar12} van Weeren, R.~J., R{\"o}ttgering, H.~J.~A., Rafferty, D.~A., et al.\ 2012, \aap, 543, A43
\bibitem{macarioetal10} Macario, G., Venturi, T., Brunetti, G., et al.\ 2010, \aap, 517, A43
\bibitem{macarioetal13} Macario, G., Venturi, T., Intema, H.~T., et al.\ 2013, \aap, 551, A141 
\bibitem{cassanobrunettisetti06} Cassano, R., Brunetti, G., \& Setti, G.\ 2006, \mnras, 369, 1577
\bibitem{cassano12} Cassano, R., Brunetti, G., Norris, R.~P., et al.\ 2012, \aap, 548, A100
\bibitem{zandanel13} Zandanel, F., Pfrommer, C., \& Prada, F.\ 2013, arXiv:1311.4795
\bibitem{cassano10b} Cassano, R., Brunetti, G., R{\"o}ttgering, H.~J.~A., \& Br{\"u}ggen, M.\ 2010, \aap, 509, A68
\bibitem{ensslinrottgering02} En{\ss}lin, T.~A., R\"ottgering, H.\ 2002, \aap, 396, 83
\bibitem{cassanogittibrunetti08} Cassano, R., Gitti, M., \& Brunetti, G.\ 2008, \aap, 486, L31
\bibitem{murgiaetal09} Murgia, M., Govoni, F., Markevitch, M., et al.\ 2009, \aap, 499, 679
\bibitem{gittibrunettisetti02} Gitti, M., Brunetti, G., \& Setti, G.\ 2002, \aap, 386, 456
\bibitem{petersonfabian06} Peterson, J.~R., \& Fabian, A.~C.\ 2006, \physrep, 427, 1
\bibitem{markevitchvikhlinin07} Markevitch, M., \& Vikhlinin, A.\ 2007, \physrep, 443, 1
\bibitem{fujita04} Fujita, Y., Matsumoto, T., \& Wada, K.\ 2004, \apjl, 612, L9
\bibitem{askasibarmarkevitch06} Ascasibar, Y., \& Markevitch, M.\ 2006, \apj, 650, 102
\bibitem{zuhone10} ZuHone, J.~A., Markevitch, M., \& Johnson, R.~E.\ 2010, \apj, 717, 908
\bibitem{mazzottagiacintucci08} Mazzotta, P., \& Giacintucci, S.\ 2008, \apjl, 675, L9
\bibitem{zuhone13} ZuHone, J.~A., Markevitch, M., Brunetti, G., \& Giacintucci, S.\ 2013, \apj, 762, 78
\bibitem{sandersetal10} Sanders, J. S., Fabian, A. C., Smith, R. K. \& Peterson, J. R.\ 2010, \mnras, 402, L11 
\bibitem{bulbuletal12} Bulbul, G. E., Smith, R. K., Foster, A., Cottam, J., Loewenstein, M., Mushotzky, R. \& Shafer, R. \ 2012, \apj, 747, 32, 2012
\bibitem{colafrancescomarchegiani08} Colafrancesco, S., \& Marchegiani, P.\ 2008, \aap, 484, 51
\bibitem{guooh08} Guo, F., \& Oh, S.~P.\ 2008, \mnras, 384, 251
\bibitem{fujitaohira12} Fujita, Y., \& Ohira, Y.\ 2012, \apj, 746, 53
\bibitem{fujitaohira13} Fujita, Y., \& Ohira, Y.\ 2013, \mnras, 428, 599
\bibitem{rossettietal13} Rossetti, M., Eckert, D., De Grandi, S., et al.\ 2013, \aap, 556, A44
\bibitem{bagchi06} Bagchi, J., Durret, F., Neto, G.~B.~L., \& Paul, S.\ 2006, Science, 314, 791
\bibitem{roettiger99b} Roettiger, K., Burns, J.~O., \& Stone, J.~M.\ 1999, \apj, 518, 603
\bibitem{ensslingopal01} En{\ss}lin, T.~A., \& Gopal-Krishna 2001, \aap, 366, 26 
\bibitem{hoeft08} Hoeft, M., Br{\"u}ggen, M., Yepes, G., Gottl{\"o}ber, S., \& Schwope, A.\ 2008, \mnras, 391, 1511
\bibitem{skillman11} Skillman, S.~W., Hallman, E.~J., O'Shea, B.~W., et al.\ 2011, \apj, 735, 96
\bibitem{markevitchetal05} Markevitch, M., Govoni, F., Brunetti, G., \& Jerius, D.\ 2005, \apj, 627, 733
\bibitem{pinzkeetal13} Pinzke, A., Oh, S.~P., \& Pfrommer, C.\ 2013, \mnras, 435, 1061
\bibitem{bonafede12} Bonafede, A., Br{\"u}ggen, M., van Weeren, R., et al.\ 2012, \mnras, 426, 40
\bibitem{clarke06} Clarke, T.~E., \& Ensslin, T.~A.\ 2006, \aj, 131, 2900
\bibitem{giacintucci08} Giacintucci, S., Venturi, T., Macario, G., et al.\ 2008, \aap, 486, 347
\bibitem{fino10} Finoguenov, A., Sarazin, C.~L., Nakazawa, K., Wik, D.~R., \& Clarke, T.~E.\ 2010, \apj, 715, 1143
\bibitem{macario11} Macario, G., Markevitch, M., Giacintucci, S., et al.\ 2011, \apj, 728, 82
\bibitem{akamatsuetal12} Akamatsu, H.,Takizawa, M., Nakazawa, K., et al.\ 2012, \pasj, 64, 67
\bibitem{akamatsukavahara13} Akamatsu, H., \& Kawahara, H.\ 2013, \pasj, 65, 16
\bibitem{bourdin13} Bourdin, H., Mazzotta, P., Markevitch, M., Giacintucci, S., \& Brunetti, G.\ 2013, \apj, 764, 82
\bibitem{ogreanshock13} Ogrean, G.~A., Br{\"u}ggen, M., van Weeren, R.~J., et al.\ 2013, \mnras, 433, 812
\bibitem{owersetal13} Owers, M.~S., Nulsen, P.~E.~J., Couch, W.~J., et al.\ 2014, \apj, 780, 163
\bibitem{vanweeren12} van Weeren, R.~J., R{\"o}ttgering, H.~J.~A., Intema, H.~T., et al.\ 2012, \aap, 546, A124
\bibitem{stroe13} Stroe, A., van Weeren, R.~J., Intema, H.~T., et al.\ 2013, \aap, 555, A110
\bibitem{ryu09} Ryu, D., \& Kang, H.\ 2009, \apss, 322, 65
\bibitem{vazzarelics12} Vazza, F., Br{\"u}ggen, M., van Weeren, R., et al.\ 2012, \mnras, 421, 1868
\bibitem{skillman13} Skillman, S.~W., Xu, H., Hallman, E.~J., et al.\ 2013, \apj, 765, 21 
\bibitem{vazzabruggen13} Vazza, F., \& Br{\"u}ggen, M.\ 2014, \mnras, 437, 2291
\bibitem{ensslin99} En{\ss}lin, T.~A., Lieu, R., \& Biermann, P.~L.\ 1999, \aap, 344, 409
\bibitem{blasi00} Blasi, P.\ 2000, \apjl, 532, L9
\bibitem{dogiel00} Dogiel, V.~A.\ 2000, \aap, 357, 66
\bibitem{petrosianbykovrephaeli08} Petrosian, V., Bykov, A., \& Rephaeli, Y.\ 2008, \ssr, 134, 191 
\bibitem{rephaeli79} Rephaeli, Y.\ 1979, \apj, 227, 364
\bibitem{rephaeli99} Rephaeli, Y., Gruber, D., \& Blanco, P.\ 1999, \apjl, 511, L21
\bibitem{rephaeli02} Rephaeli, Y., \& Gruber, D.\ 2002, \apj, 579, 587
\bibitem{fusco99} Fusco-Femiano, R., dal Fiume, D., Feretti, L., et al.\ 1999, \apjl, 513, L21
\bibitem{fusco04} Fusco-Femiano, R., Orlandini, M., Brunetti, G., et al.\ 2004, \apjl, 602, L73
\bibitem{rossetti04} Rossetti, M., \& Molendi, S.\ 2004, \aap, 414, L41
\bibitem{fusco07} Fusco-Femiano, R., Landi, R., \& Orlandini, M.\ 2007, \apjl, 654, L9
\bibitem{eckert07} Eckert, D., Neronov, A., Courvoisier, T.~J.-L., \& Produit, N.\ 2007, \aap, 470, 835
\bibitem{lutovinov08} Lutovinov, A.~A., Vikhlinin, A., Churazov, E.~M., Revnivtsev, M.~G., \& Sunyaev, R.~A.\ 2008, \apj, 687, 968
\bibitem{wik09} Wik, D.~R., Sarazin, C.~L., Finoguenov, A., et al.\ 2009, \apj, 696, 1700
\bibitem{wik11} Wik, D.~R., Sarazin, C.~L., Finoguenov, A., et al.\ 2011, \apj, 727, 119
\bibitem{fusco11} Fusco-Femiano, R., Orlandini, M., Bonamente, M., \& Lapi, A.\ 2011, \apj, 732, 85
\bibitem{eckert08} Eckert, D., Produit, N., Paltani, S., Neronov, A., \& Courvoisier, T.~J.-L.\ 2008, \aap, 479, 27
\bibitem{ajello09} Ajello, M., Rebusco, P., Cappelluti, N., et al.\ 2009, \apj, 690, 367
\bibitem{petrosian06} Petrosian, V., Madejski, G., \& Luli, K.\ 2006, \apj, 652, 948
\bibitem{ajello10} Ajello, M., Rebusco, P., Cappelluti, N., et al.\ 2010, \apj, 725, 1688
\bibitem{fuscofemiano00} Fusco-Femiano, R., Dal Fiume, D., De Grandi, S., et al.\ 2000, \apjl, 534, L7
\bibitem{rephaeli03} Rephaeli, Y., \& Gruber, D.\ 2003, \apj, 595, 137
\bibitem{nevalainen04} Nevalainen, J., Oosterbroek, T., Bonamente, M., \& Colafrancesco, S.\ 2004, \apj, 608, 166
\bibitem{nustar13} Wik, D.~R., Hornstrup, A., Molendi, S., et al.\ 2013, AAS/High Energy Astrophysics Division, 13, \#401.0
\bibitem{nakazawaetal09} Nakazawa, K., Sarazin, C.~L., Kawaharada, M., et al.\ 2009, \pasj, 61, 339 
\bibitem{finoguenovetal10}  Finoguenov, A., Sarazin, C.~L., Nakazawa, K., Wik, D.~R., \& Clarke, T.~E.\ 2010, \apj, 715, 1143 
\bibitem{melanie04} Johnston-Hollitt, M.\ 2004, in The Riddle of Cooling Flows in Galaxies and Clusters of galaxies, ed. T. H. Reiprich, J. C. Kempner \& N. Soker, 51 
\bibitem{pfrommerensslin04G} Pfrommer, C., \& En{\ss}lin, T.~A.\ 2004, \aap, 413, 17
\bibitem{gabiciblasi04} Gabici, S., \& Blasi, P.\ 2004, Astroparticle Physics, 20, 579 
\bibitem{keshet03} Keshet, U., Waxman, E., Loeb, A., Springel, V., \& Hernquist, L.\ 2003, \apj, 585, 128
\bibitem{kushnirwaxman09} Kushnir, D., \& Waxman, E.\ 2009, \jcap, 8, 2 
\bibitem{aharonian02} Aharonian, F.~A., Belyanin, A.~A., Derishev, E.~V., Kocharovsky, V.~V., \& Kocharovsky, V.~V.\ 2002, \prd, 66, 023005
\bibitem{rordford04} Rordorf, C., Grasso, D., \& Dolag, K.\ 2004, Astroparticle Physics, 22, 167
\bibitem{inoue05} Inoue, S., Aharonian, F.~A., \& Sugiyama, N.\ 2005, \apjl, 628, L9
\bibitem{vannoni11} Vannoni, G., Aharonian, F.~A., Gabici, S., Kelner, S.~R., \& Prosekin, A.\ 2011, \aap, 536, A56
\end{thebibliography}
\end{document}